# Solitons in Nonlinear Lattices

Yaroslav V. Kartashov,[1] Boris A. Malomed,[2] and Lluis Torner[1]

[1]*ICFO-Institut de Ciencies Fotoniques, and Universitat Politecnica de Catalunya, Mediterranean Technology Park, 08860 Castelldefels (Barcelona), Spain*

[2]*Department of Physical Electronics, School of Electrical Engineering, Faculty of Engineering, Tel Aviv University, Tel Aviv, 69978, Israel*

## Abstract

This article offers a comprehensive survey of results obtained for solitons and complex nonlinear wave patterns supported by purely nonlinear lattices (NLs), which represent a spatially periodic modulation of the local strength and sign of the nonlinearity, and their combinations with linear lattices. A majority of the results obtained, thus far, in this field and reviewed in this article are theoretical. Nevertheless, relevant experimental settings are surveyed too, with emphasis on perspectives for implementation of the theoretical predictions in the experiment. Physical systems discussed in the review belong to the realms of nonlinear optics (including artificial optical media, such as photonic crystals, and plasmonics) and Bose-Einstein condensation (BEC). The solitons are considered in one, two, and three dimensions (1D, 2D, and 3D). Basic properties of the solitons presented in the review are their existence, stability, and mobility. Although the field is still far from completion, general conclusions can be drawn. In particular, a novel fundamental property of 1D solitons, which does not occur in the absence of NLs, is a finite *threshold value* of the soliton norm, necessary for their existence. In multidimensional settings, the *stability* of solitons supported by the spatial modulation of the nonlinearity is a truly challenging problem, for the theoretical and experimental studies alike. In both the 1D and 2D cases, the mechanism which creates solitons in NLs is principally different from its counterpart in linear lattices, as the solitons are created *directly*, rather than bifurcating from Bloch modes of linear lattices.

## List of acronyms

1D – one-dimensional/one dimension
2D – two-dimensional/two dimensions

3D – three-dimensional/three dimensions

BEC – Bose-Einstein condensate/condensation

DNLSE – discrete nonlinear Schrödinger equation

FF – fundamental frequency

FR – Feshbach resonance

fs – femtosecond

GPE – Gross-Pitaevskii equation

GS – gap soliton

GVD – group-velocity dispersion

KP – Kronig-Penney (the piece-wise constant periodic modulation profile)

LB – light bullet (spatiotemporal optical soliton)

NL – nonlinear lattice

NLSE – nonlinear Schrödinger equation

OL – optical lattice

PCF – photonic-crystal fiber

SBB – symmetry-breaking bifurcation

SBN –Strontium-Barium Niobate (a photorefractive crystal)

SH – second harmonic

TE/TM – transverse-electric/magnetic (polarization modes of electromagnetic waves)

VA – variational approximation

VK – Vakhitov-Kolokolov (the stability criterion)

XPM – cross-phase modulation

# List of contents

# I. Introduction
## A. The subject of the review

The term *soliton*, i.e., a stable solitary wave propagating in a nonlinear medium, was coined by Zabusky and Kruskal about 50 years go. These authors were not the first to notice the remarkable properties of solitary waves, whose first known description in the scientific literature, in the form of "a large solitary elevation, a rounded, smooth and well defined heap of water", goes back to the historical observation made in a canal near Edinburgh by John Scott Russell in the 1830s. In the course of the nearly five decades that have elapsed since the publication of the paper by Zabusky and Kruskal (1965), the theoretical and experimental studies of solitary waves have seen an astonishing proliferation and penetration into many branches of science, from applied mathematics and physics to chemistry and biology. Several celebrated equations, in both their canonical and extended forms, emerge as universal models of solitons. These include Korteweg – de Vries and modified Korteweg – de Vries, nonlinear Schrödinger (with two opposite signs of the nonlinearity), sine-Gordon, Landau-Lifshitz, Kadomtsev-Petviashvili (of types I and II) and several other classical equations. The specific features of the evolution and interactions of solitons in these models are intimately related to the integrability of the above-mentioned equations. Diverse factors that in practice often break the integrability should be taken into regards, which naturally leads to the perturbation theory for solitons in nearly integrable system (Kivshar and Malomed, 1989).

A very significant contribution to the experimental and theoretical studies of solitons was the identification of various forms of robust solitary waves in nonlinear optics. Here we concentrate on *bright* solitons, which emerge as solitary pulses and/or beams. Optical solitons may be naturally subdivided into three broad categories – temporal, spatial, and spatiotemporal ones. They may exist in the form of one-dimensional (1D) or multi-dimensional objects. One-dimensional temporal solitons in optical fibers with a cubic (Kerr) nonlinearity were predicted by Hasegawa and Tappert (1973), and observed experimentally by Mollenauer, Stolen, and Gordon (1980), while stable self-trapping of light in the spatial domain was first observed in planar waveguides by Maneuf, Desailly, and Froehly (1988) [see also the paper by Maneuf and Reynaud (1988)]. Spatial two-dimensional (2D) solitary waves were first observed in photorefractive crystals, which feature a saturable nonlinearity [Duree et al., (1993)], and in optical media with a quadratic nonlinearity [Torruellas et al. (1995)]. Effectively two-dimensional spatio-temporal self-trapping of light into quasi-soliton objects was observed by Liu, Qian, and Wise (1999) also in quadratic nonlinear media.

Stable fully three-dimensional (3D) solitons, or *light bullets* (LBs) in quadratic media were predicted almost three decades ago (Kanashov and Rubenchik, (1981). However, to date, experimental generation of such long-lived 3D solitons remains elusive. In a landmark observation, the signature of 3D soliton formation was achieved recently by Minardi *et al.* (submitted) in an artificial optical material with cubic nonlinearity. Another species of robust solitary waves in optics occurs in the form of *gap solitons* (GSs), that are supported by the interplay of an appropriate lattice structure (alias *grating*), embedded into an optical medium, and nonlinearity. The observation of the first optical GSs in fiber Bragg gratings was reported by Eggleton et al. (1996).

A milestone achievement of modern physics, the creation of Bose-Einstein condensates (BECs) in ultracold vapors of alkali metals [Anderson et al. (1995); Bradley, Sackett, Tollett, and Hulet (1995); Davis et al. (1995)], was shortly followed by the creation of dark solitons of matter waves in BEC with repulsion between atoms (Burger et al., 1999) and, eventually, by the creation of bright 1D matter-wave solitons in BEC with attractive interatomic interactions [Strecker, Partridge, Truscott, and Hulet (2002); Khaykovich et al. (2002)]. This was followed by the generation of one-dimensional GSs in condensates with repulsive interactions between atoms loaded into a periodic potential induced by an optical lattice (OL), i.e., the pattern created by the interference of counter-propagating coherent laser beams illuminating the condensate (Eiermann et al., 2004).

Today experimental and theoretical studies of solitons remain an active field of research in several branches of science. A large part of this work is concentrated in the two above-mentioned fields, namely, nonlinear optics (light waves) and BEC (matter waves). There is a large gap between the theoretical and experimental studies in this area, with the theory going ahead. Experimental challenges are most often associated to the unavailability of material or metamaterial (artificially created) settings with suitable intrinsic or extrinsic properties – especially, in multidimensional geometries, where solitons supported by the common cubic nonlinearity are prone to severe instabilities. Methods for the stabilization of multidimensional solitons have been elaborated in detail theoretically, as reviewed several years ago by Malomed, Mihalache, Wise, and Torner (2005). Techniques that allow the stabilization of various species of multidimensional solitons rely on the use of settings of two types: periodic potentials, similar to those induced by the above-mentioned OLs in BEC (including multidimensional OLs), and the so-called "management", i.e., the application of external fields periodically varying in time, or the passage of solitons through periodically nonuniform media. Management methods (as well as their combination with lattices) were reviewed in the book by Malomed (2006). Studies of one- and multidimensional solitons in periodic potentials have grown into an active and large research area since seminal work of Christodoulides and Joseph (1988), who inaugurated the field of optical solitons in discrete periodic systems, and first experimental observation of optical discrete solitons in fabricated waveguide arrays performed by Eisenberg et al (1998). In the course of last decade, hundreds of original theoretical and experimental papers and several reviews have been published on the topic of discrete and continuous lattice solitons. Many results obtained in this field have been recently summarized in the reviews by Lederer *et al.* (2008), and by Kartashov, Vysloukh, and Torner (2009).

While nonlinearity is of course necessary for the existence of solitons, the usual periodic potentials represent linear ingredients of the respective physical systems. A new direction in the studies of solitons aims to predict their existence, stability, and dynamics in *nonlinear lattices* (NLs). The NLs represent spatially periodic patterns of modulation of the local strength of the nonlinearity, and in many cases they may act in a combination with usual linear lattices. In BEC, one- and multidimensional NLs may be induced by the application of spatially periodic external fields that induce an accordingly patterned modulation of the local nonlinearity through the *Feshbach-resonance* (FR) mechanism, i.e., field-induced changes of the scattering length characterizing binary collisions between atoms, which induce the nonlinearity in the BEC. In optical media, NLs may be built as material structures, represented by spatially periodic modulations of the local Kerr coefficient, or coefficients characterizing other types of the optical nonlinearity. In terms of the approach which treats solitons as quasi-particles, linear lattices give rise to the corresponding effective spatially periodic potentials. The action of the NLs may also be described in terms of an effective potential, which, however, is intrinsically nonlinear. Such effective nonlinear potentials are often called *pseudopotentials* in condensed-matter physics (Harrison, 1966).

Intensive studies of the dynamics of solitons in NLs had started only recently. However, in the course of the past five years many theoretical results have been reported. It should be stressed that the studies of the dynamics in NLs give rise to new problems, which, in many cases, are challenging in comparison with formally similar problems considered a few years earlier (and, in some cases, realized experimentally) in linear lattices. In particular, a salient feature of this topic is that it is difficult (although possible) to find the conditions for the stabilization of 2D solitons by means of NL pseudopotentials [Sivan, Fibich, and Weinstein, (2006); Sakaguchi and Malomed (2006a); Kartashov et al., (2009a); Hung, Zin, Trippenbach, and Malomed (2010)].

As mentioned above, thus far the progress in the study of the soliton formation in NLs has been made primarily in theoretical studies. As concerns the experiment, a setting which may be described, to a certain extent, as a combination of linear and nonlinear lattices, and in which solitons have been created and studied in detail, is represented by *photoinduced* *lattices* in photorefractive crystals [Efremidis et al. (2003); Neshev et al., (2003); Fleischer et al. (2003a, 2003b, and 2004); Martin et al., (2004); Neshev et al. (2004); Chen et al. (2004 and 2005); Cohen et al. (2005); Wang et al. (2007a and 2007b); Alfassi et al. (2007)]. A very promising medium for the formation of combinations of linear and nonlinear lattices is provided by *photonic-crystal fibers* (PCFs). The use of NLs in BEC, especially in 2D and 3D configurations, may also become an important tool in experimental studies of solitons. In that connection, it is necessary to stress that, thus far, no examples of self-supporting multidimensional matter-wave solitons have been reported – primarily because of difficulties with the stabilization of such solitons in non-1D settings. The above-mentioned works by Sivan, Fibich, and Weinstein (2006), Sakaguchi and Malomed (2006a), Kartashov et al., (2009a), and Hung, Zin, Trippenbach, and Malomed (2010) actually predict that the use of effective NLs induced in BEC may provide a novel mechanism for the stabilization of multidimensional matter-wave solitons.

## B. Objectives and structure of the review

One objective of this review is to summarize theoretical and, whenever possible, experimental results obtained in the field of NLs for fundamental solitons and more complex nonlinear-wave patterns, such as vortices and periodic waves. The ultimate purpose of this part of the review is to formulate generic features of the solitons and nonlinear patterns in these settings, highlighting novelties revealed by the analysis, in comparison to previously studied nonlinear media. We present the findings for the solitons in one and two dimensions; some results are also reported for 3D models. The existence, stability, and mobility of the solitons are considered in continuous, discrete, and semi-discrete media. Together with NLs, the review includes the dynamics of solitons in combined nonlinear and linear lattices and some related settings, such as the spontaneous symmetry breaking in nonlinear pseudopotentials of the double-well type.

In the description of the particular physical problems comprised in this review, and results produced by the theoretical analysis of the problems, we include both the summary of the results, and a description of the core part of the technical analysis, in those cases when the techniques (analytical and/or numerical ones) may be useful for the consideration of similar problems, in the same or other problems. Different technical items included into the review can be read, in most cases, independently from each other. This is for the benefit of those readers who may be interested in the information about particular problems. Nevertheless, all sections and subsections are linked throughout the review. The list of contents is given in a sufficiently detailed form, so as to help the interested reader in finding results on particular topics.

The bulk of the theoretical results, including the most fundamental issues, are presented in Sections IV and V, which deal with 1D and 2D settings, respectively. In each of these two basic sections, we start the presentation with summaries of *core results* and *core techniques*, which are dealing with the most fundamental systems belonging to the realm of NL models. Then, in each section we add more special and/or straightforward results which are physically relevant too, in the respective contexts. More specific topics are considered separately in Sections VI, VII, and VIII.

Another part of the review, presented in Section III, contains a description of potentially relevant experimental settings, since another major objective of this article is to motivate experimental implementation of basic predictions revealed by the theoretical analysis. Those experimental results on the topic of solitons in NLs, which have already been published, are represented in particular sections and subsections in the review, which are dealing with the respective physical settings (an example is the experimental creation of solitons in PCF infiltrated with an index-matching liquid, see subsection V.B.2). In the concluding Section IX we try to single out those new theoretical predictions whose implementation in the experiment seems most plausible, in the short run. We also indicate the predicted effects which are more challenging to the experiment. In the same section, perspectives for the further development of the theoretical analysis in this field are briefly discussed too. Some particular theoretical problems which still have to be tackled are also discussed in sections of the review dealing with topic in which these new problems emerge.

## II. Basic models

The consideration of physical settings that give rise to NLs and wave patterns linked to them, in optics, nanophotonics and BEC, make it possible to identify a few fundamental models. These models are actually *universal* ones, as they are relevant to all these physical systems, in the 1D and 2D geometries alike (in some cases, they may also be extended to 3D). In most cases, the models amount to extended versions of the celebrated and ubiquitous nonlinear Schrödinger equation (NLSE), with various additional terms, explicit spatial and/or temporal dependencies of the nonlinear coefficients, and in different dimensions. This fact is essentially responsible for the possibility to identify a few key models that play the universal role in this field. Multicomponent settings are described, accordingly, by the coupled systems of the NLSEs.

These basic models are introduced in the present section. Experimental realizations of the models are considered in Section III. In subsequent parts of the review, it will be demonstrated that main types of NL-supported wave patterns can be found, and their dynamics can be analyzed, within the framework of these basic models. It will be demonstrated too what particular features of the models are crucially important for the stability of solitons and other patterns supported by them (in particular, as concerns the challenging issue of the stability of 2D solitons, the factor crucial to the stability of the solitons is the *sharpness* of the NL).

## A. Optics
### 1. Models with permanent material lattices

Two basic types of physical systems which constitute the subject of the review are transversally inhomogeneous nonlinear optical media, and Bose-Einstein condensates (BECs) in which an effective inhomogeneity is induced by external fields. The fundamental equation which governs the transmission of electromagnetic waves in dispersive nonlinear media is the nonlinear Schrödinger equation (NLSE), which is derived from the full wave equation (that, in turn, can be obtained from the underlying system of the Maxwell's equations, combined with material equations that account for the nonlinearity and inhomogeneity of the medium). The derivation is based on the assumption that the wave can be factorized into a rapidly varying monochromatic carrier and a slowly varying envelope amplitude, which is a function of time and coordinates with the characteristic scales much larger than, respectively, the temporal period and wavelength of the carrier wave. The derivation of the NLSE in this context, including the nonlinear contribution from the Kerr effect and material group-velocity dispersion (GVD), was first developed by Hasegawa and Tappert (1973). The main result of their analysis was the prediction of temporal solitons in nonlinear optical fibers featuring the anomalous GVD. A consistent derivation of the NLSE for the propagation of optical waves in both optical and spatial, as well as spatiotemporal, domains can be found in several books [see, e.g., Agrawal (1995) and Kivshar and Agrawal (2003)].

For the purposes of the present review, the most relevant variety of the NLSE in optics is the one in the *spatial domain*, which assumes that the electromagnetic wave is strictly monochromatic in terms of the frequency, but allows the wave's amplitude, $q(x,y,z)$, to be a slowly varying function of the propagation distance, $z$, and the transverse coordinates, $x,y$, in the general case of the propagation in the bulk medium. In physical units, the corresponding (2+1)-dimensional nonlinear Schrödinger equation takes the following form:

$$i\frac{\partial q}{\partial z} + \frac{\lambda_0}{4\pi}\left(\frac{\partial^2 q}{\partial x^2} + \frac{\partial^2 q}{\partial y^2}\right) + \frac{2\pi}{n_0\lambda_0}[\delta n(x,y,z) + n_2(x,y,z)|q|^2]q = 0, \qquad (1)$$

where $\lambda_0$ is the carrier wavelength, $n_0$ is the background value of the refractive index, and $\delta n(x,y,z)$ is a local perturbation of the refractive index, which accounts for the optical inhomogeneity of the medium [in other words, $\delta n(x,y,z)$ represents the linear grating (or lattice) written in the medium to control the linear transmission of the optical beams in it]. Further, $n_2(x,y,z)$ in Eq. (1) is the Kerr coefficient, the spatial dependence of which eventually implies the existence of the *nonlinear lattice* (NL) in the inhomogeneous optical medium, which is the central theme of this review.

In the case of the long-scale periodic modulation of the linear and nonlinear refractive indices along the propagation distance ("long-scale" implies a modulation period much larger than $\lambda_0$), the gratings may be used to control the transmission of beams by means of "management" mechanisms (a survey of that topic was given in the book by Malomed, 2006). In particular, the transmission of the cylindrical beam in the medium built as a periodic concatenation of transversally uniform self-focusing and self-defocusing layers, with $n_2(z)$ periodically jumping between positive and negative values, is described, in a scaled form, by the following version of Eq. (1):

$$i\frac{\partial q}{\partial z} + \frac{1}{2}\left(\frac{\partial^2 q}{\partial x^2} + \frac{\partial^2 q}{\partial y^2}\right) + n_2(z)|q|^2 q = 0. \qquad (2)$$

A commonly known fact is that Eq. (2) with $n_2 \equiv 1$ gives rise to axially symmetric *Townes solitons*, in the form of $q(x,y,z) = w(r)\exp(ibz)$, with $r \equiv (x^2+y^2)^{1/2}$, arbitrary $b > 0$, and real function $w(r)$. The total power (i.e., norm) of the Townes solitons does not depend on $b$, being $N_{\text{Townes}} = 2\pi\int_0^\infty w^2(r)rdr \approx 5.85$ [a simple variational approximation, developed by Desaix, Anderson, and Lisak (1991), yields $N_{\text{Townes}} = 2\pi$]. While the entire family of the Townes solitons is *unstable* due to the possibility of the collapse in the two-dimensional NLSE with the focusing cubic nonlinearity (Bergé, 1998), it can be shown that the application of the *nonlinearity-management* mode, represented by the piecewise-constant $n_2(z)$ in Eq. (2), which periodically changes the sign, produces stable periodically pulsating fundamental axisymmetric solitons for positive average values of $n_2$ [Towers and Malomed (2002)]. A similar layered setting was implemented in the experiment by Centurion, Porter, Kevrekidis, and Psaltis (2006) and by Centurion et al., (2006). They had demonstrated a partial stabilization of solitary beams in a set of ten layers of silica alternating with empty gaps. Note that nonlinearity management is a particular application of the concept of the soliton formation in *tandem* material settings, i.e., layered media engineered so as to provide overall properties suitable for the formation of stable multi-dimensional solitons (Torner, Carrasco, Torres, Crasovan, Mihalache, 2001).

The subject of this review is the formation of solitons by means of transverse NLs. This means that we aim to consider models based on Eq. (1) with linear and nonlinear refractive indices periodically modulated in the transverse plane, while the medium remains uniform along the propagation direction. The accordingly modified normalized version of Eq. (1) is

$$i\frac{\partial q}{\partial z} + \frac{1}{2}\left(\frac{\partial^2 q}{\partial x^2} + \frac{\partial^2 q}{\partial y^2}\right) + [\delta n(x,y) + n_2(x,y)|q|^2]q = 0, \qquad (3)$$

where functions $\delta n(x,y)$ and $n_2(x,y)$ are periodic in both coordinates $x$ and $y$ (or, in a special case, only in one of them, being independent of the other; quasi-periodic and quasi-random generalizations of such structures are interesting too). The linear term in Eq. (3) proportional to $\delta n(x,y)$, represents the linear potential, while the nonlinear term may be regarded as an additional potential, in the form of $n_2(x,y)|q|^2$, which depends on the solution itself. As mentioned above, this type of the nonlinear potential function is often called a *pseudopotential* (Harrison, 1966).

A straightforward realization of the model based on Eq. (3) is possible in photonic-crystal fibers (PFCs), where both $\delta n(x,y)$ and $n_2(x,y)$ are determined by the transverse structure of the PCF. This actually implies that $\delta n(x,y)$ and $n_2(x,y)$ are piecewise-constant functions, with jumps at interfaces between the bulk medium (silica) and the material filling the lattice of voids running parallel to the axis of the fiber [the filling substance may be air, a fluid, or another kind of glass, in the case of *all-solid* PCF, as shown by Luan et al. (2004)].

Equation (1) derived for the light propagation in the bulk spatial domain has an effectively 2D form, with propagation coordinate $z$ playing the role of the evolutional variable. The reduction of the equation to a 1D model, which describes the transmission of beams in planar nonlinear waveguides, is straightforward, leading to the evolution equation for the light beam amplitude with a single transverse coordinate:

$$i\frac{\partial q}{\partial z} + \frac{1}{2}\frac{\partial^2 q}{\partial x^2} + [\delta n(x) + n_2(x)|q|^2]q = 0 \qquad (4)$$

As well as the 2D model, its 1D counterpart is relevant to the description of layered planar optical waveguides, where $\delta n(x)$ and $n_2(x)$ can be piecewise-constant functions. The periodic 1D modulation functions of this type are usually referred to as Kronig-Penney (KP) lattices. One-dimensional models based on the interplay of various types of linear and nonlinear KP lattices were studied in detail by Kominis (2006), Kominis and Hizanidis (2006 and 2008), Kominis, Papadopoulos, and Hizanidis (2007), and Mayteevarunyoo and Malomed (2008). Two-dimensional versions of the KP lattice, i.e., models featuring the 2D *checkerboard structure*, were elaborated too, by Maes, Bientsman, and Baets (2005), Driben *et al.* (2007), and Driben and Malomed (2008).

## 2. Models with photoinduced lattices

The above discussion was dealing with models of optical media in which linear and nonlinear lattices are created as permanent material structures. On the other hand, *virtual* optical lattices can be induced in photorefractive crystals as interference patterns, by illuminating the crystal with pairs of coherent laser beams in the ordinary polarization, for which the medium is nearly linear. Then, through the effect of the cross-phase modulation, the interference pattern induces an effective grating for the probe beam, launched in the extraordinary polarization, for which the photorefractive medium features a *saturable nonlinearity*. The full 2D equation for the amplitude of the probe beam in this setting is (see the papers by Efremidis *et al.*, 2002 and 2003):

$$i\frac{\partial q}{\partial z} + \frac{1}{2}\left(\frac{\partial^2 q}{\partial x^2} + \frac{\partial^2 q}{\partial y^2}\right) - \frac{E_0 q}{1+|q|^2 + R(x,y)} = 0, \qquad (5)$$

where function $R(x,y)$ describes the intensity distribution in the lattice-creating ordinarily polarized beams, and $E_0$ is the dc electric field which induces the saturable nonlinearity, with $E_0 > 0$ and $E_0 < 0$ corresponding to the focusing and defocusing nonlinearities, respectively. For example, $R(x,y) = I_0 \cos^2(x/L)\cos^2(y/L)$ in the practically important case of the photoinduced lattice produced by the interference of four plane waves with intensity $I_0$ and effective wavelength $L$.

In the limit case of the weak probe beam, the saturable nonlinearity in Eq. (5) may be expanded, which gives rise to the NLSE with the cubic nonlinearity and specific forms of the functions describing linear and nonlinear lattices in the respective model:

$$i\frac{\partial q}{\partial z}+\frac{1}{2}\left(\frac{\partial^2 q}{\partial x^2}+\frac{\partial^2 q}{\partial y^2}\right)-\frac{E_0 q}{1+R(x,y)}+\frac{E_0 |q|^2 q}{[1+R(x,y)]^2}=0. \qquad (6)$$

A peculiarity of the nonlinearity-modulation coefficient in Eq. (6) is that it cannot change its sign. Lastly, both equations (5) and (6) have their obvious 1D counterparts, which also apply to the description of various experimentally relevant settings, see reviews by Fleischer et al. (2005), Lederer et al. (2008), and by Kartashov, Vysloukh, and Torner (2009). Lastly, it is relevant to mention that basic elements of the model of photorefractive media outlined above were first introduced by Vinetskii and Kukhtarev (1974).

## B. Bose-Einstein condensates

The fundamental model which provides for an accurate description of the BEC in rarefied degenerate gases of bosonic atoms is the Gross-Pitaevskii equation (GPE) for the single-particle wave function, $\Psi$ ("degenerate" means that the de Broglie wavelength of atoms in the rarefied gas is comparable to the mean inter-atomic distance, see a detailed discussion in the book by Pitaevskii and Stringari, 2003). The 3D form of this equation is

$$i\hbar\frac{\partial \Psi}{\partial t}=-\frac{\hbar^2}{2m}\left(\frac{\partial^2 \Psi}{\partial x^2}+\frac{\partial^2 \Psi}{\partial y^2}+\frac{\partial^2 \Psi}{\partial z^2}\right)+U(x,y,z)\Psi+\frac{4\pi\hbar^2 N}{m}a_s(x,y,z)|\Psi|^2\Psi=0 \qquad (7)$$

where $m$ is the atomic mass, $U(x,y,z)$ is the external potential acting on individual atoms (it may depend on time too), $N$ is the total number of atoms in the condensate, and $a_s$ is the scattering length which determines collisions between the atoms, $a_s > 0$ and $a_s < 0$ corresponding to the repulsive and attractive interactions, respectively. The wave function is subject to the normalization condition, $\iiint |\Psi(x,y,z)|^2 dxdydz = 1$.

The action of the linear lattice is accounted for by the choice of a spatially periodic potential $U(x,y,z)$. Usually, the periodic potential is created as an *optical lattice*, OL, which is induced by the interference of coherent laser beams illuminating the condensate, as proposed by Jaksch et al. (1998) and demonstrated by Greiner et al. (2002). More recently, it was demonstrated by Ghanbari, Kieu, Sidorov, and Hannaford (2006), and by Abdelrahman, Hannaford, and Alameh (2009) that 2D and 1D periodic potentials may also be induced by *magnetic lattices*, created by a plate made of a permanent magnet with a lattice of holes drilled in it, or more sophisticated variants of this setting.

The NL should be induced by the introduction of a proper spatial inhomogeneity of the scattering length, therefore this coefficient is written as a function of coordinates in Eq. (7). The spatial (and, if relevant, temporal) dependence of $a_s$ may be induced by means of the *Feshbach-resonance (FR) management* technique. The direct control of the scattering length in BEC by the external magnetic field through the Feshbach resonance (this effect implies the formation of a quasi-bound state of two atoms in the course of the collision between them) was first demonstrated experimentally by Inouye et al. (1998). The spatial pattern of the FR management, which gives rise to the NL, may be induced by an appropriately structured magnetic field. Potentially, the above-mentioned magnetic lattices may also be used to induce an effective periodic NL in both 2D and 1D settings.

In addition to that, the FR can be controlled by optical fields with a properly tuned frequency, as predicted by Fedichev, Kagan, Shlyapnikov, and Walraven (1996), and demonstrated in the experiment by Theis et al. (2004), and also by electrostatic fields, according to the prediction by Marinescu and You (1998). Therefore, NLs can also be induced by means of properly patterned optical or dc electric fields.

Usually, the linear OL potentials and nonlinearity-modulation functions in BEC models, being produced as interference patterns, have a smooth sinusoidal profile. A number of particular physically relevant examples of such profiles are considered below in Sections IV-VI.

Actually, all examples of NLs which were studied, thus far, in BEC models pertain to 1D and 2D geometries. In particular, the nearly 1D setting corresponds to potential $U(x,y,z) = V(x) + (m/2)\Omega^2(y^2 + z^2)$ in GPE (7), where the strong transverse confinement is achieved due to the large trapping frequency, $\Omega$. Then, the reduction of the 3D equation (7) to the 1D limit is performed by means of the factorized *ansatz*,

$$\Psi(x,y,z,t) = \frac{1}{\pi^{1/2}\sigma} \exp\left(-\frac{i\hbar\Omega t}{2} - \frac{y^2 + z^2}{2\sigma^2}\right) q(x,t), \qquad (8)$$

where the transverse part is actually the ground-state's wave function of the isotropic harmonic-oscillator potential acting in the plane $(y,z)$, and $q(x,t)$ is an effective 1D wave function, which is subject to normalization $\int_{-\infty}^{+\infty} |q|^2 \, dx = 1$. The substitution of ansatz (8) into Eq. (7), and introduction of scaled variables, namely, the coordinates measured in units of the harmonic-oscillator length, $a_\perp = (\hbar/m\Omega)^{1/2}$, and time measured in units of $1/\Omega$, makes it possible to eliminate the transverse width, $\sigma$, in favor of the 1D density, $\sigma^4 = 1 + R(x)|q|^2$, where $R(x) \equiv 2a_s(x)N/a_\perp$, thus leading to the derivation of the 1D equation with the *nonpolynomial nonlinearity* (Salasnich, Parola, and Reatto, 2002; see also an alternative form of the nonpolynomial nonlinearity, derived by Muñoz Mateo and Delgado, 2008, which is relevant in the case of the repulsive interactions between atoms, $a_s > 0$):

$$i\frac{\partial q}{\partial t} = -\frac{1}{2}\frac{\partial^2 q}{\partial x^2} + V(x)q + \frac{1+(3/2)R(x)|q|^2}{[1+R(x)|q|^2]^{1/2}}q = 0. \quad (9)$$

Here, it is taken into regard that the scattering length may be a function of the longitudinal coordinate, $x$, which is necessary to introduce the NL, eventually. A noteworthy peculiarity of Eq. (9) is that, unlike the usual one-dimensional NLSE with the cubic nonlinearity, this equation admits the onset of the *wave collapse* in the 1D setting, in the case of the attractive nonlinearity $(R<0)$, thus keeping this important property of the full 3D version of the GPE. The dynamics of 1D solitons under the combined action of the OL potential and nonpolynomial nonlinearity was studied in detail by Salasnich, Cetoli, Malomed, and Toigo (2007).

In the limit of low density, $|Rq|^2 \ll 1$, the nonpolynomial nonlinearity in Eq. (9) may be expanded, thus casting this equation into the form of the usual one-dimensional NLSE with the cubic inhomogeneous nonlinearity,

$$i\frac{\partial q}{\partial t} = -\frac{1}{2}\frac{\partial^2 q}{\partial x^2} + V(x)q + R(x)|q|^2 q = 0 \quad (10)$$

which is tantamount to Eq. (4). The general form of the GPE reduced to two dimensions, in the case of the strong confinement in the transverse direction, is nonpolynomial too [see the recent paper by Salasnich and Malomed (2009) and references therein], but in the low-density limit it is similar to Eq. (3).

It is also possible to derive an effective 1D equation starting with the transverse wave function which corresponds not to the ground state of the 2D harmonic oscillator, but rather to its higher-order state with integer *vorticity*, $S \geq 1$ (Salasnich, Malomed, and Toigo, 2007). In this case, an additional factor must be added to the factorization of the 3D wave function in Eq. (8), *viz.*, $[(y^2+z^2)/\sigma^2]^{|S|/2}\exp(iS\theta)$, where $\theta$ is the azimuthal coordinate in the $(y,z)$ plane. Eventually, one arrives at the 1D equation in the same form as Eq. (9), but with $R$ replaced by $R_S \equiv (2S)![2^{2S}(S+1)(S!)^2]^{-1}R$, and with an additional factor, $(S+1)$, added in front of the nonpolynomial term.

Finally, it is relevant to mention that the reduction of the underlying three-dimensional GPE to 1D makes it possible to induce effective lattices without using addi-tional external fields, but rather making the trapping frequency a function of coordinate $x$, $\Omega = \Omega(x)$ [as proposed by De Nicola, Malomed, and Fedele (2006), and analyzed in detail by Salasnich *et al.* (2007)]. In particular, in the limit of the low density, the modulation of the trapping frequency induces the effective linear potential and the nonlinearity-modulation function [i.e., $\delta n(x)$ and $n_2(x)$, in terms of Eq. (4)] which are both proportional to $\Omega(x)$.

### C. Discrete systems

The models with NLs of the KP type include, as a limit case, the nonlinearity-modulation function in the form a periodic array (*comb*) of Dirac's delta-functions. The respective model, which was introduced by Sukhorukov and Kivshar (2002a and 2002b), has the form of Eq. (4) with $\delta n(x) = 0$ and $n_2(x) = n_{20} \sum_{m=-\infty}^{+\infty} \delta(x - Lm)$, where $n_{20}$ and $L$ are the strength and period of the respective NL, and $\delta(x)$ is the delta-function. Similar two-component *semi-discrete* systems were introduced by Panoiu, Malomed, and Osgood, 2008. A prototype of such models is the NLSE with the attractive (alias self-focusing) nonlinearity concentrated at a single point, i.e., Eq. (4) with $\delta n(x) = 0$ and $n_2(x) = n_{20} \delta(x)$. The latter model was proposed by Malomed and Azbel (1993) for the description of tunneling of interacting particles through a junction.

In the limit case of strong localization of light in narrow guiding channels, the models of the KP type may be also reduced to discrete nonlinear Schrödinger equations (DNLSEs), in various forms, the simplest among which is

$$i \frac{dq_n}{dz} + \frac{1}{2}(q_{n+1} + q_{n-1} - 2q_n) + |q_n|^2 q_n = 0. \tag{11}$$

It should be stressed that solitons in discrete systems governed by Eq. (11) were for the first time introduced by Christodoulides and Joseph (1988). Stationary solutions in the above-mentioned model with the nonlinearity represented by the comb of delta-functions can be mapped into usual solitons of the DNLSE (Sukhorukov and Kivshar, 2002a and 2002b). Also widely used is the 2D version of Eq. (11), i.e.,

$$i \frac{dq_{m,n}}{dz} + \frac{1}{2}(q_{m+1,n} + q_{m-1,n} + q_{m,n+1} + q_{m,n-1} - 4q_{m,n}) + |q_{m,n}|^2 q_{m,n} = 0. \tag{12}$$

DNLSEs and discrete solitons of diverse types generated by them (in particular, discrete vortices) in 1D, 2D, and 3D settings have grown into a large area of theoretical and experi-mental studies. Many results obtained in this area have been recently summarized in the book by Kevrekidis (2009).

An important physical application of Eq. (12) is its use as an accurate model for the light propagation in 2D arrays of parallel fiber-like waveguides fabricated in bulk samples of silica by means of the technique based on femtosecond (fs) pulses which write permanent guiding cores in silica (Szameit et al, 2006). Note that, unlike their continuous counterparts (3) and (4), the DNLSEs with opposite signs in front of the nonlinear terms may be transformed into each other by means of the *staggering transformation*, $q_n \to (-1)^n q_n$.

A discrete model for an array of parallel waveguides embedded into a Kerr medium was developed by Öster, Johansson, and Eriksson (2003), and by Öster and Johansson (2005), in the form of a one-dimensional DNLSE including not only on-site cubic terms, but also complex combinations of their inter-site counterparts (i.e., nonlinear terms providing for couplings between adjacent sites of the discrete lattice, see the corresponding discussion in Section IV.G). The inter-site nonlinearities give rise to various effects, such as an enhanced mobility of discrete solitons and the existence of stable twisted modes. Earlier, a similar model was proposed as a phenomenological one by Claude et al. (1993). More recently, DNLSEs with general combinations of nonlinear inter-site terms were considered by Smerzi and Trombettoni (2003) (in that work, a quantum counterpart of the system, in the form of an extended Bose-Hubbard model, was considered too), and by Abdullaev *et al.* (2008), as a model which originates, in the framework of the tight-binding approximation, from the description of BEC trapped in a combination of linear and nonlinear lattices. Belmonte-Beitia and Pelinovsky (2009) showed that starting from Eq. (4) and assuming a specific symmetry of the periodic linear potential and nonlinearity profile [i.e., $\delta n(-x) = \delta n(x)$ and $n_2(-x) = -n_2(x)$, where $\delta n, n_2$ have a common period] one can derive a one-dimensional DNLSE with the quintic on-site attractive nonlinearity and without inter-site coupling cubic terms [the reduction of the original GPE to an equation with an effective quintic nonlinearity under similar conditions was demonstrated by Sakaguchi and Malomed (2005a)]. On the other hand, if $n_2(x)$ is not antisymmetric, one still arrives at the simple DNLSE with the cubic on-site nonlinearity. Notice that solutions of the DNLSE with the cubic-quintic on-site nonlinearity were investigated by Carretero-González et al., (2006) and by Chong et al., (2009), in the 1D and 2D settings, respectively.

It is relevant to mention that the models with nonlinear inter-site coupling terms of the general type may be considered as variations of the *Salerno model*, which, in turn, was originally introduced, in the 1D case, as a combination of the integrable Ablowitz-Ladik equation and non-integrable equation (11) (Salerno, 1992):

$$i\frac{dq_n}{dt} + (q_{n+1} + q_{n-1})(1 + \lambda |q_n|^2) + 2|q_n|^2 q_n, \qquad (13)$$

where coefficient $\lambda$ accounts for the Ablowitz-Ladik coupling between adjacent sites. Here, the evolutional variable is denoted $t$, rather than $z$ [cf. Eq. (11)], as Eq. (13) is more relevant to the description of BEC. In fact, Eq. (13) is the crudest model for *dipolar condensates* trapped in a strong lattice potential, which feature the interplay of the contact interactions, accounted for by the on-site cubic term, and the long-range interaction between atomic dipoles aligned by an external magnetic field, that may be approximated by the nonlinear inter-site couplings [Li et al. (2005); Gomez-Gardeñes et al., (2006a and 2006b)].

Discrete solitons of Eq. (13) were studied in detail, for both cases of the like ($\lambda > 0$) and *competing* ($\lambda < 0$) contact and long-range interactions [see, respectively, the papers by Cai, Bishop, and Grønbech-Jensen (1996), and Gomez-Gardeñes et al., (2006a)]. In the latter case, the competition gives rise to new stable soliton species, such as cuspons (super-exponentially localized modes). Discrete solitons (including solitary vortices) were also studied in the 2D version of the Salerno model, with the like nonlinearities, by Christiansen et al. (1996), and in the case of competing nonlinearities, by Gomez-Gardeñes et al. (2006b).

As for the model with the nonpolynomial nonlinearity based on Eq. (9), its discrete version was introduced and studied in details by Maluckov et al. (2008), and Gligori et al. (2009). Discrete models with the quadratic nonlinearity, which describe the second-harmonic generation in arrays of waveguides, were introduced too, by T. Peschel, U. Peschel, and Lederer (1998), and by Darmanyan, Kobyakov, and Lederer (1998) (see section VII).

### D. Photonic nanostructures

The optical models considered above are based on the paraxial approximation, which assumes that the characteristic transverse size of the beams is much larger than the wavelength of light. In this approximation, the evolution of light field obeys equations of the NLSE type, such as Eq. (3), with the weak (*paraxial*) diffraction accurately represented by the transverse Laplacian. A different situation takes place in the case of the transmission of light through structures with the characteristic transverse sizes on the subwavelength scale, a typical example being an array of *nanowires*, with both the diameter of the guiding cores and separation between them being $\sim 100$ nanometers or even smaller. In this case, the NLSE models are irrelevant, and one should use the full system of the Maxwell's equations. A relevant example is provided by equations derived by Gorbach and Skryabin (2009) for the transverse-magnetic (TM) and transverse-electric (TE) modes in a planar-waveguide counterpart of the array of nonlinear nanowires. In these two cases, the equations for non-zero components of the complex electric and magnetic fields, **E** and **H**, and displacement **D** are, respectively,

$$\begin{aligned}&\partial^2 E_x / \partial z^2 - \partial^2 E_z / \partial x \partial z = -k^2 D_x, \\ &\partial^2 E_x / \partial x \partial z - \partial^2 E_z / \partial x^2 = +k^2 D_z, \\ &\partial H_y / \partial z = ikc\varepsilon_0 D_y;\end{aligned} \quad (14)$$

and

$$\begin{aligned}&\partial^2 E_y / \partial z^2 = -k^2 D_y, \\ &(k / c\varepsilon_0) H_x = +i\partial E_y / \partial z, \\ &(k / c\varepsilon_0) H_z = -i\partial E_y / \partial z,\end{aligned} \quad (15)$$

where $z$ and $x$ are the propagation and transverse coordinates, $\lambda, c$, and $\varepsilon_0$ are the wavelength, speed of light, and dielectric permittivity of vacuum, respectively, and $k = 2\pi / \lambda$. Both systems (14) and (15) for the TM and TE waves are supplemented by the nonlinear relation between the displacement and electric field:

$$\mathbf{D} = \varepsilon(x)\mathbf{E} + (1/2)\chi_3(x)[(\mathbf{EE}^*)\mathbf{E} + (1/2)(\mathbf{EE})\mathbf{E}^*], \quad (16)$$

where the transverse layered structure is described by the modulation of the permittivity, $\varepsilon(x)$, and it is assumed that $\chi_3(x) = (4/3)cn_2\varepsilon_0\varepsilon(x)$ is proportional to $\varepsilon(x)$.

Recently, another model for the description of plasmonic solitons in an array of metallic nanowires was developed by Ye, Mihalache, Hu, and Panoiu (2010). Unlike the approach outlined above, they aimed to reduce the effective model to a discrete form. In the general case, an extended form of the DNLSE was derived, which, in addition to the usual linear inter-site couplings – the same as in Eq. (11) – also includes couplings through $z$-derivatives of the discrete field. However, it was demonstrated that the extra coupling terms are negligible under physically relevant conditions, thus reducing the model to the usual form of the DNLSE.

Another model for subwavelength solitons trapped in a planar layered nanostructure was developed by Liu, Bartal, Genov, and Zhang (2007), in the form of an array of alternating metallic and dielectric strips, which represents a metamaterial. Unlike the model based on Eqs. (14)-(16), this system takes into regard the loss in the metallic components, therefore the respective transmission distance of nonlinear beams is finite (actually, it is short).

The models for the transmission of light in nanowire arrays outlined above were formulated in the spatial domain. Models describing the temporal and spatiotemporal transmission in the arrays were developed too, see papers by El-Ganainy *et al.* (2006) and Benton, Gorbach, and Skryabin (2008).

## III. Materials, experimental settings and findings

The availability and development of suitable materials and fabrication techniques for the generation of nonlinear or mixed linear-nonlinear lattices is a key ingredient for the advancement of the field. In this section we describe basic settings and materials where NLs may be created, sometimes in parallel with the modulation of the linear refractive index. We assume further progress in the experimental studies of solitons and related wave patterns supported by NLs is possible, first of all, in those experimental settings which are outlined in this section. As said above, thus far only a few experimental results have been reported on this topic, therefore the discussion of the relevant experimental systems is an essential part of the review, the intention being to highlight the possibilities for the progress of the work in this direction.

In particular, we discuss photonic crystals and PCFs, in which modern fabrication technologies allow the creation of arbitrary periodic (or aperiodic) structures, infiltration of crystal holes with suitable fluids offering additional options for the design of NL structures. For reviews on photonic crystals and PCFs, see Yablonovitch, 1994; Yablonovitch, 2001; Russell, 2003; Dudley, Genty, and Coen, 2006. We also discuss the following techniques and settings which offer a real potential for the advancement of the experiments on NLs: (i) writing waveguide arrays in transparent materials by means of fs laser pulses, which cause irreversible damage of the material, accompanied by simultaneous increase of the local refractive index and decrease of the nonlinearity coefficient, resulting in the appearance of out-of-phase linear and nonlinear lattices, (ii) optical induction, i.e., fabrication of lattices in photorefractive materials, where the modulation of the nonlinearity is achieved by the application of an inhomogeneous background illumination, or by the indiffusion of different dopants into the sample (recent reviews on properties of solitons in linear lattices imprinted into such media were given by Lederer et al., 2008, and by Kartashov, Vysloukh, and Torner, 2009a); and (iii) liquid crystals, where the application of a spatially inhomogeneous external voltage notably modifies both the linear refractive index and nonlinearity coeffi-cient. Finally, we also address BECs, where the FR can be used for the spatial modulation of the sign of the interatomic interactions and the creation of NLs for matter-wave excitations (reviews on matter-wave solitons in linear lattices were given by Brazhnyi and Konotop, 2004, and by Morsch and Oberthaler, 2006; applications of the FR to BEC were recently reviewed by Chin, Grimm, Julienne, and Tiesinga, 2009).

## A. Photonic crystals and photonic-crystal fibers

The concept of the photonic crystal, designed as a bulk waveguiding structure with the periodic transverse modulation of the local value of the refractive index, that should give rise to the photonic bandgap structure, emulating that for electrons in ionic crystals, was put forward in Yablonovitch (1987), see also reviews by Yablonovitch (1993 and 2001). Photonic crystals, featuring *full photonic bandgaps*, were elaborated theoretically by Yablonovitch, Gmitter, and Leung (1991), and have been built in 2D (Krauss, DeLaRue, and Brand, 1996; Johnson et al., 1999) and 3D geometries (Noda et al., 2000), see also a detailed account of the topic given in the book by Joannopoulos, Johson, Winn, and Meade (2008). Photonic crystals with the quadratic intrinsic nonlinearity were analyzed too (Centini et al., 1999).

Photonic-crystal fibers (PCFs), which are made of a transparent material with a lattice of voids running parallel to the axis of the fiber, offer a direct realization of combined linear and nonlinear lattices in the cross-section plane, as values of both the refractive index and Kerr coefficient jump between the material and the voids filled with air (or with different substances, such as liquid crystals). PCFs were created in silica (Knight et al., 1996), which was followed by the demonstration of the single-mode character of the bandgap-guided light transmission in such structures (Birks, Knight, and Russell, 1997), see also the paper by Russell (2003) and a recent review by Cerqueira (2010). Later, the creation of all-solid PCFs, with the voids filled by another material, rather than air, have been reported by Luan et al. (2004). As concerns the creation of solitons in PCFs, these media make it possible not only to engineer desired dispersion properties, but also open a way to enhance the effective nonlinearity, through confining the beam to a silica wire with a small cross section (Soljacic and Joannopoulos, 2004).

One of the important nonlinear-optical effect observed in the PCFs was the generation of the supercontinuum (see Ranka, Windeler, and Stentz, 2000, and a reviews by Dudley, Genty, and Coen, 2006, and Skryabin and Gorbach, 2010). This effect is interpreted as a result of the fission of higher-order temporal solitons carried by the PCF (Herrmann et al, 2002). In addition, remarkable possibilities offered by the PCFs for the direct creation and control of fundamental solitons have been demonstrated too – in particular, by Ouzounov et al. (2003) and Reeves et al. (2003). As for spatial solitons in photonic crystals and PCFs addressed in this review, they have not yet been reported in experimental works, to the best of our knowledge.

## B. Waveguiding arrays in bulk media

Among the most flexible and elaborated techniques for the creation of permanent periodic guiding structures in bulk materials is the direct material processing by femtosecond laser pulses. When intense ultra-short laser pulses are focused inside transparent materials, the nonlinear absorption in the focal volume causes the optical breakdown and formation of micro-plasma bubbles, which, eventually, lead to permanent structural and refractive-index modifications (Davis et al., 1996; Itoh et al., 2006). By moving the focus of the laser beam through the sample, one can write waveguides along arbitrary paths, thus creating both one- and two-dimensional lattices. One of the most suitable materials for the direct laser writing of waveguide lattices is fused silica. In such materials, low-loss 20-mm long waveguides with a spacing down to 14 $\mu$m and transverse dimensions of 3 $\mu$m $\times$ 12 $\mu$m were created (Szameit et al., 2005; Blömer et al., 2006). The emerging local change of the refractive index in the medium is a function of the writing speed, increasing exponentially with the decrease of the speed (a maximum change of the refractive index achieved by means of this technique was $\sim 1.3 \times 10^{-3}$). Importantly, the material damage produced by the femtosecond pulses results in the simultaneous *increase* of the linear refractive index and *decrease* of the nonlinearity coefficient. For writing velocity $\sim 500$ $\mu$m/s, the effective nonlinearity coefficient inside the waveguide may drop down to $n_2 \approx 0.7 \times 10^{-20}$ m$^2$/W (to be compared with $n_2 \approx 2.7 \times 10^{-20}$ m$^2$/W in unprocessed silica). Moreover, the nonlinearity strength changes with the writing velocity faster than the linear refractive index. This suggests the possibility to create mixed lattices with out-of-phase modulations of the linear refractive index and nonlinearity, adjusting the nonlinear properties of the material as per the designed pattern. The discrete nonlinear localization of light in 1D femtosecond-written waveguide arrays was reported by Szameit et al. (2005), and the formation of 2D solitons in square-shaped waveguide arrays was observed by Szameit et al. (2006), at typical peak powers $\sim 1$ MW for 100 fs pulses at wavelength 800 nm. The laser-written waveguide arrays in fused silica have been used for the observation of a number of interesting phenomena, including optical surface waves (Szameit et al, 2007), polychromatic dynamic localization in curved lattices (Szameit et al, 2009a), and the inhibition of light tunneling in longitudinally modulated arrays (Szameit et al, 2009b), to mention a few.

Photorefractive crystals such as $LiNbO_3$, $BaTiO_3$, $KNbO_3$, or SBN offer fascinating possibilities for optical information processing, holographic volume storage, phase conjugation, and interferometric holography. The photorefractive effect in this type of crystals results from a light-induced redistribution of charges released by impurities or intrinsic centers. Electrons or holes are optically excited and trapped at different sites, resulting in the buildup of internal space-charge field. This field, in turn, causes modifications in the refractive index of the material via the electrooptic effect (Buse, 1997a and 1997b). Such materials allow the fabrication of high-quality permanent waveguiding arrays, in particular by dint of the titanium in-diffusion (Schmidt and Kaminov, 1974) or proton exchange (Jackel, Rice, and Veselka, 1983). Thus, lattices fabricated in copper-doped $LiNbO_3$ crystals combine high saturable defocusing nonlinearity (arising from the photovoltaic effect) with the adjustable lattice strength. Typical linear lattices imprinted in $LiNbO_3$ crystals consist of $4\,\mu m$-wide titanium-doped stripes separated by the same distance. Each channel with refractive-index modulation depth $\sim 3\times 10^{-3}$ forms a single-mode waveguide for light at $\lambda = 514.5\,\mathrm{nm}$, while the inter-channel coupling constant is $\sim 1\,\mathrm{mm}^{-1}$. The experimental observation of bulk and surface gap solitons (GSs) in such arrays at $\mu W$ power levels was reported by Chen et al. (2005), Smirnov et al. (2006), and Rosberg et al. (2006). In addition, the conductivity of such materials, their response time, and maximal light-induced nonlinear contribution to the refractive index may be dramatically enhanced by doping with appropriate elements (for instance, $LiNbO_3$ is usually doped with Cu or Fe, while for SBN one can use Ce, Rh, or Pr as dopants). Since the diffusion time and depth (hence also the concentration of dopants inside the photorefractive material for a given annealing time) strongly differ for different dopants, or even different thickness of the dopant stripes etched on top of the crystal (Hukriede, Runde, Kip, 2003), one can potentially use inhomogeneous surface doping of photorefractive crystals to produce permanent lattices which provide for strong modulation not only of the refractive index, but also of the local nonlinearity.

## C. Optical induction in photorefractive media

Externally biased photorefractive media allow reversible optical induction of reconfigurable lattices. This method has proved to be an extremely powerful tool for the creation of various linear refractive-index landscapes. The idea of the OL induction was put forward by Efremidis *et al.* (2002). This method is especially attractive since the resulting landscapes can be adjusted to a required form by varying parameters of the lattice-creating waves, and easily erased, in contrast to the permanent technologically fabricated guiding structures described above. One or two-dimensional photoinduced lattices have to remain uniform along the propagation distance (typically, up to several centimeters). In the pioneering experiments, the periodic refractive index profile was induced by the interference of two ordinarily polarized plane waves in a biased Strontium Barium Niobate (SBN) crystal (Fleischer et al., 2003a and 2003b). The refractive index of the material was modified through the linear electro-optic effect. In SBN crystals, orthogonally polarized waves feature dramatically different electro-optic coefficients ( $r_{33} \simeq 1340$ pm/ V , $r_{13} \simeq 67$ pm/ V ), so that ordinarily-polarized interfering plane waves propagate almost linearly and create the stable 1D periodic lattice, while the extraordinarily polarized probe beam experiences strong nonlinear self-action, described by the nonlinear change of the refractive index, $\delta n \sim -r_{33}E_0 I_{\text{dark}} / [I_{\text{dark}} + I_{\text{latt}}(x) + I]$, where $E_0$ is the external electric dc field (bias), $I_{\text{dark}}$ characterizes the dark-irradiance level that can be modified with the aid of an additional background illumination, $I_{\text{latt}}$ is the intensity of the lattice-creating beam, $I$ is the intensity of the probe beam, and $x$ is the transverse coordinate along which the lattice is induced. The intersection angle between plane waves determines the lattice periodicity, while the experimentally achievable refractive-index modulation depth is such lattices is $\sim 10^{-3}$. Notice that the sign of the nonlinearity (focusing/defocusing) can be altered by reversing the sign of the bias electric field. The OL may be also made partially incoherent and created by means of the amplitude modulation, rather than by coherent interference of multiple plane waves (Chen et al., 2004). Such lattices are exceptionally stable and persist even in the weakly nonlinear regime, due to the suppression of the modulation instability by the lack of coherence. The use of the optically induced lattices has led to the observation of many different types of solitons (see works by Neshev *et al.*, 2003; Martin *et al.*, 2004; Neshev *at al.*, 2004; Fleischer *et al.*, 2004; Cohen et al., 2005; Fischer *et al.*, 2006; Wang *et al.*, 2007a).

The technique of the OL induction with some modifications can be potentially applied to the creation of nonlinear and mixed lattices. In particular, instead of launching the ordinarily-polarized lattice-creating beam into the photorefractive material, one can employ a spatially modulated background illumination. The resulting nonlinear change of the refractive index for the probe beam amounts to $\delta n \sim -r_{33}E_0 I_{\text{dark}}(x) / [I_{\text{dark}}(x) + I]$, i.e., the nonlinearity coefficient becomes spatially modulated, as seen from expression $\delta n \sim -r_{33}E_0[1 - I / I_{\text{dark}}(x)]$, obtained in the limiting case of $I \ll I_{\text{dark}}$. Alternatively, the application of a spatially-inhomogeneous external electric field to a relatively thin photorefractive crystal in the absence of any additional lattice-creating pattern translates into a strong modulation of the linear refractive index and nonlinearity coefficient, according to expression $\delta n \sim -r_{33}E_0(x) I_{\text{dark}} / [I_{\text{dark}} + I]$, that for $I \ll I_{\text{dark}}$ takes the form of $\delta n \sim -r_{33}E_0(x)(1 - I / I_{\text{dark}})$. The above-mentioned results suggest that photorefractive ma-terials may indeed be used not only for the induction of flexible linear refractive-index landscapes, but also for the creation of reconfigurable NLs.

## D. Liquid crystals

Nematic liquid crystals have emerged as suitable materials for experimentation with optical solitons in spatially inhomogeneous environments due to their strong reorientational nonlinearity, that may exceed the nonlinearity of standard materials (such as semiconductors) by several orders of magnitude. Liquid crystals are characterized by a significant degree of molecular order and birefringence under proper anchoring. An incident extraordinary polarized optical beam, whose electric field is not orthogonal to the director (main axis) of the crystal, can interact with induced dipoles in molecules and cause the molecular reorientation, which is strongest in the region of the highest intensity, but it may extend well beyond the spatial region illuminated by the laser beam. For this reason, the nonlinearity of nematic liquid crystals is nonlocal, with the nonlocality degree depending, among other factors, on the thickness of the liquid-crystal cell. The light-induced director reorientation results in a variation of the refractive index of the material, so that, with the choice of appropriate light intensities, spatial solitons (*nematicons*) may be excited. A typical nematicon forms at power levels $\sim 2$ mW for $\lambda = 1.064$ $\mu$m in a 75 $\mu$m-thick planar-aligned E7 liquid-crystal cell, with an externally applied low-frequency biasing voltage $V \sim 1.2$ V (Conti, Peccianti, and Assanto, 2004; Peccianti et al., 2004). Since the reorientation of liquid crystal molecules can be driven by a low-frequency electric field through a voltage applied across the thickness of the cell, linear, nonlinear, and nonlocal properties of this medium can be flexibly adjusted. In particular, Peccianti, Conti, and Assanto (2005) had demonstrated, studying the 1D modulational instability in liquid crystals, that the nonlocality can be tuned, versus the optical nonlinearity, by the external voltage (the nonlocality and nonlinearity simultaneously decrease with the increase of the externally applied voltage, but the nonlocality becomes negligible faster than the nonlinearity). If the cover slide of the planar cell filled with a *pre-anchored* liquid crystal incorporates two separate electrodes with distinct applied voltages (with respect to the common ground-plane electrode at the bottom slide), a graded index interface may be created, characterized by distinct nonlinear and linear refractive indices at different sides of the interface. In such a geometry, the lower biasing voltage corresponds to a lower extraordinary refractive index, along with a higher nonlinearity coefficient $n_2$; for instance, in the setting reported by Peccianti et al. (2007), $n_2$ changes approximately 10 times between regions with the applied voltage of 0 and 1.5 V. Such linear-nonlinear graded-index interfaces in liquid crystals were recently used for the demon-stration of tunable refraction and reflection of spatial solitons (Peccianti et al., 2006). Further, Beeckman, Azarinia, and Haelterman (2009) recently utilized nonuniform biasing of liquid crystals in the longitudinal direction, resulting in a gradual increase of the nonlinearity, to counterbalance the broadening of the soliton caused by intrinsic losses. This suggests that transverse lattices with the simultaneous modulation of the nonlinearity and linear refractive index can be created in properly biased liquid crystals. Such periodic voltage-controlled lattices were indeed produced by Fratalocchi et al., 2004. In these lattices, a set of periodically spaced electrodes (with typical spacing $\sim 6~\mu$m) allows the bias to be applied across a $\sim 5~\mu$m thick crystal cell, thereby periodically modulating the refractive index and nonlinearity through the molecular reorientation. This experimental setup had enabled the observation of discrete optical solitons at rather low power levels.

## E. Thermal nonlinearity

Thermal nonlinearity is featured by various materials possessing sufficiently large thermo-optic coefficients. Light propagation in such media is affected by the geometry of the sample and by the temperature distribution at its boundaries. A light beam propagating in thermal media experiences slight absorption, thus acting as a heat source. Diffusion of the heat creates a non-uniform temperature distribution, that causes a local refractive-index variation proportional to the temperature change at each point. In materials with positive thermo-optic coefficients, the heat diffusion results in a local increase of the refractive index in heated regions, that may lead to the formation of localized states, while in materials with negative thermo-optic coefficients the refractive index decreases in heated regions, typically resulting in expulsion of light toward boundaries of the sample (Rotschild et al., 2005, 2006a, and 2006b; Alfassi et al., 2007; Kartashov et al., 2007). A typical nonlinear contribution to the refractive index in such thermal nonlinear media as lead glass amounts to $\sim 5 \times 10^{-5}$ for laser beams of width $\sim 50~\mu$m, carrying total power $\sim 1$ W. While it may be difficult to generate a periodic thermal nonlinearity in a single material, this nonlinearity may be realized in composite materials, such as PCFs with holes infiltrated with suitable index-matching slightly absorbing liquids. To this end, one can fill the holes of the PCFs with suitable liquids, using capillary forces or hydrostatic pressure. This integrated technique opens the way to combine the guiding properties of the PCF, resulting from its internal structure, and the nonlinearity which is determined by properties of the liquid filling the holes (see e.g., Eggleton et al., 2001; Larsen et al., 2003; Fuerbach et al., 2005). In particular, Rosberg et al. (2007) and Rasmussen et al. (2009) utilized high-index weakly absorbing oil, featuring defocusing thermal nonlinearity, for the infiltration of holes in a hexagonal

PCF, which made it possible to create a material with a periodic modulation of the linear refractive index (the refractive index of the oil is slightly higher than that of silica) and thermal nonlinearity. This setting was used to demonstrate highly tunable beam diffraction, nonlinear self-action, power limiting, and the formation of 2D nonlocal GSs. Carefully selecting index-matching liquids and the material of the PCF, one may fabricate NLs by means of this technique.

## F. Bose-Einstein condensates

One of fundamental results of the quantum theory was the prediction of the Bose-Einstein statistics in a gas of boson particles at zero temperature. Such a state of matter in the form of the *Bose-Einstein condensate* (BEC) remained a theoretical concept during seventy years after it had been predicted. A breakthrough, which has turned out to be, arguably, the greatest achievement in the field of fundamental physics in the course of the past fifteen years, was the creation of BEC in ultracold gases of alkali metals, which was reported in 1995 by Anderson *et al.* (in the gas of atoms of $^{87}$Rb), Bradley *et al.* (in $^{7}$Li), and Davis *et al.* (in $^{23}$Na). In these celebrated experiments, the gas composed of several thousands of atoms was chilled, by means of a combination of the laser-cooling and evaporation techniques, to temperatures on the order of fractions of nano-Kelvin.

One of milestones in the subsequent experimental work on BEC was the creation of effectively one-dimensional *bright* (i.e., localized) matter-wave solitons in the condensate of $^{7}$Li atoms trapped in "cigar-shaped" configurations (Strecker, Partridge, Truscott, and Hulet, 2002 and 2003; Khaykovich *et al*, 2002). In particular, multi-soliton trains have been created in the former work, in addition to single-soliton modes. Later, similar solitons were also created in a post-collapse state of the $^{85}$Rb condensate by Cornish, Thompson, and Wieman, 2006 (the collapse was caused by the switch of the interaction from repulsive to attractive by means of the FR).

A fundamental theoretical model of the matter-wave dynamics in BEC, including the description of solitons, is provided, with a very good accuracy, by the Gross-Pitaevskii equation (GPE), which is based on the mean-field approximation – see the book by Pitaevskii and Stringari (2003), and a more recent review of nonlinear aspects of the matter-wave dynamics and solitons by Carretero-González, Frantzeskakis, and Kevrekidis, 2008. Various aspects of the matter-wave dynamics, both theoretical and experimental, have been thoroughly reviewed in a recent collection of articles edited by Kevrekidis, Frantzeskakis, and Carretero-González, 2008.

A versatile tool which makes it possible to control the collective behavior of the BEC, including matter-wave solitons, is provided by OLs, i.e., a spatially periodic atomic potential induced by the interference pattern which is, in turn, created by mutually coherent laser beams illuminating the gas from different directions. If the carrier frequency of the lattice-forming laser beams is red- or blue-detuned with respect to the dipole transition between atomic levels, the atoms are, respectively, attracted to or repelled from local maxima of the light intensity. A great deal of interest was drawn to the use of OLs in the context of BEC after the prediction, by Jaksch *et al.* in 1998, and experimental realization by Greiner et al. (2002), of the quantum phase transition between the Bose superfluid and the "Mott insulator", i.e., the state in which atoms are pinned in the OL due to repulsive interactions between them. Some other notable results which were produced by means of OLs include the prediction (Choi and Niu, 1999) and observation (Morsch *et al.*, 2001) of *Bloch oscillations* of the atomic density (under the action of a constant external force in the combination with an OL), the theoretical analysis (Wu and Niu, 2001) and experimental realization (Burger *et al*, 2001) of the transition of the superfluid motion of the condensate to a dissipative flow when its velocity exceeds the sound velocity, and the creation of the Tonks-Giradeau gas (which is composed of strongly repelling bosonic atoms, whose behavior emulates fermions) by Paredes et al. (2004). As concerns the topics of the present review, especially interesting experimental achievements are GSs created in the gas of $^{87}$Rb atoms with the repulsive interactions between them, loaded into a ramped OL (Eiermann *et al*, 2004), and the so-called "gap waves", created by the same group (Anker *et al*., 2005), which may be considered as truncated segments of nonlinear Bloch waves (Wang *et al.*, 2009a). The BEC dynamics in OL potentials has been reviewed in several comprehensive articles, as concerns both the mean-field matter-wave dynamics (Brazhnyi and Konotop, 2004; Morsch and Oberthaler, 2006) and properties of strongly correlated ultracold atoms (Jaksch and Zoller, 2005; Lewenstein *et al.*, 2007; Bloch, Dalibard, and Zwerger, 2008).

It is also relevant to mention a recent paper by Henderson, Ryu, MacCormick, and Boshier (2009), which reports a newly developed experimental technique, which allows one to "paint" virtually any desirable 1D or 2D potential landscape by means of a focused laser beam, which quickly moves along contours of the landscape. The potential is actually induced by the temporal self-averaging of the trace left by the rapidly moving laser focus.

Among other experimental achievements reported in the vast area of BEC, a notable result is the creation of BEC in the gas of $^{52}$Cr atoms (Griesmaier et al., 2005). The chromium atoms possess the magnetic moment, therefore the dynamics of this condensate is strongly affected by the long-range dipole-dipole interactions. The consideration of the interplay of such interactions with the OL potential has lead to the prediction of quantum phases in this condensate (Goral, Santos, and Lewenstein, 2002), and of several species of solitons (Gligori *et al.*, 2008, 2009, and 2010a,b; Cuevas *et al.*, 2009). A recent comprehensive review of dipolar BECs was presented by Lahaye *et al.* (2009).

Summarizing, there are a variety of experimentally available settings suitable for the exploration of solitons in media with mixed linear and nonlinear lattices. In what follows, we discuss theoretically predicted properties of such states.

## IV. One-dimensional solitons

In this section we address the first topic where fundamental results for solitons supported by purely nonlinear and mixed linear-nonlinear have been accumulated. Naturally, the theoretical analysis is easiest in the 1D case, and this was the first case for which systematic studies of NL-supported solitons had commenced. The topic has been investigated thoroughly (unlike the essentially more difficult 2D setting, see the next section), the available theoretical knowledge making it possible to formulate general conclusions concerning the existence, stability, and dynamics of the solitons created on top of NLs or mixed linear-nonlinear lattices. In particular, it has been concluded that 1D solitons of this type exist in purely nonlinear lattices only when their norm exceeds a certain *threshold* value (this is the case at least when mean nonlinearity coefficient is zero or sufficiently small in comparison with amplitude of nonlinearity modulation), which is a drastic difference from the ordinary 1D solitons, existing in the uniform medium with attraction, or gap solitons supported by a linear lattice potential embedded into a medium with the uniform repulsive nonlinearity. On the contrary to the situation known for linear lattices, the NLs create the solitons "from nothing", rather than helping them bifurcate from Bloch modes of linear lattices. Other fundamental properties of the solitons specific to the NL systems or mixed linear-nonlinear lattices are a possibility of the *multistability*, and *enhanced mobility*, in comparison with usual solitons trapped in linear lattice potentials. It is also relevant to stress that the consideration of mixed linear-nonlinear lattices offers a novel setting for the study of effects of commensurability and incommensurability (between the linear and nonlinear lattices) on nonlinear modes that will be addressed in this section.

Other basic problems outlined in this part of the review include continuous and discrete solitons in NLs and mixed lattices of various shapes, namely, harmonic, random, and KP lattices. We address properties of both scalar and vectorial (two-component) states, outline analytical and numerical methods used for the construction of solitons in NLs, and methods for the analysis of their stability. We also discuss the evolution of solitons in NLs, making emphasis on their mobility in the lattice media, controllable *switching* of solitons in NLs, oscillations of solitons in nonlinear or mixed potentials (including *Bloch oscillations* of GSs), their interactions with nonlinear defects and traps, *symmetry breaking* of the nonlinear localized modes in dual-core settings, delocalization transitions due to the spatially inhomogeneous nonlinearity, and some other topics relevant to the solitons in one-dimensional NL systems. Periodic waves in NLs, and the generation of soliton trains by nonlinear potentials ("*soliton lasers*") are considered too. The latter topic offers a potential for diverse applications which may benefit from the availability of coherent streams of intense matter-wave pulses, the inhomogeneous nonlinearity being a key element necessary for the design of such "lasers". The dissipative dynamics in lattices with nonlinear gain and losses is considered too, which is obviously necessary for the analysis of physically realistic settings, if one is interested in the long-time nonlinear dynamics. The considered models include basic types of nonlinearities available in the underlying physical settings, including cubic, polynomial, and thermal (nonlocal) nonlinear interactions.

Some of the phenomena discussed below may occur or find their analogs in usual linear lattices. For recent detailed reviews on solitons in linear lattices, see articles by Lederer et al., 2008, and Kartashov, Vysloukh, and Torner, 2009a. A rigorous mathematical treatment of solitons in 1D linear lattice potentials was very recently reviewed by Ilan and Weinstein (2010).

## A. Solitons in nonlinear lattices
## 1. The basic model and fundamental properties of solitons

It is relevant to start the presentation of the results for 1D solitons in purely nonlinear harmonic lattices from the analysis reported by Sakaguchi and Malomed (2005a), which had revealed basic properties of the solitons of this type, later found in a number of related models (as described in this subsection below). The paradigmatic dynamical models describing 1D solitons in NLs is provided the following form of the NLSE/GPE, written in terms of the mean-field approximation for the BEC:

$$i\frac{\partial q}{\partial t} = -\frac{1}{2}\frac{\partial^2 q}{\partial x^2} + R(x)|q|^2 q. \qquad (17)$$

Here, the cubic nonlinear term is periodically modulated in the coordinate, $R(x) = r_0 - \cos(2x)$, $r_0$ being a constant part of the nonlinearity coefficient. The model describes a situation when the atomic scattering length in the BEC is spatially modulated, via the FR mechanism, by the periodically patterned magnetic or optical field (the latter one can be created by means of the usual OL induced by the interference of two laser beams). The results summarized below were chiefly obtained for the most fundamenal case, $r_0 = 0$, i.e., zero average value of the nonlinearity. As said above, Eq. (17) may also be interpreted as the NLSE governing the transmission of optical signals in a periodically inhomogeneous planar waveguide, in which case $t$ is the propagation distance.

First, it is relevant to summarize results obtained by means of analytical approximations. The VA (variational approximation) for sufficiently narrow stationary solutions, $q(x,t) = w(x)\exp(-i\mu t)$, can be derived from the Lagrangian corresponding to Eq. (1), $\mathcal{L} = \int_{-\infty}^{\infty}[2\mu w^2 - (dw/dx)^2 - R(x)w^4]dx$, using *ansatz* $w = A\exp(-x^2/2W^2)$. Then, variational equations $\partial\mathcal{L}/\partial N = \partial\mathcal{L}/\partial W = 0$ (here $N = \pi^{1/2}A^2 W$ is the norm representing the number of particles or power, in the case of the matter-wave or optical solitons, respectively) predict the existence of the minimal norm (*threshold*) necessary for the existence of stationary fundamental solitons in this model. As mentioned above, this result demonstrates a *drastic difference* of the NL-supported solitons from their counterparts in the uniform 1D media, where the solitons may exist with an arbitrarily small value of the norm. The VA equations also demonstrate that the soliton may exist only in the case when the local nonlinearity at the point of the maximum of $|w(x)|$ (the center of the soliton) is attractive.

For broad solitons, a different analytical approximation may be developed. It uses the averaging method, with the solution approximated as $q(x,t) = q_0(x,t) + q_1(x,t)\cos(2x)$, where $q_0(x,t)$ and $q_1(x,t)$ are slowly varying functions of $x$ in comparison with $\cos(2x)$. The substitution of the latter *ansatz* into Eq. (17) and the application of the averaging method allows one to eliminate $q_1$ in favor of $q_0$, *viz.*, $q_1 = (1/2)|q_0|^2 q_0$, and derive an effective equation for $q_0$ which turns out to be the NLSE with the *quintic* focusing nonlinearity (while the cubic term does not appear, as a result of the averaging),

$$i\frac{\partial q_0}{\partial t} = -\frac{1}{2}\frac{\partial^2 q_0}{\partial x^2} - \frac{3}{4}|q_0|^4 q_0. \tag{18}$$

Equation (18) admits analytical solutions, $q_0 = [2|\mu|^{1/2}\operatorname{sech}(2|\mu|^{1/2}x)]^{1/2}\exp(-i\mu t)$.

For zero average value of the nonlinearity coefficient, $r_0 = 0$, the norm of the fundamental soliton solutions of Eq. (17), obtained in a numerical form, is a non-monotonous function of the soliton's amplitude $A$ and chemical potential $\mu$ (see Fig. 1). The VK stability criterion, which is relevant in the case when solitons are supported by an attractive nonlinearity, in the space of any dimension (Vakhitov and Kolokolov, 1973; Bergé, 1998), predicts that the branches of narrow solitons in Fig. 1(a) to the right and left of point $N = N_{\min}$ are stable and unstable, respectively. Undulations in the shape of the soliton, which are due to the action of the periodic NL, are pronounced only at intermediate values of the amplitude. High-amplitude solitons gradually shrink to a single site of the NL in the region of the strongest attractive interaction, while low-amplitude solitons, covering many periods of the lattice, are unable to induce a sufficiently strong nonlinear pseudopotential. Sakaguchi and Malomed (2005a) showed that unstable solitons with moderate amplitudes spontaneously rearrange themselves into persistent *breathers*. Broad solitons with small amplitudes are, strictly speaking, unstable, but in practical terms they represent nearly stable modes, as their decay is extremely slow.

The broad solitons supported by Eq. (17) can move across the NL, and they feature quasi-elastic collisions with no visible losses. Narrow solitons, which are strongly pinned by the NL, can form stable complexes composed of several out-of-phase fundamental modes. The inclusion of a weak constant attractive nonlinearity into Eq. (17), accounted for by small $-r_0 > 0$, stabilizes small-amplitude solitons. This finding may be explained by the consideration of the respective average NLSE, which turns out to have the cubic-quintic nonlinearity, cf. Eq. (18) with the quintic nonlinearity (Sakaguchi and Malomed, 2005a).

## 2. Generalized models: lattices with higher-order nonlinearities

A natural extension of the model with cubic Kerr nonlinearity is presented by the one with a general power-function nonlinearity, $-[1+R(ax)]|q|^{p-1}q$, where function $R$ describes a particular lattice profile and $a$ denotes the ratio of the soliton's width to the characteristic scale of the lattice. The existence and stability of solitons in this model were considered by Fibich, Sivan, and Weinstein (2006). They addressed three representative situations, namely, $p < 5$ (the *subcritical* case), $p = 5$ (the *critical* case corresponding to the quintic NLSE), and $p > 5$ (the *supercritical* case). This classification is based on the fact that the quintic nonlinearity plays the critical role for the onset of the collapse in the one-dimensional NLSE with the attractive self-interaction (Bergé, 1998). Accordingly, in the subcritical case the solitons are stable in the NLSE with constant coefficients, while in the critical and supercritical cases they are unstable, developing the collapse after a finite evolution time. It is therefore interesting to elucidate the impact of the NL on the stability of soliton solutions in the supercritical and critical cases. Fibich, Sivan, and Weinstein (2006) showed that, for wide modes with $a \gg 1$, the profile of stationary solitons in the NL system, $w(x)$, coincides, at the leading order in $1/a$, with the shape of the soliton in the homogeneous medium where the nonlinear coefficient is the mean value of $1+R(ax)$ over one lattice period; corrections to the profile induced by the lattice arise only at order $a^{-2}$. The NL always reduces the norm of wide solitons. For narrow solitons with $a \ll 1$, the profile is de-termined by local properties of the NL, i.e., the value of the nonlinear coefficient in a vicinity of the soliton's peak. Similar to what happens in the case of wide beams, even when variations in the NL are not small, the lattice leads to $\mathcal{O}(a^2)$ changes in the soliton's profile for $p \neq 5$, and only $\mathcal{O}(a^4)$ changes in the critical case of $p = 5$. Two stability conditions for stationary solutions of the form $q = w(x)\exp(i\mu z)$, supported by the NLs were formulated: (i) The *spectral condition*, which requires that operator $\mathcal{L}_+ = -\partial^2/\partial x^2 + \mu + p[1 + R(ax)]w^p$, obtained upon the linearization of the corresponding nonlinear evolution equation, must have no more than one negative eigenvalue, and (ii) the *slope condition*, similar to the VK criterion, which requires $\partial N/\partial \mu > 0$. The violation of the spectral condition results in a *drift instability*, i.e., a spontaneous shift of the soliton against the underlying NL, that can be initiated only by asymmetric perturbations. In contrast, the violation of the slope condition (the VK criterion) typically results in the blowup (collapse) or spreading (decay). These conditions predict that, in the subcritical case, with $p < 5$, the solitons centered at local maxima of $R(ax)$ are (quite naturally) *stable*, whereas solitons sitting on local minima of $R(ax)$ are (naturally too) *unstable* with respect to asymmetric perturbations shifting their centers, although such modes cannot be destabilized by symmetric perturbations. In the critical case of $p = 5$, the NL can only stabilize very narrow solitons centered at a local maximum of $R(ax)$, provided that the lattice itself satisfies a certain local condition (Fibich, Sivan, and Weinstein, 2006). Even in this case, the stability region is so narrow that sufficiently strong perturbations can destabilize the soliton, and it was concluded that this very weak stability is a "mathematical", rather than "physical" property.

With regard to the general issue of the stability of solitons in settings with a spatially modulated nonlinearity in the case of the *critical nonlinearity*, it has an interesting ramification in the 2D case, where the cubic nonlinearity plays the critical role. In that case, all the solitons existing in the uniform space are unstable, being tantamount to the above-mentioned *Townes solitons* (Bergé, 1998), and a challenging problem is the search for modes of the spatial modulation of the attractive cubic nonlinearity that may give rise to *stable* 2D solitons. A solution to this fundamental problem, which is considered in some detail in the next section, has been found, but under rather stringent conditions imposed on the form of the nonlinearity modulation (Sakaguchi and Malomed, 2006a; Kartashov *et al.*, 2009a; Hung, Zi , Trippenbach, and Malomed, 2010).

## 3. Exact solutions in specially designed models

Proceeding to more technical aspects of the theoretical studies of 1D solitons in NL settings, a noteworthy fact is that explicit solutions of the NLSE with spatially inhomoge-neous nonlinearity can be constructed using the Lie-group theory and canonical transformations, as illustrated in the work by Belmonte-Beitia, Perez-Garcia, and Vekslerchik (2007). Their method utilizes the fact that equation $-\partial^2 w/\partial x^2 + V(x)w + R(x)w^3 - \mu w = 0$, describing profiles $w$ of solitons with chemical potential (or propagation constant) $\mu$ in materials with the modulation of the nonlinearity and linear refractive index, can be reduced, by dint of a canonical transformation, $W = b^{-1/2}(x)w$, $X \equiv \int_0^x b^{-1}(s)ds$, to the well-known equation

$$-\frac{d^2W}{dX^2} + r_0 W^3 = EW \tag{19}$$

where $E = [\mu - V(x)]b^2(x) - (1/4)[b'(x)]^2 + (1/2)b(x)b''(x)$ is a constant, while arguments provided by the Lie-group theory may be used to establish a relation between $b(x)$ and functions $V(x), R(x)$ describing the linear and nonlinear lattices which admit the Lie symmetry:

$$\begin{aligned} R(x) &= r_0/b^3(x), \\ b'''(x) &- 2b(x)V'(x) + 4b'(x)\mu - 4b'(x)V(x) = 0. \end{aligned} \tag{20}$$

In particular, in the absence of the linear potential $[V(x) = 0]$ for $\mu > 0$, the solution of these equations can be written as $b(x) = C_1 \sin(\omega x) + C_2 \cos(\omega x) + C_3$, while for $\mu < 0$ one gets $b(x) = C_1 \exp(\omega x) + C_2 \exp(-\omega x) + C_3$, where $\omega = 2|\mu|^{1/2}$. This may give periodic nonlinearity coefficients, such as $R(x) = r_0[1 + \alpha \cos(\omega x)]^{-3}$, or localized nonlinearity coefficients, such as $R(x) = r_0/\cosh^3(\omega x)$. The latter case may be especially interesting since, under appropriate conditions, it allows mapping of the well-known periodic solutions of Eq. (19), $W(X) = C \operatorname{sn}(\mu X, k)/\operatorname{dn}(\mu X, k)$, with specifically selected values of modulus $k$ of elliptic functions sn and dn, into higher-order spatially localized solitons of the original NLSE with inhomogeneous nonlinearity coefficient, $R(x) = r_0/\cosh^3(x)$ (see Fig. 2 for examples of profiles of such localized solitons obtained at different values of $k$). The Lie-group theory was also used by Belmonte-Beitia, Perez-Garcia, and Brazhnyi, 2009, to construct explicit solitary-wave solutions of coupled nonlinear Schrödinger equations with spatially inhomogeneous nonlinearities.

Below (in subsection IV.B.3), another version of the NLSE including the nonlinear and linear lattices will be described, which also admits classes of exact solutions for specially chosen modulation functions, but those exact solutions do *not* reduce to a deformation of solutions of the NLSE with constant coefficients (Tsang, Malomed, and Chow, 2009). It is also relevant to mention models based on GPEs with specially devised modulations of the nonlinearity coefficient and linear potentials, for which exact solutions can be found by means of direct substitutions (Belmonte-Beitia, Konotop, Pérez-García, and Vekslerchik, 2009; Yan and Konotop, 2009; Yan, 2010). Very recently, similar approaches, which allow one to transform one-dimensional NLSEs with variable coefficients in front of the nonlinear terms into a system with constant coefficients, were elaborated by Cardoso, Avelar, Bazeia, and Hussein (2010), Cardoso, Avelar, and Bazeia (2010), and Rajendran, Muruganandam, and Lakshmanan (2010).

A different approach which makes it possible to devise models of the mixed NL-OL type with vast families of exact soliton and nonlinear Bloch-wave solutions was very recently elaborated by Zhang *et al.* (2010). Introducing specially designed *localized* profiles of the spatial modulation of the attractive nonlinearity, an infinite number of exact soliton solutions was produced in terms of the Mathieu and elliptic functions, with the respective chemical potential belonging to the semi-infinite bandgap of the OL-induced spectrum. Starting from the exact wave forms for solitons, which, naturally, are not generic solutions, generic families of soliton solutions were constructed in a numerical form. The stability of the solitons was investigated through of the computation of eigenvalues for small perturbations, and also via direct simulations. The same work has demonstrated a virtually exact (in the numerical sense) composition relation, which allows one to build nonlinear Bloch waves as chains of solitons.

### 4. Solitons in nonlinear lattices of the Kronig-Penney type

One-dimensional bright and dark matter-wave solitons in BEC models with periodic piecewise-constant NLs [i.e., those of the Kronig-Penney (KP) type] were analyzed by Rodrigues *et al.*, 2008. For optical solitons supported by a combination of linear and nonlinear lattices of the KP type, a detailed analysis of soliton modes was reported by Kominis (2006), Kominis and Hizanidis (2006, 2008), and Kominis, Papadopoulos, and Hizanidis (2007). The results, although they are somewhat cumbersome, are interesting as they furnish quasi-analytical results for the fundamentally important case of the KP modulation.

The KP lattice corresponds to the following modulation format of the nonlinearity coefficient $R(x) = r_0 + (r_1 - r_0)\sum_{n=-\infty}^{n=+\infty}\{\theta[x-(nL+L_1)] - \theta[x-(n+1)L]\}$, where $\theta$ is the Heaviside's step function, $L$ is the periodicity of the lattice, $r_0$ is the average value of $R(x)$, and $r_1$ determines the depth of the nonlinearity modulation. Using the fact that in such a nonlinearity landscape one can explicitly construct the solutions for each of the regions where the nonlinearity coefficient is constant, an analytical approach was developed by matching functions $w_m(x) = A_m r_m^{-1/2}\text{sech}[A_m(x-x_m)]$, centered at $x_m$, with amplitudes $A_m$ borrowed from the solutions of the respective homogeneous NLSEs with different nonlinearity coefficients $r_m$. The matching was performed by requiring $w$ and its derivative $dw/dx$ to be continuous across the boundaries. These conditions result in a system of equations:

$$\frac{A_m}{\sqrt{r_m}}\text{sech}[A_m(X_m - x_m)] = \frac{A_{m+1}}{\sqrt{r_{m+1}}}\text{sech}[A_{m+1}(X_m - x_{m+1})], \\ A_{m+1}(X_m - x_{m+1}) = \text{arctanh}\left\{\frac{A_m}{A_{m+1}}\tanh[A_m(X_m - x_m)]\right\}, \quad (21)$$

where $X_m$ is the coordinate of the interface between regions $m$ and $m+1$. Equations (21) allow one to obtain iteratively all values $x_m$ and $A_m$. This method does not produce exact solutions, but it gives rise to accurate approximations when the width of soliton is smaller than or comparable to the period of the NL. As in the case of harmonic lattices, the solitons were found to be stable and unstable, respectively, when they reside in the region with the strongest and weakest attractive interactions (the development of the instability in the latter states leads, as one may expect, to the shift into the region with stronger interactions).

Dark-soliton solutions, featuring nonvanishing asymptotic values at $x \to \pm\infty$, were obtained too, for the case of repulsive interactions, with $R(x) > 0$ at all $x$. Such dark solitons feature an intensity dip at the center, conjugate to the $\pi$-phase jump and a strongly modulated (due to the inhomogeneous nonlinearity) intensity distribution. Irrespective of being centered at minima or maxima of $R(x)$, the dark solitons develop an instability (in the latter case the instability develops much faster) and start to move across the NL, which is accompanied by strong radiative losses.

## 5. Solitons in spatiotemporal nonlinear potentials

The possibility of using the FR to control the nonlinearity in the BEC opens the way to create not only static, but also *dynamical* NLs, that are modulated both in space and time. Under specially chosen conditions, exact solutions can be constructed in this complex setting too. In particular, Belmonte-Beitia *et al.* (2008) have implemented similarity transformations to construct explicit solutions to the NLSE with the linear potential and nonlinearity depending on the time and spatial coordinate. The method is based on the transformation of the equation with $x,t$-dependent coefficients, $iq_t = -q_{xx} + V(x,t)q + R(x,t)|q|^2 q$, into the standard stationary NLSE, $\mu W = -W_{XX} - |W|^2 W$, with constant coefficients.

This approach is similar to methodology originally developed by Serkin and Hasegawa (2000 and 2002), that allows one to generate, in a systematic way, an infinite number of novel bright and dark soliton solutions in the temporal-NLSE model with the GVD coefficient, nonlinearity, and gain or absorption depending on the propagation distance. This was done by searching for a transformation of the original NLSE into its version with constant coefficients. Naturally, in the framework of this method the functions describing the local GVD, nonlinearity, and gain (or loss) cannot be chosen independently and are connected by certain relations that are imposed by the form of the transformation for field amplitude that was used in order to reduce the initial NLSE into its counterpart with constant coefficients. The method was also successfully generalized for the case of nonautonomous NLSEs with external linear and parabolic potentials by Serkin, Hasegawa, and Belyaeva (2007).

To perform the necessary transformation of the original NLSE/GPE, Belmonte-Beitia *et al.* (2008) introduced a new function, $q(x,t) = \rho(x,t)\exp[i\phi(x,t)]W[X(\gamma(t)x)]$, where $X$ is an arbitrary function of $\gamma(t)x$ and $\gamma(t)$ characterizes the temporal evolution of the width of solutions. The direct substitution of this *ansatz* into the original NLSE leads to the target equation, $\mu W = -W_{XX} - |W|^2 W$, provided that functions $\rho, \phi, X, \gamma$ and $V(x,t), R(x,t)$ are connected by the following relations:

$$\begin{aligned}
V(x,t) &= (\rho_{xx}/\rho) - \phi_t - \phi_x^2 - \mu(\gamma/\rho)^4, \\
R(x,t) &= \gamma^4/\rho^6, \\
\rho(x,t) &= [\gamma/X'(\gamma x)], \\
\phi(x,t) &= -(\gamma_t/4\gamma)x^2 + \alpha,
\end{aligned} \quad (22)$$

where $\alpha(t)$ is an arbitrary function of time. One can see that selection of arbitrary functions $\alpha(t), \gamma(t)$ and $X(\gamma x)$ fully determines $\rho, \phi$ and also $V, R$, while the shape of function $W(X)$ can be obtained from the NLSE with constant coefficients. In this way, one can generate, e.g., solitons in the time-dependent harmonic trapping potential, for a Gaussian profile of the nonlinearity modulation. Using system (22), solutions have been produced for *breathers*, i.e., solitons exhibiting quasiperiodic oscillations of the amplitude and width, and solitons with a nontrivial motion of the center of mass.

Very recently, nonautonomous matter-wave solitons near the Feshbach resonance in the 1D model of the BEC confined by the harmonic potential with a varying trapping frequency were considered by Serkin, Hasegawa, and Belyaeva (2010). They addressed physically important examples when the amplitude of applied magnetic field that determines the nonlinearity strength varies in time either linearly or periodically, and derived relations between the trapping frequency and resulting nonlinearity coefficient that are required for the integrability of the NLSE in its final form. Thus, it was found that, for the integrability of the NLSE with the periodically varying scattering length, the *reversal* of the sign of coefficient in front of trapping potential is necessary (in other words, the shape of the harmonic potential periodically switches between confining and expulsive configurations). The validity of the 1D description of the BEC by means of the latter model was tested, considering the case when, under the combined action of the time-dependent nonlinearity and confining potential, the compression of the atom cloud from the initial cigar-shaped shape into a quasi-spherical 3D configuration was achieved, leading to the collapse of the soliton.

A similar approach utilizing the transformation of the NLSE with inhomogeneous coefficients was elaborated by Tang and Shukla (2007) for the stationary 1D equation with the cubic-quintic nonlinearity, $\mu w + d^2 w / dx^2 + R(x)w^3 + G(x)w^5 - V(x)w = 0$. They have found particular forms of the periodic modulation of the coefficients in front of the cubic and quintic terms that allow one to transform this equation into its counterpart with $R = $ const and $G = $ const [exact soliton solutions for the latter equation are well known since the work by Pushkarov, Pushkarov, and Tomov (1979)]. In particular, the equation with $R(x) = R_0 \left[ c_1 + c_2 \cos(qx) \right]^{-3}$ and $G(x) = G_0 \left[ c_1 + c_2 \cos(qx) \right]^{-4}$ may be mapped into the constant-coefficient form for $R_0 < 0$, $G_0 > 0$ i.e., in the case of the cubic self-repulsion and quintic self-attraction (in that case, the solitons are actually unstable). A similar analysis, i.e., the transformation into the equation with constant coefficients, which admits exact soliton solutions, was recently reported by Belmonte-Beitia, and Calvo (2009) for the 1D equation with the *purely quintic* nonlinearity and linear potential that may be both $x$- and $t$-dependent.

## 6. Nonlocal nonlinear lattices

Another interesting variation of the topic of NLs in 1D systems is provided by the consideration of periodic NLs imprinted into *nonlocal* nonlinear media. In this case, it was found that the NLs may support solitons with unusual properties. Kartashov, Vysloukh, and Torner (2008a) studied the soliton propagation in layered thermal media made of alternating focusing and defocusing layers. The evolution of the light beam in such an environment is described by coupled equations for field amplitude $q$ and normalized temperature variation $T$:

$$i\frac{\partial q}{\partial z} = -\frac{1}{2}\frac{\partial^2 q}{\partial x^2} - \sigma(x)qT, \quad \frac{\partial^2 T}{\partial x^2} = -|q|^2. \tag{23}$$

where function $\sigma(x) = \sigma_a \, \text{sgn}[\cos(\pi x / d)]$ describes the periodic profile of the nonlinearity. When boundaries of the sample are maintained at equal fixed temperatures, system (23) can be solved with boundary conditions $q, T|_{x=\pm L/2} = 0$, since the temperature distribution depends on the sample's width, $L$. In this setting, the laser beam heats the material, and the released heat diffuses across the entire sample, resulting in local modifications of the refractive index via the thermo-optic effect. Since thermo-optic coefficients are different in different layers, a strong NL is induced whose shape is described by coefficient $\sigma(x)$. Such lattices support a variety of nonlinear excitations, including fundamental, even, dipole and tripole solitons residing at the center of the sample, see Figs. 3(a)-3(d). It is relevant to mention that, while counterparts of dipole and tripole solitons are known in strongly nonlocal uniform materials, even states, which represent in-phase combinations of bright spots, do not exist in uniform nonlocal media.

The beams propagating in nonuniform thermal media induce spatially modulated lattices that depend on the beam's width and peak intensity. Such NLs immobilize solitons and may suppress their transverse drifts. Due to this effect, multipoles and fundamental solitons may form not only in central domains, but also in any focusing domain of the layered medium, even if it is located close to the boundary of the sample [see Fig. 3(e)]. This is in contrast to the uniform focusing thermal medium, where boundaries, if maintained at equal fixed temperatures, repel light that tends to concentrate in the center of the sample. The nonlinear contribution to the refractive index in this setting is oscillatory [see Fig. 3(f)], with the width of the refractive-index distribution by far exceeding the width of the soliton, and decaying almost linearly towards the boundaries. All solitons, including multipoles, residing at the center of the periodic sample, do not feature any threshold power necessary for their existence, but shifted solitons exist only above a power threshold. Interestingly, multipoles in NLs in the thermal media can be stable irrespective of the number of local spots building the soliton, in contrast to the uniform thermal medium, where solitons built of more than four spots are always unstable.

## 7. Dynamical effects: mobility of solitons and splitting of bound states in nonlinear lattices

As mentioned above, the mobility of broad solitons in one-dimensional NLs was studied by Sakaguchi and Malomed (2005a). Additional results concerning the mobility were reported by Zhou *et al.* (2008), who considered the dynamics of *tilted solitons* in lattices represented by a shallow harmonic modulation of the nonlinearity coefficient. They had demonstrated that there exists a certain critical value of the tilt, above which the soliton leaves the original lattice channel and starts travelling across the NL (a similar phenomenon for solitons in linear lattices was analyzed by Kartashov *et al.*, 2004). The critical tilt grows with the increase of the initial soliton's amplitude and the depth of the nonlinearity modulation in the lattice. As in the case of the linear lattices, moving solitons in NLs suffer losses through the emission of radiation waves. Due to this effect, the soliton can be eventually trapped in one of the channels of the NL. The number of the channel where the trapping happens increases with the increase of the original tile, and decreases with the soliton's amplitude.

A dynamical effect of another type, which was studied too, is splitting of $N$-soliton *bound states* in weak NLs. In physical systems modeled by completely integrable evolution equations, including the NLSE with constant coefficients, such multisoliton states are made of sets of several individual solitons, with different amplitudes, which form a nonlinear superposition whose binding energy is exactly zero. The amplitudes of the solitons hidden inside the superposition are given by the corresponding Zakharov-Shabat eigenvalues. Such bound states, corresponding to inputs in the form of $N\operatorname{sech}(x)$, oscillate indefinitely in the NLSE with constant coefficients, as long as perturbations are absent. However, because the bound state made of the set of fundamental solitons has no binding energy, small perturbations, such as those induced by a weak NL, can split the bound state into its fundamental-soliton constituents, as demonstrated by Zhou *et al.* (2008). A similar splitting effect may be produced by a very weak time-periodic modulation of the nonlinearity coefficient, which established a link of the dynamical effects of this type to the models of the "management" (Sakaguchi and Malomed, 2004a; Yanay, Khaykovich, and Malomed, 2009).

Imposing transverse displacements on solitons in materials with the inhomogeneous nonlinearity may generate a number interesting dynamics. Niarchou et al., 2007, have studied soliton oscillations excited when the nonlinearity is parabolically modulated along the coordinate, $R(x) = -1 + \varepsilon x^2$, while the soliton is initially displaced from the center of the parabolic nonlinear trap. It has been shown that the excitation of persistent oscillations of solitons in such a nonlinear trap are associated with the existence of a discrete eigenvalue, $\Omega$, and associated eigenmode, which is actually the translational mode of the soliton, in the corresponding linearization problem for perturbed NLSE, where the perturbation is given by $\varepsilon x^2 |q|^2 q$. The adiabatic perturbation theory for solitons yields expression $\Omega = [(4/3)\varepsilon\mu]^{1/2}$ (here $\mu$ is the chemical potential) for the frequency of the oscillations of the soliton's center, $x_0$, that are governed by equation $d^2 x_0 / dz^2 = -\Omega^2 x_0$ (in fact, this frequency coincides with the eigenvalue of the translational mode, in this case).

## B. Solitons in mixed linear-nonlinear lattices

After the consideration of 1D solitons in models based on the NLs in their pure form, the next natural step is to study the solitons in the systems incorporating mixed lattices with linear and nonlinear components. The theoretical interest in this general setting is explained by the fact that, as discussed in the previous subsection, the fundamental properties of the 1D solitons are conspicuously different in the purely nonlinear lattices and in the free space, or in the presence of the linear lattice potentials, therefore it is natural to address the issue of the competition between the lattices of the linear and nonlinear types. On the other hand, in many cases the physical settings realizing the NLs, such as those provided by photonic crystals and PCFs in optics, give rise to structure that may be naturally classified as mixed linear-nonlinear lattices (first of all, this is the intrinsic structure of the PCFs, as explained above in subsection III.A). This circumstance provides another strong incentive for the detailed analysis of the mixed lattices.

### 1. Basic models of the mixed lattices

Solitons in periodic mixed linear-nonlinear lattices were first studied by Bludov and Konotop (2006). Their starting point was the mean-field description of a boson-fermion mixture with a dominating fermionic component, loaded into a one-dimensional OL. However, the fermions were assumed to be in the spin-polarized state, hence the Pauli principle prevents their direct interaction. It was demonstrated that, under appropriate conditions, this system may be reduced to the NLSE with a periodic linear lattice and periodically modulated nonlinearity. The main features of these systems stem from the fact that the fermionic component is effectively linear, and, at the same time, it modifies linear and nonlinear properties of the effective medium for the bosons. When the Fermi energy is of the order of the amplitude of the lattice potential, it becomes strongly dependent on the spatial coordinate. If, in such a situation, the boson-fermion interaction is not negligible compared to the boson-boson interactions, then, in the mean-field approximation, the fermionic component significantly affects not only the linear potential, but also periodically modifies the effective two-body interactions among bosons [a somewhat similar situation was recently studied by Adhikari, Malomed, Salasnich, and Toigo (2010), who analyzed the spontaneous symmetry breaking of a Bose-Fermi mixture in a symmetric double-well potential; it was concluded that, under appropriate conditions, the fermionic component (a spin-balanced one, in that case) also modified the effective boson-boson interactions, and thus affected the character of spontaneous symmetry breaking]. By taking into account that the fermionic distribution itself is determined by the trap potential, the existence of intrinsic localized modes was pre-dicted in the boson-fermion mixture. It was found that solitons in the semi-infinite gap of the spectrum of the linear lattice exist as long as the nonlinearity is attractive, $R(x)<0$, at least in certain narrow regions. However, the behavior of the solitons in a vicinity of the gap's edge may be dramatically different, depending on whether the average nonlinearity coefficient, $\chi = \int_0^\pi R(x)\phi^4(x)dx$ (here $\phi$ stands for the profile of the Bloch wave with the symmetry corresponding to the selected edge of the gap, and it is assumed that nonlinear and linear lattices are $\pi$-periodic), is negative or positive. For example, when $\chi<0$, the norm of the soliton in the mixed lattice monotonically decreases towards the edge of the gap, and the soliton dramatically broadens, as occurs too with solitons in usual linear lattices. However, for $\chi>0$ the norm acquires a minimum in a vicinity of the gap's edge, and then starts to grow, as the chemical potential approaches the edge of the gap, thus resulting in the existence of a nonzero minimal number of bosons necessary for the creation of the localized mode. A similar effect was encountered for solitons in the first finite bandgap, when the negative value of the average nonlinearity coefficient near the respective gap's edge leads to the existence of a minimal number of bosons necessary for the creation of the GS (gap soliton). Among interesting effects reported in this model is an unusual zigzag dependence of the bosonic norm on the chemical potential in the first finite bandgap, which is a result of successive bifurcations of the soliton branches (see Fig. 4). It was shown that the motion along this dependence is accompanied by the redistribution of the atomic density among minima of the linear potential and that only solitons residing on the lowest loop of this dependence are stable [see profiles corresponding to points A,B in Fig. 4(a)].

Properties of GSs in linear-nonlinear OLs were also analyzed by Abdullaev, Abdumalikov, and Galimzyanov (2007). They derived a coupled-mode system for a shallow lattice that allows one to obtain the profile of GSs in the mixed linear-nonlinear lattice *explicitly*. The coupled-mode equations were derived starting from the NLSE of the form $iq_t + q_{xx} + [r_0 + r_1\cos(2x)]|q|^2 q - \varepsilon\cos(2x)q = 0$, where $r_0>0$ ($r_0<0$) corresponds to the attractive (repulsive) condensate, and $r_1>0$ ($r_1<0$) corresponds to the out-of-phase (in-phase) linear and nonlinear lattices, respectively. In the course of the derivation, the field was represented as a sum of backward- and forward-propagating waves, $A(x,t)\exp(ix-it)$ and $B(x,t)\exp(-ix-it)$. Then, in the case of the shallow linear lattice, $\varepsilon \ll 1$, one can derive a system of coupled-mode equations that describe the interaction between the forward- and backward-propagating waves. Remarkably, the resulting system admits an exact analytical solution, $q(x,t) = 2U^{1/2}(x)\cos[x-\theta(x)/2]\exp[-i(\mu+1)t]$, for the profile of the GS in the mixed linear-nonlinear lattice, where

$$U = -\frac{2\mu - \varepsilon \cos\theta}{3r_0 + 2r_1 \cos\theta}, \quad \cos\theta = \frac{1 - \gamma^2 \tanh^2(\beta x)}{1 + \gamma^2 \tanh^2(\beta x)}, \tag{24}$$

while $\gamma = [(\varepsilon - 2\mu)/(\varepsilon + 2\mu)]^{1/2}$, $\beta = (\varepsilon^2 - 4\mu^2)^{1/2}/4$, and $\mu$ is the chemical potential. The width of this soliton is thus defined by $\mu$ and is inversely proportional to the amplitude of the linear lattice, while the soliton's amplitude is inversely proportional to the strength of the NL. Such GSs can exist and may be stable even when the constant part of the nonlinearity is absent, $r_0 = 0$.

A two-component 1D model for the binary BEC trapped in a combined linear-nonlinear lattice, with a common spatial period of both sublattices, was recently introduced by Golam Ali, Roy, and Talukdar (2010). The consideration of the solitons in the model was actually reported for symmetric solitons only (with equal numbers of atoms in the two components), and their stability was analyzed within the framework of the simplest version of the VK criterion.

As well as in the case of purely nonlinear lattice, solitons in the mixed model with the modulation functions of the KP type is a natural object for the consideration, both because of its relevance to the description of experimentally available systems in optics, and due to the possibility to construct relevant solutions in a semi-explicit form. Pursuing this line of the analysis, Kominis (2006) had constructed analytical soliton solutions in the periodic nonlinear KP system, built as a periodic concatenation of linear and nonlinear layers. Stationary soliton profiles in the system of this type are described by the following stationary equation $d^2w/dx^2 + [V(x) - \mu]w + R(x, w^2)w = 0$, where $\{V(x), R(x, w^2)\} = \{\varepsilon_n, \mathcal{N}(w^2)\}$ in nonlinear layers with a general nonlinearity law, $\mathcal{N}(w^2)$, and $\{V(x), R(x, w^2)\} = \{\varepsilon_l, 0\}$ in the linear layers. This equation was solved by analyzing $(w, dw/dx)$ diagrams in the respective phase space. The full solution was obtained by matching partial solutions found inside the linear and nonlinear layers, under the condition of the continuity of $w$ and $dw/dx$, for the case when the propagation constant is such that in the linear layers the equation admits sinusoidal solutions, while the nonlinear equation, by itself, gives rise to the usual solitons. It was shown that, for $\mu$ corresponding to the case when linear layers of width $L$ contain an integer number of half-periods of the sinusoidal solution, i.e., $\mu_n = \varepsilon_l - (n\pi/L)^2$, $n = 1, 2, \ldots$, the continuity conditions are met at all boundaries, and any solution to the NLSE starting from a point of the homoclinic orbit inside the nonlinear part at some $x$ returns to the homoclinic orbit after passing the linear part. Then, it follows the homoclinic orbit again. Thus, although periodically interrupted by passing the linear segments, the solution asymptotically approaches the origin at $x \to \pm\infty$. This is shown in Fig. 5, where the phase-space representations of the homoclinic orbit and linear system are superimposed. The branches of the so constructed solutions coincide with parts of the soliton profile and parts of the periodic orbits generated by the linear equation. The information about the full shapes of the solutions can be obtained from the phase-space portrait: For odd $n$ the solutions lie in both parts of the homoclinic orbit and thus change their sign between the nonlinear layers, while modes with a constant sign of $w$, lying only on one branch of the homoclinic orbit, are obtained for even $n$. Using this method, not only localized but also periodic solutions can be constructed. The shapes of the solutions are given by $w(x) = (-1)^{nk} v(x - kL, \mu_n)$ in nonlinear layers, and $w(x) = a_k \sin[(\varepsilon_l - \mu_n)^{1/2} x + \phi_k]$ in linear ones, where $v(x)$ is the soliton solution of the corresponding nonlinear equation with constant coefficients corresponding to propagation constant $\mu_n$, while $a_k, \phi_k$ are directly obtained from the continuity conditions. Typical examples of the solitons for the self-focusing cubic nonlinear layers with $\varepsilon_n < \varepsilon_l$, where $v = \pm(\mu - \varepsilon_n)^{1/2} \text{sech}[(\mu - \varepsilon_n)^{1/2} x]$, existing for *discrete* values of the propagation constant, $\varepsilon_n < \mu_n < \varepsilon_l$, are depicted in Figs. 5(c) and 5(d). The modes corresponding to $n = 2$ and $3$, obtained by dint of this method, may be stable upon the propagation. An extended discussion of this method of the construction of analytical solutions in nonlinear systems with piecewise-constant parameters can be found in the paper by Kominis and Bountis (2010).

A similar approach was utilized by Kominis and Hizanidis (2006) to construct spatially extended dark solitons (i.e., localized excitations on a finite periodic background) in the KP model with defocusing layers. In this case, the solutions inside the layers with the cubic nonlinearity are $v = \pm[(\varepsilon_n - \mu)/2]^{1/2} \tanh([(\varepsilon_n - \mu)/2]^{1/2} x)$, with $\mu < \varepsilon_n$. The difference against the previous situation is that the system is supposed to have a heteroclinic orbit (on the contrary to its homoclinic counterpart corresponding to the bright soliton), and the solution returns to this heteroclinic orbit, after passing a linear layer, and asymptotically approaches saddle points as $x \to \pm\infty$. It is worth to mention that, in contrast to the case of the self-focusing nonlinearity, this method predicts the existence of an *infinite set* of solutions for the defocusing nonlinearity, corresponding to values of the propagation constant $\mu_n < \varepsilon_n, \varepsilon_l$ that can be found even for $\varepsilon_n = \varepsilon_l$. Some of the dark solitons obtained by means of this method were shown to be stable.

In a similar vein, Rapti et al. (2007) studied the competition of shallow linear and nonlinear lattices and its effect on the stability and dynamics of bright solitons. Both lattices were considered in a perturbative framework, and the technique of the Hamiltonian perturbation theory was utilized to obtain the information about the existence of solutions and conditions for their linear stability. A particularly interesting result obtained in that context is a tunable cancellation of the pinning potential induced by the weak linear and nonlinear lattices, which gives rise to an increased mobility of the solitons. Starting from the respective NLSE, $iq_t = -(1/2)q_{xx} - [1 + \varepsilon R(x)]|q|^2 q + \varepsilon V(x)q$ with linear, $V(x) = B\cos(k_2 x + \phi)$, and nonlinear, $R(x) = A\cos(k_1 x)$, harmonic lattices and $\varepsilon \ll 1$, the dynamics of the excitation was considered in the form of $q(t=0) = \mu^{1/2}\mathrm{sech}[\mu^{1/2}(x-\xi)]$, with a center at point $x = \xi$, that would be an exact stable soliton solution for $\varepsilon = 0$. If $\varepsilon \neq 0$ the translational invariance of the equation is broken, which may naturally lead to the drift mode of the destabilization of localized states. It was shown that the evolution of the soliton's center obeys equation $d^2\xi/dt^2 = -(1/N)\partial V_{\mathrm{eff}}/\partial \xi$, where $N$ is the norm of the soliton, and the effective potential can be easily evaluated from the perturbing part of system's Hamiltonian, as $V_{\mathrm{eff}} = \varepsilon \int_{-\infty}^{\infty} [V(x)|q|^2 - (1/2)R(x)|q|^4]dx$. After the substitution of the presumed sech-soliton shape, this yields

$$V_{\mathrm{eff}} = -(\varepsilon\pi/12)Ak_1(k_1^2 + 4\mu)\frac{\cos(k_1\xi)}{\sinh(\pi k_1/2\mu^{1/2})} + \varepsilon\pi B k_2 \frac{\cos(k_2\xi + \phi)}{\sinh(\pi k_2/2\mu^{1/2})}. \quad (25)$$

Equation (25) provides for a basis for the understanding of the dynamics of the soliton for different values of the parameters of the linear and nonlinear lattices. Using this approach and also calculating eigenvalues of the associated linearized problem, it was demonstrated, *inter alia*, that a gradual increase of the NL amplitude $A$ can stabilize otherwise unstable solitons residing at maxima of the linear potential (which corresponds to minima of the refractive index in optics). A similar effect can be achieved by increasing wavenumber $k_1$ of the NL. The variation of $A$ and $k_1$ is accompanied by the corresponding deformations of effective potential (25), so that a minimum develops around the position of the input soliton at ceratin parameter values, resulting in the stabilization of the soliton. For $k_1 = k_2$ and $\phi = 0$, Eq. (25) predicts the mutual cancellation of the effective potentials induced by the linear and nonlinear lattices at $A_{\mathrm{cr}} = 12B/(k_1^2 + 4\mu)$, thereby restoring a regime of the effective translational invariance. In this regime, *tilted solitons* propagate without trapping, keeping their initial velocities and undergoing only small amplitude modulations under the action of the effective potential, which is very weak in this case.

The stability and drift of soliton modes in the presence of competing linear and nonlinear harmonic lattices with an arbitrary amplitude of the modulation of the refractive index and nonlinearity (but still in the regime of the continual medium) were studied by Kartashov, Vysloukh, and Torner (2008b). They considered the competition between out-of-phase linear and nonlinear lattices, with the beam's dynamics obeying (in the optical notation) equation $iq_z = -(1/2)q_{xx} - [1 - \sigma R(x)]|q|^2 q - pR(x)q$. Such lattices, with different depths of nonlinearity modulation $\sigma$, support a variety of solutions including odd, even, di-

pole, and triple-mode solitons. The power of the simplest odd soliton increases monotonously with the increase of the peak amplitude, while the intensity maximum remains in the same channel only if the nonlinearity modulation is not too deep, otherwise the spatially non-uniform self-focusing dominates over the linear refraction, resulting in the deflection of light towards regions where the nonlinearity is stronger. Eventually, this results in the development of two peaks located around minima of $R(x)$, while the soliton's norm becomes a non-monotonous function of the propagation constant. In contrast, even (symmetric) solitons which have two intensity maxima at low amplitudes, may fuse into a single peak located between the maxima of $R$. Therefore, the nonlinearity modulation gives rise to unusual power-controlled shape transformations of the lattice solitons. Multipole-mode solitons may cease to exist in this setting if the nonlinearity modulation depth exceeds a certain critical value, which is again in contrast to properties of multipoles in the linear lattice. While the VK stability criterion is satisfied for fundamental odd solitons, they may become unstable due to the nonlinearity modulation above a critical power. This is accompanied by a violation of the spectral stability criterion (Fibich, Sivan, and Weinstein, 2006). The resulting drift instability causes a rapid displacement of the soliton into regions where $R$ attains a local minimum. The stability domain for odd solitons vanishes completely for sufficiently large $\sigma$, while the entire family of even solitons may become stable at the same point. The soliton mobility in this setting is intimately related to their stability. To set a soliton in motion across the lattice, one has to kick it by the application of the phase tilt, i.e., multiplying the soliton by $\exp(i\alpha x)$. While in the usual linear lattices the *critical tilt*, at which the soliton starts to move, grows rapidly and monotonically with the power, in NLs the critical tilt turns out to be a non-monotonous function of the propagation constant, see Fig. 6(a). It completely vanishes not only in the linear limit, but also exactly at the point where the odd solitons become unstable. Very small tilts may result in an almost radiationless motion of odd solitons across the lattice in the region of their drift instability. Even for tilts slightly exceeding the critical value, the solitons move across the lattice almost without losses and do not experience trapping, even in the stability region [Fig. 6(c)]. The situation is similar for even solitons that become mobile at the edge of their stability domain [see Fig. 6(b) for the corresponding critical tilt]. Note that similar enhancement of the mobility due to the stability inversion is possible in discrete systems, as discussed by Öster, Johansson, and Eriksson, 2003 (see subsection IV.F), and by Susanto et al. (2007).

The power-dependent location of stationary solitons and their stability in linear-nonlinear lattices was analyzed by Kominis and Hizanidis (2008). While it is known that in the simplest 1D linear lattices with the harmonic spatial modulation of the refractive index the soliton's position and stability do not depend on its power, it was shown that in more complex structures, where the refractive index, nonlinearity, or both of them are modulated with multiple wavenumbers, the position and stability of the soliton become functions of the power [see also results of Sakaguchi and Malomed (2010) presented below]. Melnikov's theory was used to study the respective power-dependent bifurcations and to determine specific positions, with respect to the spatial structure, where solitons can be located. This theory allows a simple analytical treatment of a large variety of the refractive-index and nonlinearity landscapes, even shifted or with incommensurate spatially periodic modulations. The respective NLSE was taken as $iq_z + q_{xx} + 2|q|^2 q + \varepsilon[V(x)q + R(x)|q|^2 q] = 0$, giving rise to soliton solutions $q = w(x)\exp(i\mu z)$. The resulting equation for stationary profiles $w(x)$ corresponds to a one-degree of freedom dynamical system with Hamiltonian $H = (1/2)(p^2 - \mu q^2 + q^4 + \varepsilon[V(x)q^2 + R(x)q^4])$, written in terms of variables $(q, p) = (w, dw/dx)$. The last term in the Hamiltonian, representing linear and nonlinear lattices, was treated as a first-order perturbation. The case of $\mu > 0$ was considered, for which a homoclinic (bright) soliton solution of unperturbed translationally invariant system exists. This solution is formed by the merger of the stable and unstable manifolds of the hyperbolic saddle point located at the origin in the phase space. This highly degenerate structure is expected to break under perturbations, and may perhaps yield transverse homoclinic orbits or no orbits at all. Thus, the solution to the unperturbed NLSE, $\{q_0, p_0\} = \{\pm\mu^{1/2}\text{sech}[\mu^{1/2}(x-x_0)], \mp\mu\,\text{sech}[\mu^{1/2}(x-x_0)]\tanh[\mu^{1/2}(x-x_0)]\}$, describes, as a matter of fact, an infinite number of solutions homoclinic to the origin, $(q, p) = (0, 0)$, which correspond to different $x_0$. The Melnikov's theory predicts that, in the presence of perturbations, only a discrete set of such solutions may persist, corresponding to values of $x_0$ given by zeros of the corresponding Melnikov's function, $M(x_0) = -\varepsilon \int_{-\infty}^{+\infty} p_0(x)[V(x)q_0(x) + R(x)q_0^3(x)]dx$. For the generic form of the modulation of the linear refractive index, $V = \sum_m a_m \cos(k_m x + \phi_m)$, and nonlinearity, $R = \sum_m b_m \cos(r_m x + \varphi_m)$, one thus obtains

$$M(x_0) = \frac{\varepsilon\pi\mu^{1/2}}{2}\sum_m \frac{a_m k_m^2}{\sinh(\pi k_m / 2\mu^{1/2})}\sin(k_m x_0 + \phi_m)$$
$$+ \frac{\varepsilon\pi\mu^{1/2}}{24}\sum_m \frac{b_m r_m^2(r_m^2 + 4\mu)}{\sinh(\pi r_m / 2\mu^{1/2})}\sin(r_m x_0 + \varphi_m). \tag{26}$$

Notice that the structure of the Melnikov's function (26) closely resembles the effective potential (25). By using function (26), one can determine not only stationary positions of solitons that correspond to zeros of (26), apparently depending on $\mu$ (hence, on the soliton power), but also make conclusions about the stability of the solitons, that depends on the sign of derivative $\partial M / \partial x_0$.

The KP system may also be used as a natural model for 1D photonic crystals, built as a periodic lattices of waveguiding nonlinear channels of width $D$ separated by empty channels of width $(L-D)$. Mayteevarunyoo and Malomed (2008) analyzed spatial solitons in the model of this type with *defocusing* nonlinearity in the waveguiding channels. In that setting, several interesting effects were predicted due to the competition between the linear trapping potential and the defocusing nonlinear pseudopotential. The impact of ratio $D/L$, which determines the bandgap structure of the lattice's spectrum, on the properties of solitons emerging in different finite bandgaps was studied. It was found that, for $D/L \to 1$, properties of solitons in this version of the KP model approach those of usual GSs, but for $D/L \sim 1/2$ they are quite different. For a fixed peak value of the refractive index, the solitons cease to exist when $D/L$ becomes smaller than a certain critical value. Besides the fundamental single-peak solitons, families of spatially symmetric (even) modes with two, three or four peaks were obtained. For all such states, the norm is a decreasing function of propagation constant $\mu$, and all the modes get strongly stretched near gap edges. It was found that, while for $D/L \sim 0.75$ such solitons are stable, for the intermediate case, $D/L \sim 0.50$, there exists an *intrinsic stability border* in the middle of the second bandgap. The transition from stable to unstable solutions goes through a specific flat-top shape, see Fig. 7. Deeper into the instability region, where the self-defocusing nonlinear pseudopotential becomes stronger than the trapping linear potential, the higher-order solitons develop inverted shapes: Peaks emerge above the flat-top background, placing themselves in empty spaces between the guiding channels. In the model with narrow channels, $D/L \sim 0.25$, fundamental and higher-order solitons exist only in the first finite bandgap, where they are stable, despite the fact that they also feature inverted shapes with peaks in linear layers.

The materials with transversally inhomogeneous nonlinearity and linear refractive-index landscapes can support not only localized soliton solutions, but also periodic or modulated amplitude waves (Porter et al., 2007). Such nonlinear waves are also physically interesting objects. For the landscapes where nonlinearity $R(x)$ does not change its sign, the introduction of the wave amplitude as $\rho = R^{1/2} q$ transforms the original NLSE with inhomogeneous nonlinearity $R(x)$ and linear potential $V(x)$ into equation $i\rho_z = -(1/2)\rho_{xx} + |\rho|^2 \rho + V(x)\rho + V_{\text{eff}}(x)\rho$, where the inhomogeneity of the nonlinearity is mapped into the effective potential, $V_{\text{eff}}(x) = (\alpha''/2\alpha) - (\alpha'^2/\alpha^2) + (\alpha'/\alpha)\partial/\partial x$, with $\alpha = R^{1/2}$. For $R(x) = r_0 + r_1 \sin^2(kx)$ and $r_1 \ll r_0$, this can be approximated by a *superlattice* potential plus a first-derivative operator term. Approximate harmonic analytical solutions to the above-mentioned equation were found in the small-amplitude limit. Further investiga-tions had shown that such solutions are weakly unstable, although the on-site periodic waves, whose maxima coincide with maxima of the NL, are more robust than their off-site counterparts.

Besides models with periodically modulated nonlinearity and local refractive index, several settings were considered with parabolic linear potentials and unusual nonlinearity profiles. These results constitute a relevant addition to the studies of solitons in the mixed linear-nonlinear lattices. In particular, Theocharis et al. (2005) investigated the dynamics of dark and bright matter-wave solitons for the nonlinearity coefficient linearly varying in the transverse direction, *viz.* $R(x) = 1 + \delta x$. In this case, the spatially-dependent nonlinearity leads to a *quasi-gravitational* potential, as well as to a renormalization of coefficient $\Omega^2$ of the parabolic potential, $V(x) = \Omega^2 x^2 / 2$, a feature that allows one to control the motion of fundamental and higher-order solitons. By treating the linear and nonlinear potentials as perturbations and considering the motion of the input fundamental soliton, $q(z=0) = \mu^{1/2} \text{sech}[\mu^{1/2}(x-x_0)]\exp(ikx)$, with width $\mu^{-1}$ much smaller than the characteristic spatial scale of the trapping potential, $\Omega^{-1/2}$, and the scale of the nonlinearity variation, $\delta^{-1}$, the following equation was derived for the motion of the soliton's center:

$$\frac{d^2 x_0}{dz^2} = -\frac{\partial V(x_0)}{\partial x_0} + \frac{\mu}{6R^2(x_0)}\frac{\partial R^2(x_0)}{\partial x_0}. \tag{27}$$

This equation describes the motion of a unit-mass particle in the presence of the respective potential, $V_{\text{eff}} = (1/2)(\Omega^2 - \delta\beta)x_0^2 - \beta x_0$ with $\beta = \mu\delta / 3(1+\delta x_0)^2$. The potential includes an effective gravitational potential that induces an acceleration of the soliton towards larger values of $x_0$, a modification in the frequency of free oscillations of the soliton, and also shows that the effective potential may vanish or become expulsive for $\delta\beta \geq \Omega^2$, as shown in Fig. 8. Thus, depending on nonlinearity gradient $\delta$, the soliton may either undergo periodic oscillations or accelerate indefinitely. It was demonstrated that a bound soliton state may split into its constituents (which will perform oscillations inside the trap, interacting with each other) due to the inhomogeneous nonlinearity, and particular solitons with highest amplitudes may be released from the trap, which may be controlled by properly selecting its parameters. Oscillations of dark solitons in traps with the inhomogeneous nonlinearity were addressed too.

A related analysis was reported by Zezyulin *et al.* (2007), who investigated the stability of higher-order nonlinear modes of BEC loaded into parabolic trapping potentials, and showed that a local variation of the nonlinearity strength allows one to create multistable configurations. Stationary solutions, in the usual form of $q = w(x)\exp(-i\mu z)$, were consid-ered for the corresponding GPE, $iq_t = -q_{xx} + x^2 q + R|q|^2 q$. In the case of the uniform nonlinearity, several branches of the solutions were found. Each branch displays a one-to-one correspondence between the chemical potential and the norm, bifurcating from states of the linear harmonic oscillator. The solutions corresponding to the first two branches (fundamental solitons and dipoles) are linearly stable, while higher-order solutions are linearly unstable for both attractive and repulsive nonlinearities (in particular, third-order solutions with the repulsive nonlinearity are stabilized above a critical value of the norm). This picture drastically changes for the step-like nonlinearity, with $R = R_+$ for $x > 0$ and $R = R_-$ for $x < 0$. In this case, the branches of solutions that were monotonic for constant $R$ exhibit a snake-like behavior with a number of turning points which increases with the number of the branch, $n$, i.e., the number of poles (constituents) in the soliton (Fig. 9). The increase of the norm is accompanied by a gradual localization inside the attractive part of the space (at $x < 0$). The stability of the solutions on each branch depends on the parameter range, as indicated by bolder regions (stable solitons) against lighter ones (unstable solitons) in Fig. 9. Regions of the stability and instability in this setting may alternate several times along each branch of the solutions, and there exist regions of multistability (simultaneous existence of several stable solitons with equal values of $\mu$, belonging to a common branch), a phenomenon that is not present in the case of the spatially uniform nonlinearity. The evolution of bright solitons in parabolic trapping potentials and Gaussian nonlinearity landscapes, as well as the evolution of dark solitons in periodic nonlinearity landscapes, were also studied numerically by Hao *et al.*, 2008.

Perez-Garcia and Pardo, 2009, investigated properties of fundamental solitons in the NLSE with spatially inhomogeneous interactions, trapped in strong box-shaped and parabolic potentials. They showed that, when the nonlinearity is repulsive and vanishes (or its coefficient takes smaller values) in a certain spatial region, the localization of the atomic density, $n = w^2$, occurs in the regions where the nonlinearity vanishes. This localization becomes more and more pronounced with the increase of the soliton's norm. The chemical potential has a cutoff value in such systems (the norm diverges at the cutoff point), hence it takes values in a finite interval. When the norm of the soliton becomes sufficiently large, the density grows only in regions with vanishingly weak interactions, while in regions with nonzero interactions the density remains virtually unchanged. By tuning the control (magnetic or optical) fields, in terms of BEC, this phenomenon can be used to design regions with large particle densities in various geometries.

A related idea of the creation of BEC configurations with unusual spatial density distributions, when the nonlinearity is tuned from attractive (at the periphery of the BEC cloud) to repulsive (in the center of the cloud) by a far-off-resonant optical field in an ex-ternal parabolic linear potential was proposed by Dong, Hu, and Lu (2006). It was found that this setting is characterized by the existence of a certain maximal soliton's norm, above which one cannot find localized soliton solutions.

## 2. Effects of commensurability between linear and nonlinear lattices

An interesting aspect of the soliton dynamics in the 1D model combining the linear and nonlinear lattices is a possibility to study effects induced by the commensurability and incommensurability between the two lattices. This problem was recently studied by Sakaguchi and Malomed (2010). The analysis was based on the following variety of the GPE,

$$i\frac{\partial q}{\partial t} = -\frac{1}{2}\frac{\partial^2 q}{\partial x^2} - [\varepsilon \cos(2\pi x) - g\cos(\pi qx)|q|^2]q, \qquad (28)$$

where $g = -1$ or $g = +1$, the period of the linear lattice is scaled to be $L_{\mathrm{lin}} = 1$, and *commensurability index* $q$ determines the period of its nonlinear counterpart, $L_{\mathrm{nonlin}} = 2/q$. Three basic cases were considered, *viz.*, the direct commensurability between the lattices, $L_{\mathrm{lin}} = L_{\mathrm{nonlin}}$, i.e., $q = 2$, subharmonic commensurability, corresponding to $L_{\mathrm{lin}} = L_{\mathrm{nonlin}}/2$, i.e., $q = 1$, and $q = \sqrt{5} - 1$, which represents the case of incommensurability.

Fixing $\varepsilon > 0$ and placing the center of the solitons at $x = 0$, a family of solitons (termed ordinary ones), that resemble usual odd (*on-site-centered*) lattice solitons, was found for $g = +1$, while for $g = -1$ one obtains a GS family. The general shape of the families of both the ordinary solitons and GSs in the case of the direct commensurability, $q = 2$, are similar to their counterparts in the usual models with the uniform nonlinearity: There is no existence threshold, and the entire families are stable [the case of commensurate linear and nonlinear lattices with a nonzero average value of the nonlinearity coefficient, corresponding to the repulsion, was considered by Bludov, Brazhnyi, and Konotop (2007), who had concluded that a finite existence threshold in terms of the norm may exist in that case].

However, properties of both the ordinary and GS families are *completely different* in the cases of the subharmonic commensurability, $q = 1$, and incommensurability, $q = \sqrt{5} - 1$. Namely, in both these cases there is a *finite threshold norm* necessary for the existence of the solitons, similar to the case of the purely nonlinear lattice (Sakaguchi and Malomed, 2005a), which was considered abobe in subsection IV.A.1, and only *parts* of the soliton families are stable, *viz.*, those with $d\mu/dN < 0$ and $d\mu/dN > 0$, as concerns, respectively, the ordinary solitons and GSs. The former stability condition is tantamount to the usual VK criterion, while the latter one, termed *anti-VK criterion* by Sakaguchi and Malomed (2010), is specific to GSs, and may be justified with the help of the averaging approximation. The corresponding dependencies $\mu(N)$ are shown in Figs. 10(a) and 10(c). Notice that $\mu(N)$ curves for the GS families in Fig. 10(c) feature turning points, except for the case of the direct commensurability $(q=2)$. The presence of the turning points makes it possible to actually test the validity of the above-mentioned anti-VK criterion. Stability borders for ordinary and gap solitons are depicted in Figs. 10(b) and 10(d), respectively. All the ordinary solitons are stable above the border shown in Fig. 10(b), while in the case of GSs there is a single stable GS above the upper and beneath lower lines in Fig. 10(d), and three solitons - two stable and one unstable - in the *bistability region* between the two lines.

For the analytical consideration of the broad GSs whose chemical potential, $\mu$, is close to the edge of the first finite bandgap, one can adopt *ansatz* $q(x,t) = \Phi(x,t)\cos(\pi x)$, where the second multiplier emulates the respective Bloch function, and $\Phi(x,t)$ is a slowly varying function, for which an effective equation can be derived by means of the averaging method:

$$i\frac{\partial \Phi}{\partial t} = -\frac{1}{2m_{\text{eff}}}\frac{\partial^2 \Phi}{\partial x^2} + g_{\text{eff}}|\Phi|^2\Phi. \tag{29}$$

Here the effective mass is $m_{\text{eff}} = -\varepsilon/(2\pi^2 - \varepsilon)$, and $g_{\text{eff}} = \langle \cos^4(\pi x)\cos(\pi q x)\rangle$ is the spatially averaged nonlinearity coefficient, which is different from zero in the following cases: $g_{\text{eff}}(q=0) = 3/4$, $g_{\text{eff}}(q=2) = 1/2$, and $g_{\text{eff}}(q=4) = 1/8$. The description of the GSs in the present form makes sense if, as usual, the approximation yields $m_{\text{eff}} < 0$, as the GSs are supported by the interplay of the repulsive nonlinearity, $g_{\text{eff}} > 0$, and the *negative effective mass* (see reviews by Brazhnyi and Konotop, 2004, and Morsch and Oberthaler, 2006). Equation (29) predicts a width-amplitude relation, $W \sim 1/A$, for broad gap solitons at $q=2$, i.e., in the case of the direct commensurability.

In the case of the subharmonic commensurability, $q=1$, where the previous approach yields $g_{\text{eff}} = 0$, one may use an approximation with two slowly varying amplitudes, $q(x,t) = [\Phi_1(x,t)\cos(\pi x) + \Phi_4(x)\cos^4(\pi x)]\exp(-i\mu t)$. After the elimination of amplitude $\Phi_4(x)$ in favor of $\Phi_1(x)$, this approximation leads to the stationary equation with an effective *quintic nonlinearity*, while the cubic term does not appear:

$$\mu\Phi_1 = -\frac{1}{2m_{\text{eff}}}\frac{d^2\Phi_1}{dx^2} - \frac{15m_{\text{eff}}}{8(\pi^2 - 2m_{\text{eff}}\mu)}|\Phi_1|^4\Phi_1. \tag{30}$$

Equation (3) admits exact soliton solutions, $\Phi_1(x) = A/\cosh^{1/2}(kx)$, with $A^2 = (4\pi^2 + k^2)^{1/2}k/(20^{1/2}m_{\text{eff}})$ and $\mu = -k^2/8m_{\text{eff}}$, where inverse width $k$ is an arbitrary parameter of the soliton family. These solutions are characterized by scaling $W \sim 1/A^2$.

As shown by Sakaguchi and Malomed (2010), the analytical approach based on the averaging method may also be applied to the ordinary broad solitons. Besides that, direct simulations demonstrate that broad solitons of both types are mobile, as the application of a *kick* to them, i.e., multiplication by $\exp(iPx)$ with arbitrary momentum $P$, can readily set them in stable motion. On the other hand, collisions between such moving solitons are essentially inelastic, giving rise to generation of additional solitons.

## 3. Models admitting exact solutions

A possibility of designing special models including nonlinear and linear lattices which admit exact solutions for trapped states is a subject of obvious interest, as exact solutions provide for specific insight into properties of models of the present type. As mentioned above, an approach to this problem was elaborated by Belmonte-Beitia, Pérez-Garcia, and Vekslerchik (2007), who constructed models that, together with appropriate solutions, could be transformed into the NLSE with constant coefficients. Another class of models which support particular solutions for exact periodic and solitary modes, and which *cannot* be reduced to the usual NLSE, was recently introduced by Tsang, Malomed, and Chow (2009). The respective GPE was taken as

$$i\frac{\partial \Psi}{\partial t} + \frac{1}{2}\frac{\partial^2 \Psi}{\partial x^2} + R(x)|\Psi|^2 \Psi - V(x)\Psi = 0 \tag{31}$$

Its exact solutions have been obtained on the basis of the Jacobi's elliptic functions of three types, *viz.*, cn, dn, and sn. In the absence of the linear potential $(V = 0)$, the cn-type waves were constructed in the form of

$$\begin{aligned}\Psi(x,t) &= \frac{A_0 \text{cn}(x,k)}{\sqrt{1+b\text{cn}^2(x,k)}}\exp(-i\mu t),\\ A_0^2 &= g_0[-b(2+3b)+(1+b)(1+3b)k^2],\\ \mu &= [1+3b-(2+3b)k^2]/2.\end{aligned} \tag{32}$$

These exact solutions are supported by the following form of the nonlinearity-modulation function:

$$R(x) = \frac{g_0 + g_1 \text{cn}^2(x,k)}{1 + b\,\text{cn}^2(x,k)}, \quad g_1 = \frac{g_0 b}{2} \frac{b(1+3b) + (1+b)(1-3b)k^2}{b(2+3b) - (1+b)(1+3b)k^2}. \tag{33}$$

Here the modulus of the elliptic cosine, $k$, and constant $b$, which take values, respectively, $0 < k \leq 1$ and $b > -1$, are two *free parameters* of the solution family. An additional sign parameter, $g_0 = \pm 1$, corresponds to the nonlinearity which is, on the average, attractive or repulsive. The above cn-type waves may be stable at $g_0 = -1$, $-1 < b < -1/2$. An example of such a stable wave is shown in Fig. 11(a). Notice that, in the entire stability area of these waves, the nonlinearity modulation function $R(x)$ in Eq. (31) is a *sign-changing* one. Exact periodic solutions to NLSE (31) with $R = \text{const}$ were found earlier by Carr, Clark, and Reinhardt (2000), and Bronski, Carr, Deconinck, and Kutz (2001).

Two other families of solutions to Eq. (31), based on elliptic functions dn and sn, were constructed too. Unlike the cn-type solutions, they may be stable only when the periodic modulation of the nonlinearity coefficient, $R(x)$, is combined with the action of a specially chosen linear potential, $V(x)$, i.e., in the case of the mixed linear-nonlinear lattice. In particular, the dn-type solution can be written as

$$\begin{aligned} \Psi(x,t) &= \frac{A_0 \text{dn}(rx,k)}{\sqrt{1 + b\,\text{dn}^2(rx,k)}} \exp(-i\mu t), \\ A_0^2 &= \frac{3br^2(1+b)[bk^2 - (1+b)]}{2(g_1 - g_0 b)}, \\ \mu &= r^2[(1+3b)k^2 - (2+3b)]/2, \end{aligned} \tag{34}$$

with the corresponding modulated nonlinearity coefficient and linear potential given by expressions

$$\begin{aligned} R(x) &= \frac{g_0 + g_1 \text{dn}^2(rx,k)}{1 + b\,\text{dn}^2(rx,k)}, \quad V(x) = \frac{V_0 \text{dn}^2(rx,k)}{1 + b\,\text{dn}^2(rx,k)}, \\ V_0 &= \frac{r^2}{2} + \frac{br^2}{2}[(3b+1)k^2 - (3b+2)] - \frac{3g_1(b+1)}{2(g_0 b - g_1)}[bk^2 - (1+b)]. \end{aligned} \tag{35}$$

Unlike the solutions obtained for $V = 0$, which include two free continuous parameters, the family based on Eqs. (34) and (35) depends on *four* independent parameters, *viz.*, $g_1, r, b, k$, while $g_0 = \pm 1$ is the additional sign coefficient, as before. Unlike the family of the cn-type waves, which is stable only with $g_0 = -1$, the dn-type waves are stable at $g_0 = +1$, when the average nonlinearity is attractive (exact solutions of the third type, based on functions sn, may be stable only for $g_0 = -1$, like their cn counterparts, but solely if an appropriate linear potential is added). A typical example of the stable evolution of the dn-type wave is shown in Fig. 11(b). A noteworthy feature of this solutions is that maxima of the density, $|\Psi(x)|^2$, coincide with maxima of $V(x)$ and with minima of $R(x)$ (i.e., linear and nonlinear potentials are *competing* ones, in this case). The above- mentioned periodic solutions reduce to solitons in the limit of $k \to 1$.

### 4. Lattices with a local nonlinear defect

A simple physical system where the periodic modulation of the linear refractive index is combined with a spatially inhomogeneous nonlinearity is represented by a periodic KP lattice with a single nonlinear defect, which represents a thin-layer nonlinear waveguide. Solitons supported by such lattices were analyzed by Sukhorukov and Kivshar, 2001. In that work, linear lattice $V(x)$ was approximated by a piecewise-constant function, while the defect was accounted for by a specific term, $\delta(x)(\alpha + \beta|q|^2)q$, where $\delta(x)$ is the delta-function, while positive/negative $\beta$ corresponds to the self-focusing/defocusing. Different situations were analyzed, corresponding to possible combinations of the signs of coefficients $\beta$ and $\alpha$ (the latter defines the defect in the linear limit). In particular, for positive defects with $\alpha > 0$, localized modes exist already in the linear regime. The modes originating in both semi-infinite and first finite bandgaps, bifurcating from the corresponding linear states, are stable in the quasi-linear regime, but get destabilized as their amplitudes increase. In the case of the focusing nonlinearity and negative defect, with $\alpha < 0, \beta > 0$, all modes with symmetries identical to those of modes that were stable in linear regime for $\alpha, \beta > 0$ become unstable, but a new stable (anti-waveguiding) mode appears in the first finite gap. In the case of the defocusing nonlinearity and negative defect with $\alpha, \beta < 0$, only modes of this latter type may exist and may be stable in the first finite gap. Finally, positive defects with the defocusing nonlinearity ($\alpha > 0, \beta < 0$) support localized waves bifurcating from linear modes in both semi-infinite and first finite gaps, and also anti-waveguiding modes that exist above a certain threshold power, which is necessary to change the overall response of the defect from focusing to defocusing. Modes of all the three types can be stable in a part of their existence domain.

### C. Two-component (vectorial) models

The generalization of the concept of solitons in NLs to multicompoment settings is of obvious interest, since the interaction between several optical fields (or atomic species in BEC) may result in stabilization of otherwise unstable states and appearance of much more complicated soliton families with components featuring different symmetries. The properties of localized states of two-component BECs confined in a nonlinear periodic lattice were investigated by Abdullaev *et al.* (2008). They studied the symmetry of localized states with respect to the underlying NL, and concluded that such lattices can support bright solitons with the same symmetry in both components, as well as bright solitons of mixed symmetries, in the form of combinations of odd and even states, and also dark-bright solitons and bright modes placed on top of a periodic background. The evolution of the BEC under the action of the NL was described by a coupled system of equations for wave functions $q_{1,2}$:

$$\begin{aligned} i\frac{\partial q_1}{\partial t} &= -\frac{\partial^2 q_1}{\partial x^2} + [\beta_1(x)|q_1|^2 + \sigma_{12}(x)|q_2|^2]q_1, \\ i\frac{\partial q_2}{\partial t} &= -\frac{\partial^2 q_2}{\partial x^2} + [\beta_2(x)|q_2|^2 + \sigma_{12}(x)|q_1|^2]q_2, \end{aligned} \qquad (36)$$

where the inter-species interaction strength, $\sigma_{12}(x) = g_0 + g_1 \cos(2x)$, and the intra-species nonlinearity strengths, $\beta_n(x) = \gamma_{n0} + \gamma_n \cos(2x)$, $n = 1, 2$, are periodic functions of the transverse coordinate. Such lattices may support two-component (alias *vectorial*) solitons with equal or different norms in the two components. In the simplest soliton solutions, maxima of atomic densities are symmetric around the minimum of the corresponding pseudopotential induced by the NL. Such *odd-odd* modes (here we use the same classification of states which is commonly utilized for solitons in linear lattices, see, e.g., the review by Kartashov, Vysloukh and Torner, 2009a) with equal norms of the components, have also equal chemical potentials, $\mu_1 = \mu_2$, while for different norms the component with smaller norm and amplitude has a lower chemical potential. Such modes are exceptionally robust in sufficiently strong NLs. Besides odd-odd modes, states were also found that are even (i.e., symmetric around the maximum of the nonlinear pseudopotential) in only one or in both components. Modes of this type, including odd-even and even-even solitons, turn out to be unstable. It was found too that it is possible to couple a localized mode in one component to an extended mode in the other, so that the extended state will act as a periodic potential for the localized mode. This is possible, in particular, in a binary mixture with the average repulsive interaction in the first component, i.e., $\gamma_{10} > |\gamma_1| > 0$, and average attractive interaction in the second component, i.e., $\gamma_{20} < -|\gamma_2| < 0$. The resulting dark-bright soliton can be stable, as well as a bright-bright soliton existing for the same parameters, but with one bright compo-nent existing on top of the background (such a mode is sometimes also called an "*antidark soliton*").

If the strength of the NL accounting for the inter-species interactions is varied in time, the odd-odd vectorial solitons may undergo a delocalizing transition, despite the fact that the strengths of the intra-species nonlinearities are kept constant. In this case, upon an adiabatic decrease of $g_1$ to lower values and its subsequent return to the original level, the mode, instead of following adiabatically the modifications of the NL, suffers a complete decay. The latter effect is related to the existence of an unstable localized solution which is extended over many sites of the NL, and exhibits shrinkage (decay) for slightly overcritical (under-critical) values of the norm. For a recent comprehensive survey of results on delocalization transitions in linear and nonlinear lattices, see the paper by Kruz *et al.* (2009).

The spatially periodic modulation of the nonlinearity enables the existence of complex multi-hump vectorial states with different symmetries of the two components, as discussed by Kartashov *et al.* (2009b). In particular, the vector solitons composed of dipole and fundamental, or dipole and even components, exist and may be stable. This suggests that families of scalar solitons that are unstable in NLs may be *stabilized* in the vectorial form, due to the coupling to a stable second component. In that connection, the impact of cross-modulation coefficient $C$ on the existence and stability of complex vector soliton solutions was considered within the framework of the model based on coupled equations, $i\partial q_{1,2}/\partial z = -(1/2)\partial^2 q_{1,2}/\partial \eta^2 + \sigma(\eta)q_{1,2}(|q_{1,2}|^2 + C|q_{2,1}|^2)$, which describe the two-component field in the NL. The simplest vectorial solitons may appear, for instance, as a result of the coupling of odd and dipole components (i.e., featuring two out-of-phase humps), or of even and dipole ones. The power sharing between the components strongly depends on the propagation constants, $\mu_1$ and $\mu_2$. In particular, at $C=1$, with the increase of $\mu_1$ in the even-dipole soliton depicted in Fig. 12(a), the dipole component becomes stronger, while its even counterpart is vanishing, and at $\mu_1 = \mu_1^{\text{upp}}$ one observes a transformation of the vectorial mode into a scalar dipole soliton. In contrast to that, the even component becomes more pronounced with the decrease of $\mu_1$, and one observes the transformation into an even scalar soliton at $\mu_1 = \mu_1^{\text{low}}$, see Fig. 12(b). The power sharing between the components strongly depends on the cross-modulation coefficient, $C$. Namely, the situation described above takes place at $C \leq 1.02$, while for $C > 1.02$ the picture is just the opposite, with the $q_1$ component vanishing with the increase of $\mu_1$. Even though the even component is unstable in the scalar case, the cross-modulation coupling to a stable dipole component may result in the *stabilization* of the vectorial complex as a whole. At $C=1$, one gets such stabilization for values of $\mu_1$ close to $\mu_1^{\text{upp}}$, where the dipole component is sufficiently strong [see Fig. 12(c), where the stability domain for such solitons is depicted in the $(\mu_1, \mu_2)$ plane, and Fig.

12(d) where the stability domain is shown in the $(C,\mu_1)$ plane]. Although at $C=1$ even-dipole solitons are stable only in a rather narrow part of their existence domain, their odd-dipole counterparts are stable almost in the entire existence domain. The stabilization due to the cross-modulation coupling between the two components under the action of the NLs is expected to occur also for more complex multi-hump types of vector solitons.

Explicit solitary-wave solutions of coupled NLSEs with spatially inhomogeneous nonlinearities were constructed by Belmonte-Beitia, Perez-Garcia, and Brazhnyi (2009), using the Lie-group theory. Another approach, which allows one to transform a system of coupled one-dimensional NLSEs with variable coefficients in front of the nonlinear terms into a system with constant coefficients, was proposed by Cardoso, Avelar, Bazeia, and Hussein (2010). Finally, it is relevant to mention that Cheng (2009) studied the interaction of two coupled binary (two-component) matter-wave bright solitons in the presence of a spatially varying nonlinearity and derived the corresponding effective potential characterizing the interaction.

## D. Symmetry breaking in dual-core nonlinear potentials

Double-well potentials support a wide variety of localized modes, which include, in addition to the obvious symmetric and antisymmetric states, *asymmetric* ones. It is commonly known from standard quantum mechanics that, without nonlinearity, the full set of eigenstates supported by double-well potentials splits into alternating symmetric and antisymmetric subsets. The addition of the nonlinearity changes the situation through the *symmetry-breaking bifurcation* (SBB). In the presence of the focusing nonlinearity, the SBB gives rise, at some critical value of the nonlinearity strength, to an asymmetric state which bifurcates from the symmetric one. In the simplest case, the SBB may be described by the two-mode approximation, which replaces the underlying partial differential equation for the wave's amplitude by a system of two linearly coupled ordinary differential equations for amplitudes of the waves trapped in the two deep wells, the linear coupling representing the linear mixing between them due to the tunneling across the potential barrier which separates the wells. Probably, the SBB was first studied in this approximation in the context of the general self-trapping problem by Eilbeck, Lomdahl, and Scott (1985), and then by Snyder et al. (1991) in the framework of the model of dual-core nonlinear optical fibers (alias optical couplers); it is also relevant to mention an early work by Davies (1979), which introduced the problem of the SBB in equations of the NLSE type in an abstract context.

For the BEC loaded into a double-well potential, the two-mode approximation was developed by Milburn, Corney, Wright, and Walls (1997), in both the mean-field approxi-mation and within the framework of the fully quantum description. Independently, the mean-field analysis of the double-mode system for the BEC was reported by Smerzi, Fantoni, Giovanazzi, and Shenoy (1997) [see also the paper by Raghavan, Smerzi, Fantoni, and Shenoy (1999)]. It is relevant to stress that, in the case of the repulsive nonlinearity, which is most relevant to the BEC, symmetric eigenmodes are not subject to the bifurcation, but another bifurcation generates asymmetric states from antisymmetric ones (which do not bifurcate in the case of the self-attraction).

Symmetry-breaking effects were also studied, by Mayteevarunyoo, Malomed and Dong (2008), in the 1D model based on the nonlinear *pseudopotential* of the double-well type. This situation corresponds to a system with two sharp symmetric maxima of the nonlinearity coefficient, $R(x) = -\{\exp[-(x+1)^2/a^2] + \exp[-(x-1)^2/a^2]\}/a\pi^{1/2}$ (here $a$ is the width of each well), which may be realized in the BEC by means of accordingly applied spatially nonuniform FR management, or in optics in a planar linear waveguide with two narrow nonlinear channels embedded into it. In the limit of $a \to 0$, one gets $R(x) = -[\delta(x+1) + \delta(x-1)]$ (note that a model based on the NLSE with the self-focusing nonlinearity concentrated in the form of a single delta-function was introduced earlier by Malomed and Azbel (1993)]. In the framework of model with the two delta-functions, one can obtain stationary analytical solutions, which are continuous everywhere and feature a jump of the first derivative at the locations of the embedded nonlinear channels, as per conditions $\partial q/\partial x|_{x=\pm 1+0} - \partial q/\partial x|_{x=\pm 1-0} = -2(q|_{x=\pm 1})^3$. Such solutions, with chemical potential $\mu$, can be written in the following form:

$$q(x,t) = e^{-i\mu t} \begin{cases} B_1 e^{\sqrt{2|\mu|}(x+1)}, \text{ at } x < -1, \\ A_0 e^{-\sqrt{2|\mu|}(x-1)} + B_0 e^{\sqrt{2|\mu|}(x+1)}, \text{ at } -1 < x < +1, \\ A_1 e^{-\sqrt{2|\mu|}(x-1)}, \text{ at } x > +1. \end{cases} \quad (37)$$

Using the continuity conditions, one has $B_0 = (e^{2\sqrt{2|\mu|}}A_1 - B_1)/(e^{4\sqrt{2|\mu|}} - 1)$. Further, three types of exact solutions can be found: symmetric, antisymmetric, and asymmetric ones. These are given, respectively, by $A_1 = B_1 = (2|\mu|)^{1/4}(1+e^{-2\sqrt{2|\mu|}})^{-1/2}$, $A_1 = -B_1 = (2|\mu|)^{1/4}(1-e^{-2\sqrt{2|\mu|}})^{-1/2}$, and

$$(A_1, B_1) = \frac{|\mu|^{1/4}[(1+2e^{-2\sqrt{2|\mu|}})^{1/2} \pm (1-2e^{-2\sqrt{2|\mu|}})^{1/2}]}{2^{3/4}(1-e^{-4\sqrt{2|\mu|}})^{1/2}}. \quad (38)$$

The SBB happens on the branch of the symmetric solutions, with the increase of amplitude $A_1$ (i.e., with the increase of $-\mu$) at the critical point, $A_1 = 3^{-1/2}(\ln 2)^{1/2}$, which corresponds to $\mu = -(\ln 2)^2 / 8$.

Numerically, the SBB has been analyzed in this model for finite values of width $a$ of the two wells in the $R(x)$ structure. A set of typical bifurcation diagrams, which show the effective asymmetry of the pinned pattern, $\Theta = N^{-1}\left(\int_0^\infty |q|^2 dx - \int_{-\infty}^0 |q|^2 dx\right)$, versus total norm $N$, are shown in Fig. 13. At $a=0$, a very peculiar feature of the bifurcation is its *entirely subcritical* character: The branches representing the pair of the asymmetric solutions go *backward* as unstable ones and *never* turn forward, that indicates their full instability. Another peculiarity of this limit case is that the symmetric branch (which is destabilized by the bifurcation) ends at a finite value of the norm, $N_{\max} \approx 2.08$. Antisymmetric solutions, that never bifurcate, are completely unstable at $a=0$. The character of the bifurcation quickly changes with the increase of $a$: the backward-going branches turn forward at some point, which makes them stable, and, at $a > 0.2$, the bifurcation becomes *supercritical*, giving rise to a pair of forward-going stable branches. At sufficiently large values of $a$, both asymmetric and antisymmetric solutions can be stable.

The results known for the SBB in *two-component* systems with linear double-well potentials (Wang, Kevrekidis, Whitaker, and Malomed, 2008) suggest that a challenging extension of the model with the double-well nonlinear potential would be to consider its two-component extension. Moreover, the analysis of the SBB in the model with the *two-dimensional* linear potential, represented by a symmetric set of *four* potential wells (Wang *et al*, 2009b) suggests that it may be very interesting to analyze the symmetry breaking in the model with a 2D symmetric pseudopotential, represented by two or four mutually symmetric wells.

As concerns other settings which give rise to the symmetry breaking, this effect was studied by Wang *et al*. (2009a) in a model combining a linear double-well potential with a spatially inhomogeneous (step-like) nonlinearity landscape. The settings were studied where the nonlinearity was of the same or of different signs in two wells of the potential. The analysis was based on the continuation of the symmetric ground state and anti-symmetric first excited state of the non-interacting (linear) limit into their nonlinear counterparts, and it was shown that, even for the weakly inhomogeneous nonlinearity, the asymmetry (which is induced by the spatial dependence of the nonlinearity) causes a modification of the usual bifurcation picture characteristic to the double-well potential, and a change in the nature of the symmetry-breaking bifurcation, from a *pitchfork* to the *saddle-node* type.

### E. Solitons in layered nanostructures

The shape and stability of subwavelength spatial solitons of both TM and TE types, trapped in the periodic nanostructure described by Eqs. (14)-(16) were studies in detail by Gorbach and Skryabin (2009). To this end, solutions for all the fields were looked for as localized functions of $x$ times $\exp(iqkz)$, with the corresponding relative propagation constant, $q$. First, the bandgap structure of the linearized version of the equations was calculated, for the set of alternating strips made of silicon and silica. Then, the full nonlinear equations for the functions of $x$ were solved numerically, and their stability was studied through numerical solutions of the respective eigenvalue problem, generated by the linearization of the full system of equations for small perturbations. Both the bandgap structure and the nonlinear solutions for the spatial solitons critically depend on the fact whether the width of the silica strips, $s$, is larger or smaller than the special value, $s_0$, at which the so-called Brewster condition holds, implying zero reflection of the TM-polarized waves from the intrinsic interface.

Typical examples of the fundamental TM and TE solitons (with the subwavelength transverse size), found at $s > s_0$, are displayed in Figs. 14(a) and 14(b). The solitons are classified as on-site and off-site ones, according to the location of their centers relative to the strips with the higher value of the refractive index. The entire families of these solitons are presented in Figs. 14(c) and 14(d) by means of dependences of the total power of the solitons on the nonlinear shift of their propagation constants. In the case of $s < s_0$, the shape of the on-site and off-site solitons of both the TM and TE types is more complex, but the general character of the stability remains the same as in the case shown in Fig. 14., i.e., the on-site solitons are *stable*, while their off-site counterparts are *unstable*. In fact, this stability pattern is typical for solitons in discrete and quasi-discrete systems [see an earlier review by Kevrekidis, Rasmussen, and Bishop (2001), and a recent book by Kevrekidis (2009)]. Notice that coupled-mode approach for description of light propagation in an array of nonlinear plasmonic waveguides was recently developed by Marini, Gorbach, and Skryabin (2010).

## F. Interactions of solitons with defects

Among interesting aspects of soliton evolution in 1D models with inhomogeneous nonlinearity landscapes are interactions of solitons with a local inhomogeneity of the strength of the nonlinearity. A model of this type was introduced by Abdullaev, Gammal, and Tomio (2004), in the form of the accordingly modified GPE:

$$i\frac{\partial \psi}{\partial t} = -\frac{\partial^2 \psi}{\partial z^2} + \alpha z^2 \psi - [1 + \varepsilon f(z)]|\psi|^2 \psi, \qquad (39)$$

where the negative sign in front of the nonlinear term implies that the nonlinearity is attractive, $f(z)$ accounts for the local perturbation of the nonlinearity, and $\alpha$ is the strength of the external parabolic trapping potential (if any). The VA, based on the substitution of *ansatz* $\psi(z,t) = A(t)\mathrm{sech}[(z-\zeta)/a(t)]\exp\{i[\phi(t) + w(z-\zeta) + b(z-\zeta)^2]\}$ into the Lagrangian associated with Eq. (39), yields the system of evolutional equations for the soliton's amplitude, $a$, and central coordinate, $\zeta$:

$$\begin{aligned}
\frac{d^2 a}{dt^2} &= -4\alpha a + \frac{16}{\pi^2 a^3} - \frac{4n_0}{\pi^2 a^2} + \frac{3\varepsilon n_0}{\pi^2 a^2}\left(\frac{\partial F}{\partial \zeta} - 2\frac{F}{a}\right), \\
\frac{d^2 \zeta}{dt^2} &= -4\alpha\zeta + \frac{\varepsilon n_0}{4a^2}\frac{\partial F}{\partial \zeta},
\end{aligned} \qquad (40)$$

where $n_0 = \int_{-\infty}^{+\infty}|\psi(z)|^2 dz$ is the norm of the 1D wave function, and the effective pseudopotential characterizing the interaction of the soliton with the local inhomogeneity of the nonlinearity is $F(a,\zeta) = \int_{-\infty}^{+\infty} f(z)\mathrm{sech}^4[(z-\zeta)/a]dz$. The analysis was performed for a strongly localized inhomogeneity, $f(z) \sim \delta(z)$, and predictions of the VA were confirmed by direct simulations of Eq. (39). The results demonstrate three different outcomes of the collision between a freely moving soliton, with initial velocity $V_0$, and the attractive nonlinear inhomogeneity: passage, capture, and *rebound*. The latter outcome is noteworthy, as the impinging soliton may bounce back from the local defect despite the attractive sign of the interaction between them [earlier, a similar counterintuitive result was reported by Kivshar, Fei, and Vázquez (1991), who considered the collision of a kink with an attractive defect in the sine-Gordon equation]. In particular, the rebound was observed in direct simulations of Eq. (39) with $\varepsilon f(z) = 0.4\delta(z)$ in the interval $0.42 < V_0 < 1.2$ of the velocities of the incident soliton, whose norm was fixed to be $n_0 = 4$. Qualitatively, the rebound might be explained by a resonance between the oscillatory motion of the two degrees of freedom of the soliton obeying Eq. (40). Indeed, the eigenfrequencies of small oscillations predicted by these equations are $\omega_a^2 = (4n_0/\pi^2)[2n_0/(8-3\varepsilon n_0)]^3$ and $\omega_\zeta^2 = \varepsilon n_0[2n_0/(8-3\varepsilon n_0)]^4$, which shows that the resonance is indeed possible. Static solitons trapped by the attractive defects $f(z) = \delta(z)$ were obtained in an analytical form, $\psi(z,t) = 2^{1/2}a\,\mathrm{sech}(a|z|+\beta)\exp(ia^2 t)$, where $\beta = (1/2)\mathrm{sign}(\varepsilon)\ln[2|\varepsilon|a + (4\varepsilon^2 a^2 + 1)^{1/2}]$ [such states are stable only for $\varepsilon > 0$, i.e., for the attractive nonlinear defect in Eq. (39)]. It is interesting to compare this exact solution with its counterpart found, in a numerical form, from the corresponding full three-dimensional GPE (with the delta-function replaced by a proper numerical approximation), see Fig. 15.

A relevant comment concerning the 1D model based on Eq. (39) is that its limit form, without the harmonic trap ($\alpha = 0$) and with the entire nonlinearity concentrated in the form of the single delta-function, amounts to the model originally introduced by Malomed and Azbel (1993), $i\psi_t = -\psi_{xx} - \varepsilon\delta(z)]|\psi|^2\psi$. Obviously, this simplest model with $\varepsilon > 0$ supports a family of exact pinned "solitons", $\psi(x,t) = (2\lambda/\varepsilon)^{1/2}\exp(i\lambda^2 t - \lambda|x|)$, where the inverse width, $\lambda$, is an intrinsic positive parameter of the solutions. This family is *degenerate* in the same sense as 2D *Townes solitons* are in the NLSE with the uniform nonlinearity (Bergé, 1998), i.e., their norm, $N \equiv \int_{-\infty}^{+\infty}|\psi(x)|^2 dx = 2/\varepsilon$, is *the same* for all the entire family. A simple analysis demonstrates that this family is entirely unstable, which also resembles the well-known property of the Townes solitons Nevertheless, the solitons pinned by the attractive nonlinear delta-functional potential may be readily *stabilized* by adding the usual linear periodic potential (i.e., OL) to the model. Moreover, in the case of $\varepsilon < 0$, i.e., with the repulsive delta-functional nonlinearity, the model including the OL potential readily supports *stable* solitons of the gap type, pinned to the *repulsive* center (Dror and Malomed, 2010).

Another interesting resonant effect was reported by Primatarowa, Stoychev, and Kamburova (2005), who performed systematic simulations of a model based on an equation equivalent to Eq. (39) with $\alpha > 0$, but with $f(z)$ representing a long attractive rectangular box, rather than a delta-function. This pseudopotential can readily trap an incident soliton, provided that its velocity falls below a certain threshold. Then, depending on the length of the box, $L$, a nearly periodic alternation of the trapping and transmission intervals is observed, with the increase of $L$, at fixed values of the initial soliton's velocity and depth of the box. The alternation was explained by a resonance between intrinsic oscillations of the soliton, which was perturbed while passing the step between the free space and the box, and the time of the flight of the soliton through the box: if the flight time is a multiple of the period of the intrinsic vibrations of the initially perturbed soliton, its collision with the second edge of the box may help the soliton to retrieve the energy initially lost to excite the intrinsic vibrations. The energy recovery will allow the soliton to pass the step and thus get transmitted through the pseudopotential box. On the other hand, in the absence of the resonance, the soliton irreversibly loses a part of its initial kinetic energy, and thus it is not able to escape from the box. A similar analysis was performed, in the same work, also for a model combining the linear and nonlinear potentials in the form of the rectangular boxes.

The analysis of the transmission of the incident soliton through a defect represented by a combination of a local linear potential and localized inhomogeneity of the nonlinearity co-efficient was also reported by Theocharis *et al.* (2006). They used an effective quasi-particle equation of motion for the soliton, similar to the second equation in system (40), in combination with systematic simulations of the underlying one-dimensional GPE, i.e., Eq. (39) (with $\alpha = 0$). One of conclusions reported in that work was that the addition of the pseudopotential, induced by the inhomogeneity of the nonlinearity coefficient, may *enhance* the transmission of the incident soliton through a local linear potential barrier. Related to this finding is a non-monotonous dependence of the respective transmission coefficient, $T$, on the width of the barrier, with $T$ attaining a maximum at a particular value of the width.

Settings of this types were further analyzed by Garnier and Abdullaev (2006), who considered the transmission of an incident soliton through a local defect combining a Gaussian linear potential barrier and a similar Gaussian-shaped local variation of the nonlinearity coefficient. In addition to direct simulations, the analysis made use of the perturbation theory for solitons. In particular, the radiation loss due to the emission of quasi-linear waves by the soliton traversing the local inhomogeneity was calculated, using the perturbative method based on the inverse scattering transform (a comprehensive review of the method was given by Kivshar and Malomed, 1989). In fact, an essential role of the radiation losses was observed in the simulations reported in the above-mentioned works. Garnier and Abdullaev (2006) were able to explain how the radiative effects may essentially alter predictions of the simple adiabatic perturbation theory [the one based on Eqs. (40)]. Similar to the work by Theocharis *et al.* (2006), another inference was that the addition of the nonlinear pseudopotential may facilitate the transmission of the impinging soliton through the local barrier induced by the linear potential.

The analysis of the emission of radiation was also reported by Abdullaev and Garnier (2005) for a soliton moving through a regular (periodic) or random NL. Predictions of the analytical perturbation theory for this situation were compared to direct simulations of the one-dimensional GPE with the periodic or random spatial modulation of the nonlinearity coefficient.

## G. Discrete models

As mentioned above in subsection II.C, discrete systems naturally emerge, in the tight-binding approximation, as limit forms of many models which incorporate strong linear and nonlinear lattices [see Christodoulides and Joseph (1988)]. However, in most cases the resulting discrete systems seem as standard DNLSEs [see, e.g., Eqs. (11) and (12)], which have been studied thoroughly in other contexts [an extensive account of the topic can be found in the book by Kevrekidis (2009)]. By themselves, such discrete systems do not directly belong to the class of models categorized as those including NLs.

An example of a more complex variety of the one-dimensional DNLSE, which may be derived from the consideration of media with embedded linear and nonlinear lattices, was proposed and investigated by Abdullaev et al. (2008):

$$
\begin{aligned}
i\frac{dc_n}{dt} &= \omega_0 c_n + \omega_1(c_{n+1} + c_{n-1}) + W_0 |c_n|^2 c_n + \\
&\quad W_1(|c_{n-1}|^2 c_{n-1} + \sigma c^*_{n-1} c_n^2 + 2\sigma|c_n|^2 c_{n-1} + 2|c_n|^2 c_{n+1} + c^*_{n+1} c_n^2 + \sigma|c_{n+1}|^2 c_{n+1}) + \\
&\quad W_2(2|c_{n-1}|^2 c_n + 2|c_{n+1}|^2 c_n + c^*_n c^2_{n-1} + c^*_n c^2_{n+1})
\end{aligned}
\quad (41)
$$

where $\omega_0$ and $\omega_1$ are characteristics of the linear spectrum, while nonlinear coefficients $W_0, W_1$, and $W_2$ are determined by certain overlap integrals, and $\sigma$ is the parity of the nonlinearity-modulating function in the underlying continuous equation [in fact, this equation can be obtained from (4), with $\sigma = \pm 1$ corresponding to $n_2(-x) = \pm n_2(x)$, while it is assumed that $\delta n(x)$ is always an even function, i.e., $\delta n(-x) = \delta n(x)$]. It was also concluded that $W_0 = W_2 = 0$ for $\sigma = -1$.

The model (41) and the origin of inter-site terms in this model was also discussed by Belmonte-Beitia and Pelinovsky (2009) (see also Claude et al., (1993)). A set of characteristic examples of discrete modes, both localized (bright solitons) and delocalized ones ("kinks" and "anti-dark solitons"), supported by Eq. (41) in the form of $c_n = f_n \exp(-i\omega t)$, is displayed in Fig. 16. These examples of bright and kink solutions are stable, while the anti-dark solitons are completely unstable. All the modes shown in Fig. 16 are of the on-site-centered type. Their inter-cite-centered counterparts were found too, but they all turn out to be unstable. In addition to the three types of the discrete modes shown in Fig. 16, an additional type of solutions reported by Abdullaev et al. (2008), that may be stable too, represents kinks with wavy tails.

Another discrete system that originates from continuous models with NL potentials is built of two semi-infinite discrete lattices with attractive and repulsive on-site cubic terms (Machacek et al., 2006). The respective version of the one-dimensional DNLSE is

$$i\frac{du_n}{dt} = -C(u_{n+1} + u_{n-1} - 2u_n) - d_n |u_n|^2 u_n, \quad (42)$$

where $d_n = d_- = -0.9$ for $n < 0$, and $d_n = d_+ = 1.1$ for $n > 0$, while $d_0 = (d_+ + d_-)/2$. Various families of asymmetric discrete solitons supported by Eq. (42) around the interface ($n = 0$) were found. These families naturally form pairs of stable and unstable ones, which mutually annihilate, at some critical point, with the increase of coupling constant $C$ (if the intrinsic frequency of the solitons is kept constant). Examples of stationary modes belonging to two different families of the solutions are shown in Fig. 17.

A model which may be considered as an example of discrete NL was introduced by Hizanidis, Kominis, and Efremidis (2008), in the form of a 1D system with alternating linear and nonlinear sites [a similar system, with two sites only, but uniformly extended in an additional direction, was introduced by Zafrany, Malomed, and Merhasin (2005)]. The model is based on the following system of coupled equations:

$$i\frac{d\psi_{2m}}{dz} + \frac{1}{2}(\psi_{2m-1} + \psi_{2m+1} - 2\psi_{2m}) + \varepsilon_{\text{even}}\psi_{2m} + \sigma|\psi_{2m}|^2\psi_{2m} = 0,$$
$$i\frac{d\psi_{2m\pm 1}}{dz} + \frac{1}{2}(\psi_{2m\pm 2} + \psi_{2m} - 2\psi_{2m\pm 1}) + \varepsilon_{\text{odd}}\psi_{2m\pm 1} = 0,$$
(43)

where $\sigma = \pm 1$ determines the sign of the nonlinear term. The analysis performed by Hizanidis, Kominis, and Efremidis (2008) was chiefly focused on the analysis of the spectrum of the linearized version of Eq. (43), and on the study of the modulational instability of continuous-wave states in the framework of the full system. In particular, the spectrum includes [similar to other *diatomic* discrete systems, see the book by Kevrekidis (2009)] two semi-infinite gaps and a finite bandgap between them. Examples of solitons in all the three gaps were found too. The fundamental solitons appear to be stable in all the cases, while their antisymmetric bound states (dipoles) are stable only in the finite bandgap, see an example in Fig. 18. Note the staggered structure of the solitons displayed in this figure (alternating signs of the discrete field at adjacent sites).

A prototypical model with the Kerr nonlinearity, which does not reduce to the usual discrete one, was introduced by Panoiu, Malomed, and Osgood (2008). It describes a waveguide built in the form of a slab substrate, with an array of guiding ribs either mounted on top of it, or buried into the slab. Selecting parameters of this system, it is possible to get a setting in which the slab and array support the transmission of the waves with different polarizations, one TE and one TM. The model of such a system may be called a *semi-discrete* one, as it is based on coupled continuous and discrete NLSEs:

$$i\frac{d\phi_n}{dz} + \phi_{n-1} + \phi_{n+1} + |\phi_n|^2 \phi_n + \kappa \phi_n |\Psi(\eta = n)|^2 = 0,$$
$$i\frac{\partial \Psi}{\partial z} + \frac{1}{2}\frac{\partial^2 \Psi}{\partial \eta^2} + \beta\Psi + |\Psi|^2 \Psi + \kappa \Psi \sum_n |\phi_n|^2 \delta(\eta - n) = 0, \quad (44)$$

where $\delta(\eta - n)$ is the delta-function, $\eta$ is the transverse coordinate, scaled so that the spacing of the discrete array is 1, $\beta$ is the mismatch between the continuous and discrete subsystems, and $\kappa$ is the coefficient accounting for the XPM (cross-phase-modulation) nonlinear coupling between the subsystems. The signs in front of the SPM (self-phase-modulation) coefficients in Eqs. (44) imply that the nonlinear material is self-focusing, hence it is natural to set $\kappa > 0$. Equations (44) conserve the total power, $P = \sum_n |\phi_n|^2 + \int_{-\infty}^{\infty} |\Psi|^2 d\eta$. Stationary solutions may be found, in the general case, with two independent propagation constants, viz., $\phi_n = u_n \exp(i\lambda_1 z)$ and $\Psi = V(\eta)\exp(i\lambda_2 z)$. As shown in Fig. 19, three types of simplest solitons are generated by Eqs. (44), namely, odd (on-site-centered), even (inter-site-centered), and twisted (antisymmetric). The calculation of the spectrum of instability growth rates for small perturbations around the semi-discrete solitons demonstrates that the odd solitons are entirely stable, while the even and twisted ones are always unstable, even though the VK criterion does not predict the instability of the even-soliton family (the actual instability growth rate for this family is complex, which cannot be detected by the VK condition).

## H. Dynamical regimes
### 1. Matter-wave-laser models

Unlike the field of nonlinear optics, the studies of BEC have not yet led to many technological applications, being more focused on fundamental aspects. Nevertheless, matter-wave setups have a potential for the use in various technologies. In particular, the possibility of employing condensates as a reservoir for generation of coherent atomic beams (i.e., as a basis for *matter-wave lasers*) has drawn considerable attention. A straightforward idea, elaborated by Carr and Brand (2004), was to suddenly flip the sign of the scattering length in a cigar-shaped (quasi-1D) reservoir from attractive to repulsive, by means of the FR effect, and thus initiate the release of pulses from it. Another design of the soliton laser, proposed by Chen and Malomed (2005 and 2006), was based on two parallel quasi-1D traps, coupled by tunneling of atoms across a barrier separating them, with one to be used as the reservoir, and the other – as the lasing cavity.

Experimental prototypes of matter-wave lasers operating in the regime of releasing continuous atomic beams were reported by Guerin et al. (2006) and Robins et al. (2008). The latter work used a setting with the reservoir separated from the lasing element, somewhat similar to the above-mentioned proposal by Chen and Malomed (2005 and 2006).

Another scheme of matter-wave lasers capable to generate chains of solitons was developed by Rodas-Verde, Michinel, and Pérez-García (2005) and Carpentier, Michinel, Rodas-Verde, and Pérez-García (2006). It was based on the assumption that a usual axial trap for the BEC, implemented by dint of an appropriate linear potential, is combined with an adjacent region where the scattering length is made negative (the influence of three-body collisions on this scheme was recently studied by Carpentier, Michinel, Olivieri, and Novoa, 2010). The respective model is based on the following version of the one-dimensional GPE:

$$i\frac{\partial q}{\partial t} = -\frac{1}{2}\frac{\partial^2 q}{\partial x^2} + V(x)q + R(x)|q|^2 q, \qquad (45)$$

where $V = 0$ for $|x| > L$, $V = V_0 < 0$ for $|x| < L$, and $R = 0$ for $x < L$, $R = R_0 < 0$ for $x > L$. First, an equilibrium position for a soliton was predicted in this model, using the VA based on the ordinary Gaussian ansatz, $q(x,t) = A\exp[-(x-x_0)^2/2w^2]$. This allows one to derive the equation of motion for the soliton's center, $d^2 x_0 / dt^2 = -d\Pi/dx_0$, where the effective potential is given by

$$\Pi = \frac{V_0 \pi^{1/2}}{2}\left(1 - \frac{\exp[-x_0^2/(L^2 + w^2)]}{(1 + w^2/L^2)^{1/2}}\right) + \frac{R_0 N}{(2\pi)^{1/2} w}\operatorname{erfc}\left(\frac{2^{1/2}(L - x_0)}{w}\right), \qquad (46)$$

with $N$ being the soliton's norm. The second term here actually represents the *pseudopotential* induced by the nonlinearity modulation. The equilibrium position of the soliton is defined as a root of equation $d\Pi/dx_0 = 0$. Using this equation, a critical value of $R_0$ was predicted, such that at $R_0 < -|R_{\text{cr}}|$ the equilibrium position does not exist, hence the soliton cannot remain trapped in the laser cavity, and will be released. In agreement with this analytical prediction, the generation of clusters of solitary pulses was observed in direct simulations of Eq. (45). The number of the released solitons is determined by the total norm of the condensate initially stored in the cavity, see Fig. 20. The analysis of this model and its modifications with smoothed modulation functions has demonstrated that *sharp edges* in function $R(x)$ help to improve characteristics of the soliton-generation regime. It is relevant to mention, in this connection, that the sharpness of the nonlinearity-modulation function is a crucial factor determining the stability of 2D solitons pinned by the modulated self-focusing nonlinearity, see Section V below.

## 2. Oscillations of driven solitons in nonlinear lattices

GSs (gap solitons), which exist due to the interplay of the linear lattice and repulsive nonlinearity, feature unusual dynamical properties because of their affinity to the corresponding Bloch waves. In particular, they inherit the negative mass of the Bloch excitations, which gives rise to stable oscillations of the GSs in *inverted* (anti-trapping) potentials, in the 1D and 2D settings alike (Sakaguchi and Malomed, 2004b and 2004c).

It is also well known that the application of an external potential with a constant slope to linear wave packets in periodic lattice potentials induces *Bloch oscillations* of the packets. It was demonstrated by Salerno, Konotop, and Bludov (2008) that, in the model which incorporates a NL in addition to the linear lattice, GSs may perform stable Bloch oscillations under the action of a constant driving force (i.e., an extra potential with the constant slope).

In the quasi-linear approximation, the motion of the central coordinate ($\xi$) of a wave packet driven by the constant force, $F$, added to the periodic linear-lattice potential, obeys the following equations: $d\xi/dt = V(q) \equiv dE/dq$, $dq/dt = F$, where $V$ is the velocity, $q$ is the quasi-momentum, and $E(q)$ is the respective dispersion law (energy-momentum relation) within a given band. Because $E(q)$ and hence $V(q)$ are periodic functions of $q$, the linear growth of $q$ with time, $q(t) = Ft$, implies a *periodic motion* of the wave packet. Applying this argument to the GS, one should take into regard that GSs exist only close to those edges of the bands where the effective negative mass coexists with the repulsive nonlinearity. In the usual model, with the constant nonlinearity coefficient, the latter condition cannot hold everywhere, because the effective mass always has opposite signs at opposite edges of a given band. Therefore, the GS performing the Bloch oscillations is subject to a slow destruction, as it spends some time in the environment where it cannot exists as a stationary state. In the work of Salerno, Konotop, and Bludov (2008), it was proposed to add an appropriate sign-changing NL to the model, so as to synchronize the changes of the sign of the nonlinearity with the sign flips of the effective mass, thus providing the fulfillment of the GS existence condition everywhere.

The corresponding model was based on the following version of the GPE [cf. Eq. (10)]:

$$i\frac{\partial \psi}{\partial t} = -\frac{\partial^2 \psi}{\partial x^2} - [Fx + V\cos(2x)]\psi + [g + G\cos(2x)]|\psi|^2 \psi. \qquad (47)$$

As shown in Fig. 21(a), the analysis has produced a region in the plane of parameters $(V, G)$ where the above-mentioned condition necessary for the persistence of the GS performing Bloch oscillations, *viz.*, the keeping opposite signs of the effective mass and effective nonlinearity, is met, due to the inclusion of the NL into Eq. (47). Direct simulations [see Figs. 21(b) and 21(c)] demonstrate the robustness of the Bloch oscillations of the GS inside the predicted stability area, and decay of the oscillating GS outside of it.

Stable periodic oscillations of GSs between two different bands (*Rabi oscillations*) were demonstrated, by means of systematic simulations, by Bludov, Konotop, and Salerno (2009), within the framework of a model similar to that described by Eq. (47), but with a linear periodic potential whose strength is periodically modulated in time: $F(t) = F_0 \cos(\omega t)$. These oscillations are somewhat similar to those reported by Gubeskys, Malomed, and Merhasin (2005) in the 2D and 1D models with the spatially uniform nonlinearity, whose strength was subjected to the "management", i.e., periodic modulation in time. A stability region for *alternate solitons* was identified in the latter work, i.e., solitons with a periodically varying chemical potential, that regularly switches between the semi-infinite gap and the first or even second finite bandgap. This means that the localized modes periodically change their character, between ordinary solitons in the semi-infinite gap and GSs in the finite gap.

## V. Two-dimensional solitons

The study of solitons in two-dimensional NLs was a natural extension of the original work done in 1D. Quite a few theoretical papers have addressed this topic. The results of the studies of the 2D setting, summarized in this section, demonstrate a drastic difference from the 1D case. Namely, it is *very difficult* to stabilize 2D solitons by means of NL configurations, or, speaking more generally, by means of the general spatial modulation of the nonlinearity in 2D settings.

As it was mentioned in the introduction, the new problem which is posed by the 2D geometry is the instability of solitons supported by the cubic nonlinearity in the 2D free space (*Townes solitons*) against the collapse (Bergé, 1998). On the other hand, previously published results had demonstrated that, using *linear* lattice potentials, on can *easily* stabilize solitons against the collapse (see original works by Baizakov, Malomed, and Salerno, 2003, Yang and Musslimani, 2003, and reviews by Lederer *et al.*, 2008, and Kartashov, Vysloukh, and Torner, 2009a). Moreover, linear lattices make it possible to stabilize localized vortices characterized by the respective topological charge, $S$ (i.e., the *winding number* of the underlying phase pattern). It is possible too to stabilize "*supervortices*", in the form of ring-shaped chains of compact ("crater-shaped") vortices, each carrying topological charge $s = +1$, with independent global vorticity, $S = \pm 1$, imprinted onto the chain (Sakaguchi and Malomed, 2005b). Finally, it is worthy to note that, as shown by Baizakov, Malomed, and Salerno (2004), a reduced quasi-1D linear-lattice potential, which depends on one coordinate only, is sufficient for the full stabilization of 2D solitons.

On the contrary to the plethora of the predictions of the stabilization provided by the 2D and quasi-1D linear lattices, theoretical studies have demonstrated that smooth NL potentials, taken in a similar form in the 2D setting, cannot stabilize *anything*, in a practical sense. To say it more accurately, it was demonstrated, in particular, by Sivan, Fibich, and Weinstein (2006) that the 2D model with the sinusoidal quasi-1D modulation of the local nonlinearity coefficient may support stable 2D solitons, but in such a tiny region that the authors had categorized this result is "mathematical", rather than "physical". Nevertheless, other results summarized in this section demonstrate that the stabilization of 2D solitons by nonlinearity-modulation patterns is practically possible, but under the condition that the modulation pattern features *sharp edges*, rather than being sinusoidal, or featuring another smooth shape. This condition is, as a mater of fact, a novel generic property of the 2D geometry revealed by the analysis of many settings.

In this section, we summarize the results obtained for 2D continuous and discrete solitons in nonlinear and mixed linear-nonlinear lattices, making the emphasis on the most challenging issue of the stability of such solitons. We start from the consideration of the *core problems* for the 2D solitons supported by purely nonlinear periodic lattices or by localized modulations of the nonlinearity, including quasi-1D nonlinearity landscapes, and power-dependent shape transformations of vortex solitons in mixed linear-nonlinear lattices. It is relevant to stress that, as well as in the 1D situation, the solitons supported by NLs do not bifurcate from linear Bloch models, but emerge, under the action of the modulated nonlinearity, "from nothing". Then, we proceed to the description of a variety of phenomena predicted in the models of photonic crystals and PCFs. These models amount to concomitant modulations of the refractive index and nonlinearity, that is why they are amenable to a more straightforward analysis, which readily produced stable solitons and vortices, as well as multi-soliton complexes. Among the respective results are the prediction of the self-trapping of stable bright solitons, the formation of solitons on defects in PCFs, the existence of GSs and soliton trains in finite periodic and quasi-periodic photonic crystals, including the practically important setting based on liquid-infiltrated PCFs featuring thermal nonlinearities, the formation of vortex, nodal and vector solitons, as well as soliton clusters, in such media, a limitation on values of the vorticity of localized states in photonic crystals with certain discrete rotational symmetries, and the possibility to build nonlinear dual-core photonic-crystal couplers. Finally, we describe the stability and mobility of solitons in 2D discrete models of waveguide arrays with a nonlinearity modulation.

## A. Solitons in nonlinear and mixed lattices
### 1. Circular nonlinearity-modulation profiles

As said above, the stabilization of 2D solitons in materials with the self-focusing cubic nonlinearity against the collapse solely via the spatial modulation of the nonlinearity coefficient is a challenging problem, a solution to which requires a careful adjustment of the modulation landscape. For example, it is difficult (if not impossible) to achieve the stabilization using *smooth* sinusoidal or Bessel-like profiles of the modulation function. Sakaguchi and Malomed (2006a) had proposed a setting where *stable* axisymmetric solitons can be created by a spatially-localized modulation of the nonlinearity coefficient, when it is different from zero in a circle or annulus. The respective form of the two-dimensional GPE is $i\partial q/\partial t = -(1/2)\Delta q - R(r)|q|^2 q$, where $\Delta$ is the 2D Laplacian acting on coordinates $x$ and $y$, and, in the general case of the annulus, $R(r) = 1$ for $\rho_0 < r < \rho_1$, and $R(r) = 0$ for $r < \rho_0$ and $r > \rho_1$, $r \equiv (x^2 + y^2)^{1/2}$ being the radial coordinate. This choice of the modulation profile clearly demonstrates its *sharpness*, which is crucial for the stability of localized modes supported by this profile.

Typical examples of axisymmetric stationary soliton solutions in this model, $q = w(r)\exp(-i\mu t)$, are displayed in Fig. 22(a). The radial functions, $w(r)$, were found in a numerical form. A quasi-analytical solution is available in the limit case of $\rho_1 - \rho_0 \to 0$, when the modulation function amounts to $R(r) = R_0 \delta(r - \rho_0)$, and the solution itself may be expressed through the modified Bessel and Hankel functions, $I_0$ and $K_0$, at $r < \rho_0$ and $r > \rho_0$, respectively; however, in this limit case all the solutions are unstable (see below). While for $\rho_0 = 0$ (i.e., for the circle with no inner hole) the soliton has a bell-like shape, for $\rho_0 \neq 0$ it develops a shallow deep at $r = 0$, and the local field attaints a maximum at $r = \rho_0$. Families of soliton solutions are characterized by dependences $\mu(N)$ which are displayed in Fig. 22(b). These dependencies imply that the VK criterion, $dN/d\mu < 0$, is satisfied for the corresponding soliton branches. However, in the present context this criterion can only suggest the stability against perturbations that do not break the axial symmetry of the solutions. A linear stability analysis, that takes into account azimuthal perturbations, predicts that solitons in this model may be also destabilized by perturbations with azimuthal index $m = 1$. The stability diagram is shown in Fig. 22(c), where the upper curve denotes a lower border for the azimuthal instability with $m = 1$, while the lower curve de-notes the existence and stability border for the solitons, which is identified as a set of turning points of the $\mu(N)$ curves. Solitons are stable between those two curves. A notable feature of this diagram is that the stability domain shrinks to zilch at $\rho_0^{\max} \approx 0.95$ for $\rho_1 = 2$, which means that, in this case, there exists a critical ratio, $\rho_0 / \rho_1 \approx 0.47$, of the inner and outer radii above which the nonlinear annular ring cannot support stable solitons. Axisymmetric vortex solitons with various topological charges can also be found in this model, but they *all* turn out to be azimuthally unstable.

## 2. Quasi-one-dimensional nonlinearity-modulation profiles

The possibility of the stabilization of 2D solitons by a quasi-1D periodic NL was first considered by Sivan, Fibich, and Weinstein (2006). They addressed the stationary localized solutions to equation $iq_z + q_{xx} + q_{yy} + [1 + R(ax)]|q|^2 q = 0$, where $R(\xi)$ is smooth periodic modulation function, with $a$ standing for the ratio of the beam's width to the lattice period. It was found that, in this geometry, the structure of the soliton and its stability properties strongly depend on whether it is wider, of the same width, or narrower than the lattice period. Soliton solutions were found to be stable if and only if the norm-versus-propagation-constant curve describing the soliton family satisfies the slope (VK) condition against the onset of the collapse, and, simultaneously, the spectral condition guarantees the absence of the drift instability. In particular, the soliton may be unstable, in the presence of the NL, even if it satisfies the VK criterion (an example of that was given by Kartashov, Vysloukh, and Torner, 2008b). The size of the stability region depends on the magnitude of the slope of the $N(\mu)$ curve. In the work by Sivan, Fibich, and Weinstein (2006) it was concluded that solitons in the quasi-one-dimensional NLs may be stable "mathematically", in the sense that a tiny stability region exists for them, but the region is so narrow that rather weak *finite-amplitude* (rather than infinitesimal) perturbations can easily destroy such formally stable solitons. In particular, the solitons centered at minima of $R(ax)$ violate the spectral stability condition, resulting in a drift instability, as the solitons naturally tend to "roll down" from the respective maximum of the effective pseudopotential. Further, wide solitons (with $a \gg 1$), as well as those whose width is comparable to the lattice period $[a = \mathcal{O}(1)]$, are unstable (against the collapse) when they are centered at a maximum of the NL modulation function, $R(ax)$, since for them the VK criterion is not satisfied. Note the difference from the 1D case, where solitons centered at local lattice maxima may easily be stable, as the collapse instability is absent in the 1D model (Fibich, Sivan, and Weinstein, 2006). In fact, the quasi-one-dimensional NL can only stabilize *narrow* solitons, with $a \ll 1$, centered at a maximum of $R(ax)$. These narrow solitons are affected, as a matter of fact, by the local variation of the nonlinearity coefficient, but not by the global periodic NL structure. The lattice that may give rise to a stability region for the 2D solitons has to be specially designed to satisfy a certain local shape criterion, and even in this case the resulting stability domain remains very narrow. A rigorous proof of these facts, together with a quantitative analysis of the soliton's stability for the particular shapes of the linear and nonlinear lattices, was given by Sivan *et al.* (2008).

On the other hand, the conclusion about the instability of solitons with $a = \mathcal{O}(1)$, residing on lattice maxima, which is definitely valid for the smooth sinusoidal modulation of the nonlinearity coefficient, does not necessarily hold in lattices with *sharp* step-like variations of the nonlinearity. The crucial role of the sharpness was actually demonstrated by Sakaguchi and Malomed (2007), who had found stable 2D optical solitons in a model of a quasi-1D layer defined by the sharp transverse modulations of the GVD and nonlinearity coefficients. In the case of the *concomitant* localization of the GVD and nonlinearity in the stripe, this model describes the propagation of *spatiotemporal solitons* (2D "light bullets"), while the model with only the nonlinearity coefficient subjected to the transverse modulation may be realized in terms of BECs. The evolution of nonlinear excitations in such a system obeys the following NLSE $iq_z + (1/2)[q_{xx} + \beta(x)q_{tt}] + R(x)|q|^2 q = 0$, where $\beta(x) = R(x) = 1$ for $|x| < 1$ and $\beta(x) = \beta_0 \leq 1$, $R(x) = r_0 \leq 1$ for $|x| > 1$. It was found that the GVD modulation alone cannot stabilize solitons in this setting. To secure the stability, it must be combined with the nonlinearity modulation. In this notation (with the width of the modulation stripe fixed to be 2), the stabilization of 2D solitons is possible for $\beta_0 = 0$ and $r_0 \leq 0.6$, i.e., there is a certain *minimal* nonlinearity-modulation depth necessary for the stabilization. It is worthy to note that 2D solitons may be stable in the channel induced *solely* be the nonlinearity modulation, when $\beta_0 = 1$ and $r_0 = 0$, although in this case the range of energies that the stable spatiotemporal solitons may carry is narrower than in the case of the concomitant modulations of the nonlinearity and GVD. The latter result suggests that a periodic system of such purely nonlinear layers (i.e., a quasi-one-dimensional NL of the KP type) may stabilize 2D "light bullets" too. Combining the transverse modulation of the GVD and nonlinearity with the ordinary guiding structure in this model [i.e., an increased refractive index inside the modulation layer, described by additional term $V(x)q$ in the NLSE] allows one, quite naturally, to strongly expand the stability region of solitons, in terms of the energy carried by them. Finally, for $\beta_0 = 0$ and $r_0 = 1$ (the modulation of the GVD, combined with the uniform nonlinearity) the stabilization is possible only when the depth of the additional linear potential, $V(x)$ exceeds a minimal value, $v_0 \approx 0.2$.

The impact of the sharpness of the quasi-1D modulation of the local nonlinearity for the stability of the respective 2D solitons was also clearly demonstrated in a very recent work by Hung, Ziń, Trippenbach, and Malomed (2010), in the framework of the model with the nonlinearity-modulation landscape in the form of a single stripe, or a symmetric set of two parallel stripes. Using the VA and numerical methods, it was demonstrated that *all* the solitons supported by the stripe with a smooth (Gaussian) transverse profile are unstable. On the contrary, the stripe with the sharp rectangular profile gives rise to a conspicuous stability region for the 2D solitons. A set of parallel symmetric rectangular stripes may support stable 2D solitons of three types, namely, symmetric and asymmetric *single-peak* ones, and also symmetric *double-peak* solitons. The uniformity of the 2D space along the stripes in the latter model suggests to consider collisions between stable 2D solitons that may freely move in this direction. The results of direct simulations demonstrate that collisions may easily lead to the merger of the solitons with the subsequent collapse. In some cases, the colliding solitons suffer mutual destruction. Examples of quasi-elastic collisions were also found by Hung, Ziń, Trippenbach, and Malomed (2010).

## 3. Stability of solitons in two-dimensional nonlinear lattices

Stable 2D solitons in purely nonlinear 2D periodic lattices were constructed by Kartashov *et al.* (2009a). The lattice that is capable of supporting stable 2D modes can be composed of self-focusing circular regions, arranged into a square array, which is embedded into a linear medium with the same refractive index. Lattices of this type with the cubic (Kerr) nonlinearity support stable fundamental solitons, while the stability of multipoles and vortices is only possible if the nonlinearity is made *saturable*. The evolution of the light beam in such a medium is described by the equation $iq_z = -(1/2)\Delta q + R(x,y)(1+S|q|^2)^{-1}|q|^2 q$ written in the "optical" notation, where the local nonlinearity coefficient is $R=-1$ inside each nonlinear circle, and $R=0$ between them, while $S \geq 0$ accounts for the possible saturation of the self-focusing nonlinearity. Stationary solutions to this equation are looked for as $q(x,y,z) = e^{i\mu z}w(x,y)$, where $\mu$ is the real propagation constant.

Properties of fundamental solitons in such an array are presented in Fig. 23. In contrast to the uniform media with the cubic nonlinearity, where one has the well-known critical value of the norm (total power), $N_{\mathrm{T}} \approx 5.85$ for the unstable Townes solitons (Bergé, 1998) [the dashed line in Figs. 23(a) and 23(b)], the total power of the solitons supported by the NL at $S=0$ is a non-monotonous function of propagation constant $\mu$. It rapidly grows for $\mu \to 0$, as the corresponding soliton expands across the lattice. Note, however, that, in contrast to solitons in linear lattices, which bifurcate from the amplitude-modulated Bloch waves (Shi and Yang, 2008), low-power solitons in NLs remain, quite naturally, almost unmodulated, as they spread out. On the contrary, the increase of $\mu$ results in the confine-ment of the soliton to a single circle in the NL, which is accompanied by a change of the sign of slope $dN/d\mu$. Thus, solitons in the NL exist above a minimum (*threshold*) value of the power, $N_\mathrm{m}$, similar to the fundamental property of 1D solitons in one-dimensional NLs, see subsection IV.A.1. The threshold value decreases with increasing spacing $w_\mathrm{s}$ between the circles, approaching its minimum at $w_\mathrm{s}=\infty$, which corresponds to the soliton supported by a single circle, see Fig. 23(b). The non-monotonous dependence $N(\mu)$ suggests that such NLs may stabilize the fundamental solitons, as per the VK criterion. A direct linear stability analysis confirms this conjecture [Fig. 23(c) shows that the perturbation growth rate vanishes exactly at the point where $dN/d\mu$ becomes positive]. The inclusion of the *saturation* of the nonlinearity ($S>0$) results in a substantial expansion of the stability domain, as the collapse is absent even in the uniform medium with $S>0$. At $S>0$, there also exists an upper cutoff for $\mu$, where the soliton power diverges, see Fig. 23(d). Notice, that 2D solitons may be also made stable in a specific nonlinear lattice composed of alternating cubic and saturable domains provided that their centers reside on domains with cubic nonlinearity [Borovkova, Kartashov, and Torner (2010)].

Besides the fundamental solitons, NLs supports multipole and vortex states. They also feature threshold values of the total power necessary for their existence (naturally, the minimum power for the vortices exceeds that for dipoles, which, in turn, is higher than for the fundamental solitons). While the multipoles and vortices are completely unstable in the NLs with the cubic nonlinearity, they may be stabilized by the saturation. As the localization enhances with the increase of $\mu$, both dipoles and vortices get stable above a critical value of $\mu$, in the case of the saturable nonlinearity. At equal values of $w_\mathrm{s}$, the width of the stability domain for the dipoles is substantially larger than for the vortices. Other types of higher-order soliton states, such as on-site vortices and quadrupoles, may also be stable in the saturable NL.

In a related works, Hang, Konotop, and Huang (2009), and Hang and Konotop (2010) considered the propagation of a weakly nonlinear probe light beam in a resonant three-level atomic medium, where an optical lattice is induced by a standing-wave pump field, and conditions for the electromagnetically induced transparency are met. In this setting, one can achieve an effective quasi-1D modulation of the nonlinearity experienced by the 2D probe beam, and implement various dynamical regimes, either by means of simple manipulations of parameters of the induced lattice, or by varying one- and two-photon detunings, or by changing the geometry of the incident probe beam.

Very recently, Wang *et al.* (2010) have reported exact analytical solutions describing 2D matter-wave solitons trapped in the parabolic potential combined with the specially devised Gaussian nonlinearity-modulation landscape. The solutions were obtained using essen-tially the same transformations which were considered in subsection IV.A.3. Interestingly, such potentials can support an arbitrary number of nonlinear waves corresponding to discrete energy levels that can be classified by dint of two quantum numbers, following the analogy with modes of the linear parabolic potential. A somewhat similar approach was independently pursued in another recent work by Wu *et al.* (2010), who constructed exact solutions to the GPE for solitary vortices, and approximate ones for fundamental solitons, in 2D models of BEC with a spatially modulated nonlinearity of either sign (attractive or repulsive) and the parabolic trapping potential. The number of vortex-soliton modes found in the model is again determined by the discrete energy spectrum of the related linear Schrödinger equation. The vortex-soliton families found in the system with the attractive and repulsive nonlinearity turn out to be mutually complementary. *Stable* localized vortices with topological charges $S \geq 2$, and those corresponding to higher-order radial states were found, respectively, in the case of the attraction and repulsion.

Power-dependent shaping of vortex solitons in OLs featuring the modulation of both the linear refractive index and nonlinearity was addressed by Kartashov, Vysloukh and Torner (2008c). The following model with an out-of-phase modulation of the refractive index and nonlinearity was considered:

$$i\frac{\partial q}{\partial z} = -\frac{1}{2}\left(\frac{\partial^2 q}{\partial x^2} + \frac{\partial^2 q}{\partial y^2}\right) - [1 - \sigma R(x,y)]|q|^2 q - pR(x,y)q, \qquad (48)$$

where function $R(x,y) = \sin^2(\Omega x)\sin^2(\Omega y)$ describes the shape of the lattice, while parameter $\sigma$ determines the depth of the nonlinearity modulation. The nonlinear coefficient $1 - \sigma R$ in this model attains its minima at points where the linear refractive index features maxima.

Solutions for vortex solitons in such lattices can be found too. They feature four bright spots whose positions coincide with local maxima of the linear lattice at moderate power levels, when effects of the nonlinear and linear refraction are comparable. If the modulation depth of the local nonlinearity is small enough ($\sigma \lesssim 0.5$ at $p = 8$), the bright spots always stay in a vicinity of maxima of the linear lattice, but at $\sigma \gtrsim 0.5$ the competition between the linear and nonlinear refraction may result in a remarkable shape transformation, due to the concentration of the density in regions where the nonlinearity is stronger. Such shape transformation are usually accompanied by a change of the slope of the $N(\mu)$ dependence and decrease of the soliton's power with $\mu$. Thus, the nonlinearity modulation imposes a restriction on the maximal power of the vortex solitons. At $\sigma \gtrsim 0.9$, with the increase of $\mu$ the bright spots in the vortex patterns tend to fuse into modulated ring-like structures i.e., high-power vortices tend to *shrink*, rather than to expand, as it happens for smaller $\sigma$ [see eq. (48)]. The modulation of the nonlinearity profoundly affects the stability of the vortex solitons: in the NLs, they are stable only within a certain interval of the propagation constant, $\mu_{\text{low}} \leq \mu \leq \mu_{\text{upp}}$ (hence, also in a limited interval of the total power), in contrast to vortices in linear lattices, where they enjoy the strong stabilization. Since $\mu_{\text{low}}$ increases with $\sigma$ while $\mu_{\text{upp}}$ decreases, the stability domain shrinks to nil at $\sigma \gtrsim 0.79$. Thus, the off-site vortex solitons may be stable in the NL only when the nonlinearity modulation depth does not exceed the critical value.

Lastly, it is relevant to mention that a mathematically rigorous approach to the analysis of the soliton stability in 2D models including mixed linear and nonlinear lattices was very recently elaborated by Lin, Wei, and Yao (2010). They started the analysis with the case of the NL only, and then investigated the stabilization of the solitons by the addition of linear-lattice potentials.

## B. Solitons in models of photonic-crystal fibers
### 1. Theoretical considerations

To make use of the high potential of PCFs for applications, it is important to realize dynamical tunability of their properties, including the band-gap spectrum. Such tunability can be achieved in *nonlinear* PCFs featuring periodic modulations of the nonlinearity, i.e., the light beams coupled into the PCF induce a local modification of the refractive index that depends on its beam's intensity profile.

The possibility of the self-trapping and formation of the localized modes near the band's edge in a 2D nonlinear photonic crystal with a reduced symmetry was demonstrated by Mingaleev and Kivshar (2001). They employed the technique based on the Green's function to explain the physical mechanism of the mode stabilization, associated with the effective nonlinear dispersion and long-range interactions in the photonic crystal. They studied 2D photonic crystals represented by a square lattice built of two types of cylindrical rods: those of a larger radius, made from a linear material, were placed in the corners, while nonlinear rods of a smaller radius were set at the center of each cell. The evolution of the $z$-polarized slowly evolving light field in this setting is governed by equation $[\Delta + \varepsilon \omega^2 / c^2]q(x,t) = -(2i\varepsilon\omega / c^2)\partial q / \partial t$, where $\varepsilon(x,y)$ is the dielectric constant, and $\Delta$ is the transverse Laplacian. In the linear limit, when $\varepsilon(x,y)$ does not depend on the intensity, the frequency spectrum of such a crystal has a characteristic bandgap structure, so that solitons may form in the gaps, in the presence of the nonlinearity. Since in real photonic crystals the modulation of $\varepsilon(x,y)$ is comparable with its average value, making the standard NLSE description inappropriate, the nonlinear rods were considered as defects with dielec-tric constant $\varepsilon_2(x,y,E) = (\varepsilon_{20} + \chi^{(3)}|E|^2)\theta(x,y)$, where $\varepsilon_{20}$ is the linear dielectric constant, and $\theta(x,y)$ is a function describing the distribution of rods. The Green's function of the linear crystal, $G(x,y,\omega)$, was employed to transform the initial evolution equation into its discrete version that takes into account effective long-range interactions:

$$i\sigma \frac{\partial E_{nm}}{\partial t} - E_{nm} + \sum_{kl} J_{n-k,m-l}(\omega)(\varepsilon_{20} + \chi^{(3)}|E_{kl}|^2)E_{kl} = 0, \qquad (49)$$

where amplitudes $E_{nm}$ pertain to the rods, while parameters $\sigma$ and $J_{nm}$ are determined by Green's function $G(x,y,\omega)$ of the real crystal, that can be computed numerically by means of the finite-difference time-domain method for the particular geometry and dimensions of the rods. Typical GS (gap-soliton) solutions to the latter equation are characterized by multiple field oscillations and the concentration of light mainly in the nonlinear rods. Such solitons can be stable even in the low-amplitude regime (which is potentially accessible in the experiment), when they are very broad. The stability of such broad modes was attributed to nonlinear long-range interactions described by coefficients $J_{nm}$ that slowly decrease with $n,m$, and also strongly depend on radii of the rods and respective dielectric constants.

The existence of GSs and GS trains was also confirmed numerically in finite-sized 2D nonlinear photonic crystals by using the multiple-scattering approach with an iterative scheme that allowed to solve the nonlinear Helmholtz equation describing the transmission of light in such crystals (Xie, Zhang, and Zhang, 2003). They considered photonic crystals modeled by a square matrix of nonlinear cylinders with an enhanced refractive index. Assuming the cylindrical symmetry of the light field in each nonlinear cylinder allowed an iterative calculation of the total dielectric constant, $\varepsilon(x,y) = \varepsilon_0 + \chi|E(x,y)|^2$, in the framework of the Helmholtz equation (here $\chi < 0$, i.e., the nonlinearity is *defocusing*), using the multiple-scattering method. This, in turn, enables one to determine the transmission coefficient as a function of the input light-field's amplitude for a fixed frequency inside the gap, near the lower band edge. The transmission coefficient features several maxima at different frequencies, that indicate the formation of different GS modes, ranging from a single-soliton to soliton trains. A similar procedure was used to obtain symmetric and asymmetric GSs in quasiperiodic crystals (GS solutions in the model with the constant self-repulsion coefficient and quasiperiodic linear-lattice potential were reported by Sakaguchi and Malomed, 2006b). The size of the localized modes depends on the frequency at which the transmission coefficient was calculated: As this frequency moves deeper into the gap, the spatial size of structures decreases.

## 2. Experimental realization: solitons in a liquid-infiltrated photonic-crystal fiber

GSs were *experimentally created* in PCFs with holes filled with high-index nonlinear liquids featuring a thermal nonlinearity (Rosberg *et al.*, 2007, Rasmussen *et al.*, 2009). In these experiments, strongly tunable diffraction of beams was first demonstrated in a triangular array created by infiltrating holes of a standard PCF, fabricated from fused silica, with the castor oil, that allows one to reduce the refractive-index contrast with the host silica (Rosberg *et al.*, 2007). The sample was placed into a temperature-controlled oven which allows precise thermo-optic tuning of the temperature-dependent refractive index of the infiltrated oil (the corresponding thermo-optic coefficient is $\beta \approx -3 \times 10^{-4}$ $K^{-1}$). The refractive-index difference between the fused silica cladding and holes filled with oil was decreased to below $2 \times 10^{-3}$ by heating the PCF above 70 °C. In this regime, individual waveguides feature the single-mode transmission, and the coupling between neighboring sites due to the overlap between evanescent modal fields significantly increases, resulting in a dramatic enhancement of the discrete diffraction with the increase of the sample's temperature. Because of the large negative thermo-optic coefficient inherent to most liquids, the heating produced by partial absorption of the propagating beam itself causes a further decrease of the refractive index in the holes of the photonic crystal, and nonlinear self-defocusing of the beam. The latter can be used to build a tunable all-optical power limiter. By properly balancing the linear-refractive index contrast and the strength of the defocusing nonlinearity in such crystals, Rasmussen et al., 2009, achieved the formation of 2D nonlocal GSs. Mathematically, the propagation of the laser radiation in such media is described by the corresponding NLSE, $iq_z + \Delta q = -[V(x,y) - R(x,y)T]^2 q$, coupled to equation $\Delta T = -|q|^2$ describing the steady-state temperature distribution, where $V$ stands for the refractive-index field, and $R = 1$ in the holes and zero otherwise. The corresponding sample consists of high-index cylinders placed so as to form a hexagonal pattern inside a circle of radius $R_0$, surrounded by a large homogeneous silica circle of radius $R_1$ outside the last ring of the holes [Fig. 24(a)]. By analyzing eigenmodes of this finite structure, it was found that the spectrum of eigenvalues is divided into bands of closely spaced values that may be separated by a gap for a proper set of parameters, akin to gaps in the spectrum of truly periodic systems. The presence of the gap allows the formation of GSs bifurcating from the bottom of the first band, existing due to the defocusing thermal nonlinearity which reduces the refractive index in the holes, but almost does not affect the refractive index of the host medium. The representative intensity distribution in such a soliton with a staggered phase structure is shown in Fig. 24(b). It is worthy to note that such states may have propagation constants penetrating into the bands of the spectrum, which is a consequence of the finite extent of the system under consideration. GSs were observed by heating a sample to 76 °C, that reduces the refractive-index difference between the holes and cladding to an appropriate value, followed by coupling the beam into the central hole. A transition was observed from the linear diffraction at low input powers [Fig. 24(c)] to the formation of solitons at high powers [Fig. 24(d)]. Measurements of the light power trapped in the input hole as a fraction of the total input power suggest that the strongest localization of light occurs at intermediate powers, while delocalization takes place at low and high powers, in agreement with known properties of stationary GSs.

## 3. The symmetry analysis

Properties of bright solitons in PCFs with defects were theoretically analyzed by Ferrando *et al.* (2003). The guidance in such crystals in the linear regime is provided by the transverse localization of light at a defect (in fact, this is the core region, where one hole is missing), due to a complex mechanism of the interference of light in the periodic air-hole array that form the photonic crystal cladding. This mechanism is similar to that occurring in electron crystals in the presence of donor or acceptor impurities. In crystals of this type, the guided-mode's power is chiefly confined to the silica core, and the nonlinear localized solutions also form in this region, rather than on "sites" of the photonic crystal (i.e., air holes), where the nonlinearity is negligible. The modeling of the light transmission in such crystals should be based on the Helmholtz equation, due to the large refractive-index contrast: $\Delta q + k_0^2[n_0^2(x,y) + n_2^2(x,y)|q|^2]q = -\partial^2 q / \partial z^2$. Here $k_0$ is the wavenumber in vacuum, $n_0 = 1$ in the air holes, and $n_0 = n_{\text{silica}}$ in the cladding, whereas $n_2$ is different from zero only in silica. Solutions to this equation can be found by dint of the modal-expansion method, adjusted so as to include the inhomogeneous nonlinear term (for details see Ferrando *et al.*, 2003). The solution pertaining to the highest eigenvalue of $\mu^2$ corresponds to the fundamental mode of the effective fiber existing due to the combined effect of the linear and nonlinearity-induced guidance. Typical fundamental soliton profiles corresponding to different values of the normalized power , $\gamma = P n_2^2 / A_0$ (here $P$ is the total power carried by the soliton, and $A_0$ is the effective core area) are shown in Fig. 25(a). With the increase of $\gamma$, the solitons that bifurcate from linearly guided modes of the PCF shrink dramatically and transform into a narrow bright spot residing at the center of the crystal's core. Such solutions cease to feel the periodic structure of the photonic crystal and thus are almost tantamount to the *Townes solitons* in the uniform medium, which are unstable in the presence of perturbations. At the same time, broad solitons, as well as solitons with intermediate values of the width, whose shapes are notably modulated by the air-hole structure, are expected to be completely stable and can be excited by Gaussian input beams. This indicates that the photonic-crystal structure not only helps to generate solitons by reducing the power necessary for their excitation, but also facilitates their stabilization. Notice that, besides the fundamental solitons, PCFs can support nodal (or dipole) solitons characterized by nodal lines determined by the discrete symmetry of the underlying crystal. Ferrando *et al.* (2005a) used the group-theory approach to analyze the role played by nonlinearities in the realization of the discrete symmetry, and showed that the nonlinearity may cause breaking of the discrete symmetry, which is associated with the generation of a new type of solitons with a lower symmetry than that of the underlying setting. They considered the following nonlinear eigenvalue problem, that can be obtained from the Helmholtz equation: $(\mathcal{L}_0 + \mathcal{L}_{\mathrm{nl}})q = \mu^2 q$. Here $\mathcal{L}_0$ is a linear operator (that includes the Laplacian and describes the refractive-index profile in the PCF) acting on the transverse coordinates. The operator is invariant under the action of the discrete point-symmetry group, $\mathcal{G}$, while $\mathcal{L}_{\mathrm{nl}}$ is a nonlinear operator that locally depends on $|q|$. The group-symmetry arguments predict that, if the system described by this equation is invariant under some discrete-symmetry group $\mathcal{G}$, then any of its solutions either belongs to one of representations of group $\mathcal{G}$, or to one of its subgroups $\mathcal{G}'$. This means, in particular, that, if for some function $q$ the nonlinear operator $\mathcal{L}_{\mathrm{nl}}$ will be invariant under group $\mathcal{G}'$, which is a subgroup of group $\mathcal{G}$ associated with the invariance of $\mathcal{L}_0$, then the total operator, $\mathcal{L}_0 + \mathcal{L}_{\mathrm{nl}}$ is also invariant under subgroup $\mathcal{G}'$, and function $q$ featuring such a symmetry may be a solution to the full equation, $(\mathcal{L}_0 + \mathcal{L}_{\mathrm{nl}})q = \mu^2 q$. This explains why nonlinear solutions with the lower symmetry (defined by $\mathcal{G}'$) may exist in the system that in the linear case possesses the higher symmetry (defined by $\mathcal{G}$); a simpler manifestation of this general principle is represented by the spontaneous symmetry breaking in the double-well potential or pseudopotential (see subsection IV.D above) - while all eigenstates in the linear double-well system are either symmetric or antisymmetric, the nonlinearity gives rise to asymmetric modes.

A particular example of such solutions was found in a triangular PCF with symmetry group $\mathcal{C}_{6v}$, composed by discrete $\pi/3$ rotations, plus the specular reflections with respect to $x, y$ axes. Fundamental solitons supported by the PCF of this type are the simplest examples of solutions featuring the full $\mathcal{C}_{6v}$ symmetry. However, in accordance with the group-theory predictions, one can also construct, for such PCFs, nonlinear solutions belonging to subgroup $\mathcal{C}_{2v}$ (composed by $\pi$ rotations plus the specular reflections with respect to $x, y$ axes) of group $\mathcal{C}_{6v}$. The solutions, depicted in Fig. 23(b), correspond to the nodal, or dipole, solitons. Two types of the solitons were found, with different orientations of the nodal lines and different eigenvalues $\mu$. However, the analysis reveals that such solutions are subject to oscillatory instabilities.

The existence of vortex solitons in PCFs with defects was analyzed by Ferrando *et al.* (2004), and by Ferrando, Zacarés, and Garcia-March (2005). Vortex solitons with topological charges $m = 1, 2$ were obtained by Ferrando *et al.* (2004), for a triangular PCF with a defect. They represent another example of solutions with the $\mathcal{C}_{6v}$ symmetry (the PCF is characterized by the same symmetry), in addition to the fundamental solitons. Similarly to the fundamental solitons, vortices that are notably modulated by the air-hole structure of the PCF at low powers gradually shrink to the center of the defect with the increase of the power, and acquire almost axially symmetric shapes. There is no finite threshold power stipulating the generation of vortex solitons in this setting, i.e., they bifurcate from the corresponding linear defect modes carrying the vorticity. In this case, the group theory predicts the following angular dependence of the phase of the vortex soliton: $\arg(q) \approx m\theta + b_m(r)\sin(6\theta)$, for crystals with the $\mathcal{C}_{6v}$ symmetry, i.e., unlike vortices in uniform media, the additional sinusoidal modulation of the phase appears, that reflects the symmetry of the guiding structure. As in the case of the nodal solitons, vortices in PCFs with defects are unstable, and tend to decay into two filaments featuring complex dynamics at the center of the crystal. However, the group theory also allowed to make a very important prediction, *viz.*, that, unlike the uniform medium, symmetric vortices of an arbitrarily high order *cannot* be generated in the 2D system featuring a discrete-point symmetry, i.e., the symmetry imposes the restriction on the *largest possible* charge of symmetric vortex solitons (Ferrando, Zacarés, and Garcia-March, 2005). For example, the allowed lowest-order eigenfunctions of the above-mentioned operator, $\mathcal{L} = \mathcal{L}_0 + \mathcal{L}_{\mathrm{nl}}$, possessing the $\mathcal{C}_{6v}$ symmetry are presented in Fig. 24. They include the fundamental soliton, vortices with charges $\pm 1$ and $\pm 2$, but, instead of vortices with charges $\pm 3$, only a multipole state can be found. This is a manifestation of the vorticity-cutoff theorem, which asserts that, if the system is invariant under the $\mathcal{C}_n$ or $\mathcal{C}_{nv}$ point-symmetry group, then charge $m$ of the vortex solutions in such a system has an upper bound (cutoff) given by $m < n/2$ for even $n$, and $m < (n-1)/2$ for odd $n$. Note that a similar conclusion was obtained for solitons in optically-induced lattices with the $\mathcal{C}_{nv}$ symmetries, where the propagation of light can be described by the standard NLSE (Kartashov *et al.*, 2005). The group theory was used by Garcia-March *et al.*, 2009a, to develop a concept of the *angular pseudo-momentum*, that allows a simple classification of possible solutions in the discrete-symmetry media, and which is conserved upon the evolution of the optical field in such media. This concept allows one to predict a transformation of the charge of vortex solitons at the boundary between two materials with different symmetries (Ferrando *et al.*, 2005b). The impact of the discrete sym-metry of the underlying guiding structure with the inhomogeneous nonlinearity on the properties of more complicated vortex solitons with multiple off-axis phase singularities was analyzed by Garcia-March *et al.* (2009b).

Higher-order nonlinear modes with nontrivial topology in the form of vortex solitons and soliton clusters, and their bifurcations in PCFs with defects were investigated numerically by Salgueiro and Kivshar (2009). They found a variety of soliton clusters, with symmetries that may be different from the lattice symmetry, and discussed their stability. Structures with a regular triangular lattice and a central defect were considered, where the light propagation is described by equation $iq_z + \Delta q + [n_a + V(x,y)(\delta + |q|^2)]q = 0$ (i.e., the paraxial description was employed), with $V(x,y) = 0$ in the holes and $V(x,y) = 1$ in the nonlinear substrate, and $\delta = n_s - n_a$ characterizing the difference of the refractive index between the substrate and material of the holes. In addition to the simplest vortex and dipole solitons that were discussed above, this system supports azimuthally modulated multipole solitons presenting a central dislocation combined with a vortex structure. For example, such solitons can possess three or four well-resolved maxima in the azimuthal direction, and are characterized by different orientations with respect to the underlying $\mathcal{C}_{6v}$-symmetric guiding structure. Such states require a *threshold* (minimal) power for their existence, and they bifurcate from different points of the $N(\mu)$ diagram for usual vortex solitons, so that states featuring a larger number of azimuthal intensity oscillations require higher powers for their existence. Similarly to previously found excited states in PCFs, such solitons are prone to instabilities.

Salgueiro and Kivshar (2009) also constructed vortices and soliton clusters in dual-core PCF couplers that are represented by two missing holes, i.e., two defects, in the underlying periodic triangular structure. Various nonlinear combinations of different modes were found, including double-vortex structures [Fig. 27(a)], combinations of vortices and fundamental modes [Fig. 27(b)], as well as combinations of cluster-like modes [Figs. 27(c) and 27(d)]. Dual-core PCF couplers may function as conventional directional couplers, despite their complex refractive-index and nonlinearity structure (Salgueiro and Kivshar, 2005). Such twin-core PCFs support stationary symmetric, antisymmetric, and asymmetric nonlinear modes. While at low powers only symmetric and antisymmetric modes can exist, at a certain threshold power the asymmetric mode bifurcates from the symmetric one, while the symmetric mode becomes unstable above this point. Similar to other nonlinear couplers, the switching in such a coupler manifests itself as a periodic energy exchange between the cores at low powers, when only one core is excited at the input, while the exchange is completely arrested at sufficiently large powers when the system can support asymmetric modes.

Nonlinear vortex modes in such twin-core PCFs were studied by Salgueiro and Santos (2009).

PCFs may support two-component localized nonlinear waves (vectorial solitons) consisting of two mutually trapped components confined due to the specific linear refractive-index distribution in the PCF and the self-focusing nonlinearity of its material (Salgueiro *et al.*, 2005). It was demonstrated that such mutually trapped states, bifurcating from the corresponding scalar states, may be *stable*. To describe the propagation of the vectorial states, the following coupled equations were used:

$$i\frac{\partial q_{1,2}}{\partial z} + \left(\frac{\partial^2 q_{1,2}}{\partial x^2} + \frac{\partial^2 q_{1,2}}{\partial y^2}\right) + n_a q_{1,2} + V(x,y)(\delta + |q_{1,2}|^2 + C|q_{2,1}|^2)q_{1,2} = 0, \qquad (50)$$

where the meaning of parameters $\delta, n_a, V(x,y)$ is the same as indicated above. The existence domain for the vectorial solitons is symmetric with respect to values of the propagation constants, $\mu_{1,2}$, as is evident from the symmetry of Eq. (50) – see Fig. 28. The domain is bounded by two lines at which the vectorial solitons bifurcate from their scalar counterparts. When $C < 1$ is close to the lower bifurcation curve, the second component decreases, while near the upper bifurcation curve the first component gradually vanishes. When $C > 1$, one is dealing with the opposite situation, just as in the case of 1D vectorial solitons in NLs (see section IV.C). The presence of the periodic lattice of holes in the PCF suggests that the vectorial solitons may be stable in this system. It was shown that, by applying the generalized matrix stability criterion, it is possible to determine a border between the stable and unstable regions, which is defined as a set of points that fulfill the marginal stability condition, $\det J = 0$, where elements of the Jacobian are $J_{ij} = \partial P_i / \partial \mu_j$, with $\mu_i, P_i$ $(i = 1, 2)$ being the propagation constants and powers of the respective soliton components (this condition is, as a matter of fact, a vectorial generalization of the VK criterion pertaining to the scalar setting). The corresponding stability border, as well as the stability and instability regions, are depicted in Fig. 28, in the plane of $(\mu_1, \mu_2)$. The structure of the stability domain strongly depends on the cross-modulation coefficient $C$, which is a characteristic feature of multi-component systems.

## C. Discrete models

A 2D nonlinear-Schrödinger lattice with nonlinear inter-site couplings, which models a square array of evanescently coupled linear optical waveguides, embedded in a nonlinear Kerr material, was studied by Öster and Johansson, 2009. The corresponding DNLSE that describes such a system, with the out-of-phase modulation of the linear refractive index and nonlinearity, and taking into account only nearest-neighbor couplings, is

$$i\frac{dq_{nm}}{dz} = Q_1 q_{nm} + Q_2 \Delta q_{nm} + 2Q_3 q_{nm} |q_{nm}|^2 + 2Q_4(2q_{nm}\Delta|q_{nm}|^2 + q_{nm}^*\Delta q_{nm}^2) \quad (51)$$
$$+ 2Q_5[2|q_{nm}|^2 \Delta q_{nm} + q_{nm}^2 \Delta q_{nm}^* + \Delta(q_{nm}|q_{nm}|^2)],$$

where $q_{nm}$ is the complex amplitude of the electric field in the waveguide with number $n,m$, coupling parameters $Q_1$ through $Q_5$ are determined by overlap integrals of modes localized at adjacent sites, and operator $\Delta$ is defined as $\Delta q_{nm} = q_{n-1m} + q_{n+1m} + q_{nm-1} + q_{nm+1}$. An important difference between this equation and the standard DNLSE which describes the transmission of light in waveguide arrays with the uniform nonlinearity [see Eq. (12)] is the presence of terms with coefficients $Q_4$ and $Q_5$, which account for the out-of-phase modulation of the nonlinearity and linear refractive index. These nonlinear coupling terms may be of the same order as the usual on-site nonlinearity coefficient, $Q_3$. A number of discrete-soliton solutions to Eq. (51) were obtained, ranging from the simplest odd soliton to even ones (solitons centered between two lattice sites, with equal amplitudes at both sites) and dipole modes (solutions with the sign alternating between adjacent sites). Solutions of all the above-mentioned types may be stable for properly selected values of $Q_4$ and $Q_5$, which is surprising, as even solitons of the usual DNLSE with the cubic on-site nonlinearity are always unstable. Moreover, combined stability regions for even and dipole states cover the instability region of odd solitons. The stability boundaries for different solutions do not coincide exactly, in contrast to the 1D lattice with nonlinear couplings, where odd solitons become unstable almost precisely at the same locus where even solitons become stable, resulting in a stability exchange between the two species. Instead, in the presently considered model one observes simultaneous stability of at least two different species of 2D solitons in sufficiently wide parameter regions. Thus, it was concluded that, although the stability boundaries for odd and even solitons are located far apart in the parameter plane, one may still define the stability exchange between them, which is connected to the existence of an intermediate asymmetric unstable solution between the stability boundaries of the odd and even solitons. The asymmetric solutions emerge through a pitchfork bifurcation. Moreover, comparing values of the Hamiltonian for the solutions with equal values of the total power (norm) reveals the equality of the Hamiltonians of the odd and even solitons along some boundaries in the parameter space. This difference in values of the Hamiltonian characterizes the height of the respective *Peierls-Nabarro barrier*, i.e. the potential barrier that has to be overcome (for instance, by imposing a kick, in the form of a phase tilt, on the soliton) to achieve the mobility. Therefore, one may expect enhancement of the soliton mobility in the region of the stability inversion. Nevertheless, direct simulations show that, in the 2D model, the mobility of the solitons remains very poor. This fact was attributed to the fact that stability boundaries for the odd and even solitons are located too far apart in the parameter plane, in the case of the 2D lattice (in contrast to its 1D counterpart), that hampers the transition between such states (which is a basic step in the motion across the lattice). An interesting finding is that the nonlinear coupling terms may exactly cancel their linear counterparts, thus leading to the existence of exact *compacton* solutions, whose amplitude strictly vanishes outside a certain domain.

## D. Solitons in a dissipative nonlinear lattice

As discussed above, a versatile technique allowing for the creation of NLs in BEC is based on inducing the FR with the spatially modulated strength. On the other hand, it is known that, generally speaking, the FR gives rise not only to the change of the scattering length, $a_s$, but also to the nonlinear loss, which is accounted for by an imaginary part of $a_s$ (see, e.g., Fedichev *et al.*, 1996). Thus, the NL induced by means of the FR technique may also include a dissipative component. This possibility was analyzed by Abdullaev, Gammal, da Luz, and Tomio (2007) in the framework of an accordingly extended two-dimensional GPE with an *anisotropic* NL,

$$i\frac{\partial q}{\partial t} = -\left(\frac{\partial^2 q}{\partial x^2} + \frac{\partial^2 q}{\partial y^2}\right) - \{\gamma_0 + (\gamma_1 + i\gamma_2)[1 + \delta + \cos(kx) + \delta\cos(ky)]\}|q|^2 q \qquad (52)$$

where coefficient $\gamma_2 > 0$ accounts for the dissipative component of the NL. In this connection, it is relevant to mention that a model of a dissipative OL acting on BEC was very recently introduced (in the 1D form) by Bludov and Konotop (2010).

Applying to the model based on Eq. (52) a version of the VA generalized so as to include effects of the dissipative term [see, e.g., papers by Chávez Cerda, Cavalcanti, and Hickmann (1998), and Skarka and Aleksic (2006)], and using direct simulations, it was demonstrated, in the case of the attractive nonlinearity, that the nonlinear loss prevents the collapse of the condensate, replacing it by several cycles of quasi-collapse and expansion, which is followed by an eventual decay [in this respect, it is relevant to mention that, as shown by Leblond, Malomed, and Mihalache (2009), the cubic loss term prevents the collapse even in the case of the *supercritical* focusing nonlinearity, which may be accounted for by the self-attractive quintic term in the 2D equation]. Abdullaev *et al.* (2007) also analyzed a possibility to compensate the nonlinear loss by a linear "feeding" term $i\alpha$, with $\alpha > 0$, added to the right-hand side of Eq. (52).

## VI. Surface solitons in nonlinear lattices

In this section we address surface solitons that may form at the edge of nonlinear or mixed linear-nonlinear periodic lattices. Surface solitons constitute an important class of localized modes in lattice media (Lederer *et al.*, 2008), therefore it is relevant to study the surface solitons and their stability in NLs. In this section, we first describe surface solitons states in the KP model that allows one to obtain analytical expressions for soliton shapes, by properly tailoring known analytical solutions at different sides of the interface. We outline specific features of the dynamics of 1D solitons at the interface between the purely nonlinear lattice and uniform medium, the formation of surface solitons at the edge of layered thermal media, and differences in properties of such states and conventional nonlocal surface solitons in uniform thermal media.

In this section, we also address asymmetric matter-wave vortices and multipole solitons forming in external parabolic potentials in the presence of sharp boundaries between regions of different strengths of interatomic interactions, as well as power-dependent shape transformations and interactions with interfaces of truncated lattices in systems featuring out-of-phase modulations of the linear refractive index and nonlinearity. The concept of the nonlinear surface-wave formation at the interface of periodic OLs with the uniform nonlinearity was introduced by Makris *et al.* (2005), in the framework of a discrete model. Suntsov *et al.* (2006) have confirmed experimentally that the formation of surface waves at the edge of 1D waveguiding arrays is possible for moderate power levels. Formation of gap surface solitons at the edge of defocusing lattices is also possible, as was shown theoretically (Kartashov, Vysloukh, and Torner, 2006) and confirmed experimentally in defocusing LiNbO$_3$ waveguiding arrays (Rosberg *et al.*, 2006; Smirnov *et al.*, 2006). It is relevant to mention that experimental observations of 2D surface waves at the edge of usual OLs were reported too (Wang *et al.*, 2007b; Szameit *et al.*, 2007).

### A. One-dimensional models
### 1. The Kronig-Penney model

Kominis, Papadopoulos, and Hizanidis (2007) used the phase-space method for the construction of analytical soliton solutions at the interface of the NL of the KP type and a linear or nonlinear homogeneous medium, as well as at the interface between two dissimilar NLs. This method allowed them to find soliton solutions with both zero and nonzero semi-infinite backgrounds.

The evolution of light beams in the respective structure is described by the usual equation, $iq_z + q_{xx} + V(x)q + R(x)|q|^2 q = 0$, where $V(x)$ and $R(x)$ change in the step-like manner between the segments, forming the KP lattice. The phase space corresponding to NL segments (for the case of $\mu > \varepsilon_1$, where $\mu$ is the propagation constant, and $\varepsilon_1$ stands for the value of $V$ in nonlinear segments) is shown in Fig. 29(a), indicating the existence of a homoclinic solution. The phase space corresponding to the linear part is shown in Figs. 29(b) and 29(c), for $\mu < \varepsilon_{2,3}$ and $\mu > \varepsilon_{2,3}$, respectively (here $\varepsilon_{2,3}$ stand for the values of $V$ in the linear segments of the lattice, and in the uniform medium, respectively). As before, for $\mu_n = \varepsilon_2 - (n\pi/L)^2$, $n = 1,2,...$, that corresponds to an integer number of half-periods of the solution in the linear part of width $L$, the continuity conditions for the field and its derivative are met at all boundaries inside the periodic medium. This fact makes it possible to construct analytical solutions, by using known sech-type expressions for solitons in the uniform medium and sinusoidal functions in the linear one. In terms of the corresponding phase space, this corresponds to the motion along the homoclinic orbit, periodically interrupted due to the passage through the linear segments. For the case of the interface with a nonlinear medium which has the same characteristics as nonlinear segments in the lattice, the point representing the soliton in the phase space keeps moving along the same homoclinic orbit in the uniform medium, approaching the origin as $x \to -\infty$, that corresponds to fully localized surface solitons, see Fig. 29(d). For the interface with the linear medium, two situations are possible: When $\mu < \varepsilon_3$, the solution meets one of elliptical curves in the phase space and evolves periodically at $x \to -\infty$, which corresponds to solitons having a zero asymptotic value at $x \to +\infty$, and a periodic pedestal at $x \to -\infty$ [see Fig. 29(e)], while, for $\mu > \varepsilon_3$, there exists a solution for which the part of the homoclinic orbit intersects one of straight lines from Fig. 29(c), tending to the origin and giving rise to a fully localized soliton decaying in the linear medium too [see Fig. 29(f)]. Some of the modes constructed in this way may be stable, while some others undergo reshaping and transformation into stable modes with different symmetries.

## 2. The interface between the uniform medium and a nonlinear lattice

Abdullaev, Galimzyanov, Brtka, and Tomio (2009) studied the dynamical trapping and propagation of matter-wave solitons through an interface between a uniform medium and a purely nonlinear lattice in the framework of the NLSE, $iq_z + q_{xx} + R(x)|q|^2 q = 0$, with $R(x) = 1 + \theta(x)[r_0 + r_1 \sin(2x)]$, where $\theta(x)$ is the Heaviside's step function. Collisions of solitons, arriving with a certain initial velocity from the uniform medium, with the nonlinear interface located at $x = 0$ were considered too. Taking into regard that fact that the interface in this case is, in fact, self-induced (i.e., solitons with higher peak amplitudes feel stronger perturbations when they hit the interface), it was shown that the interaction an of incident soliton, $q = \sqrt{2} A \operatorname{sech}[(x-\xi)/\alpha] \exp[i\mu(x-\xi)^2 + ik(x-\xi)]$, with the nonlinear interface may be described by equations $\partial^2 \xi / \partial z^2 = -\partial V_\xi / \partial \xi$ and $\partial^2 \alpha / \partial z^2 = -\partial V_\alpha / \partial \alpha$, where the respective effective potentials are

$$V_\xi(\alpha, \xi) = -\frac{N}{4\alpha}(r_0 F_0 + r_1 F_2), \qquad (53)$$
$$V_\alpha(\alpha, \xi) = \frac{8}{\pi^2 \alpha^2} - \frac{N}{\pi^2 \alpha}[4 + 3(r_0 F_0 + r_1 F_2)].$$

Here $N$ is the soliton's norm, $F_0 \equiv (2/3) + \tanh(\xi/\alpha) - (1/3)\tanh^3(\xi/\alpha)$, and $F_2 \equiv \int_{-\xi/\alpha}^{\infty} \operatorname{sech}^4(y)\sin(2\alpha y + 2\xi)dy$. The resulting potential $V_\xi$ oscillates in the region occupied by the NL. Since the interface is nonlinear, the amplitude of this effective potential decreases with the decrease of the soliton's norm, reflecting the fact that low-amplitude solitons would be less disturbed upon passing the interface than their high-amplitude counterparts. The position and width of a stationary soliton located near the surface can be determined from equations $\partial V_\xi / \partial \xi = \partial V_\alpha / \partial \alpha = 0$, that yield a very accurate prediction, in comparison with direct simulations. The conditions of reflection or transmission of the soliton at the interface are determined by the potential-barrier's height, $V_\xi$. If the initial kinetic energy of the effective particle associated with the soliton is larger than the height of the potential barrier, the soliton passes the interface and starts to travel across the NL (where it can be eventually trapped due to radiation losses), while for low kinetic energies it is reflected (at least for $r_0 = 0$, i.e. when there is no step in the constant part of the nonlinearity). Since the height of the effective potential barrier depends on the nonlinearity-modulation depth and input norm, one can switch between regimes of the soliton transmission and reflection by tuning input conditions or the NL strength. For broad solitons, the reflection is only possible for $r_1 < 0$, i.e., it may be crucial whether the lattice has a maximum or minimum at the interface.

Another effect in the system of the same type was recently reported by He, Mihalache, and Hu (2010), *viz.*, a spontaneous drift of a soliton along the surface, provided that the soliton's norm exceeds a certain critical value. They also studied the rebound, penetration and trapping of a tilted soliton colliding with the surface.

Interesting results were also obtained by Dong and Li (2010), who studied a nonlinear interface of a different type, between two photonic lattices embedded into saturable media with different values of the saturation parameter. Surface optical solitons of dipole, quadrupole, and vortex types were found in this system. The multipole and vortex solitons are stable when their total powers exceed the corresponding threshold values.

### 3. The surface of a thermal layered medium

The existence and properties of multipole surface solitons localized at a thermally insulating interface between layered thermal media and a linear dielectric were analyzed by Kartashov, Vysloukh, and Torner (2009b). The propagation of light in such a material is described by the NLSE for the amplitude $q$ of the optical field, coupled to the equation for the temperature perturbation, $T$:

$$i\frac{\partial q}{\partial z} = -\frac{1}{2}\frac{\partial^2 q}{\partial x^2} - \sigma(x)Tq, \quad \frac{\partial^2 T}{\partial x^2} = -|q|^2 \quad \text{for } -L \leq x \leq 0,$$
$$i\frac{\partial q}{\partial z} = -\frac{1}{2}\frac{\partial^2 q}{\partial x^2} \quad \text{for } x < -L \text{ and } x > 0.$$
(54)

Here $L$ is the width of the thermal medium, $\sigma(x)$ accounts for variations of the thermo-optic coefficient (from positive to negative values) between different layers, while boundary conditions for temperature are $T|_{x=-L} = 0$ and $\partial T/\partial x|_{x=0} = 0$, i.e., the boundary at $x = 0$ is assumed to be *thermally insulating*, while the boundary at $x = -L$ is *thermally stabilized*. The surface solitons in this model may form in a vicinity of the thermally insulating interface even in the uniform focusing thermal medium, when $\sigma$ is constant. Such solitons may be built of several constituents (*poles*). Due to the specific character of the thermal nonlinearity, the perturbation of the refractive index is nonzero everywhere in the thermal medium, decreasing almost linearly toward its left border. At the edge of the uniform thermal medium, only multipoles with number of poles $n \leq 2$ may be stable (this resembles constraint $n \leq 4$ on the number of poles in stable multipoles in the ordinary bulk nonlocal materials). In the case of the layered thermal medium composed of alternating focusing and defocusing layers, a light beam entering the medium self-induces a NL, that becomes more pronounced with the increase of the peak amplitude of the beam. It is strongly asymmetric because of the boundary conditions, see Fig. 30(d). In that case, the soliton's peak may be localized in any focusing layer. Fundamental surface solitons residing in the first layer exhibit pronounced oscillations in their left wings [Fig. 30(a)]. Multipoles with $n > 1$, centered at intermediate and high powers in the second, third, etc., focusing layers carry the number of poles equal to the number of the layer where the most pronounced peak is located, see Figs. 30(b) and 30(c). When the power decreases, the left outermost pole shifts into the bulk of the thermal medium, gradually jumping between adjacent focusing layers. In contrast, when the power increases, light tends to concentrate almost within a single layer of number $n$. Periodic modulations of the thermo-optic coefficient dramatically affect the stability of the surface solitons. In such periodic media, there is *no restriction* on the number of poles in stable solitons (the multipoles with $n$ up to 10, at least, can be stable).

## 4. Gap solitons at the surface

Many families of 1D surface gap solitons at a nonlinearity interface (i.e. the interface created by a jump of the nonlinearity coefficient in a system where perfectly periodic linear lattice is imprinted) were constructed by Dohnal and Pelinovsky (2008), and by Blank and Dohnal (2009). The linear stability of such surface gap solitons was studied with the aid of the Evans-function method. The results show the existence of both unstable and stable surface GSs. In this system even some solitons centered in the domain with the weaker focusing nonlinearity may be stable.

## B. Two-dimensional models

The basic properties of strongly asymmetric 2D matter-wave solitons that form at the interface produced by regions with different interatomic interaction strengths in a pancake-shaped BECs were studied by Ye, Kartashov, and Torner (2006). They considered several types of solitons featuring topologically complex structures, including vortex and dipole solitons placed into an external parabolic potential, where the nonlinearity strength changes in one direction in a step-like fashion, so that in one half of the space the repulsive interatomic interactions are weaker than in other half, or the nonlinearity even switches from repulsion to attraction. The confinement of the condensate in this case is achieved due to the external parabolic potential, while the presence of the nonlinearity interface causes severe distortions of vortex and multipole solitons bifurcating from the corresponding eigenmodes of the parabolic potential. While in the absence of the nonlinear interface vortex-soliton profiles are axially symmetric, and the vortex core is located at the center of the parabolic potential, in the case of the inhomogeneous nonlinearity the core of the vortex shifts into the region of weaker repulsive interactions, and the soliton's amplitude in this region substantially increases, which results in a strong asymmetry of the vortex' shape. The asymmetry becomes more pronounced with the increase of the soliton's norm. Notice that a similar effect of the soliton's ejection into the region with weaker interactions was reported by Perez-García and Pardo (2009), and by Zezyulin *et al.* (2007). For fixed chemical potential $\mu$ and radial-confinement frequency $\Omega$, the asymmetric vortex solitons can exist only when the strength of interatomic interactions exceeds a certain critical value. Despite the strong shape asymmetry, such vortex solitons are stable also in the strongly nonlinear regime, when the norm exceeds a certain critical value. Dipole solitons were found in this setting too. They feature asymmetric shapes and curved nodal lines due to the presence of the nonlinearity interface. Such solitons are stable in the region adjacent to the cutoff, while the maximal norm of the stable dipole increases with the increase of the nonlinearity step at the interface, i.e., the interface acts as a stabilizer for such solitons.

Two-dimensional solitons were also investigated at the edge of a truncated lattice with out-of-phase modulations of the nonlinearity and refractive index (Kartashov *et al.*, 2008). In such a setting, which is governed by equation $iq_z = -(1/2)\Delta q - [1 - \sigma R(x,y)]|q|^2 q - p R(x,y) q$, where $R(x,y)$ describes the profile of the truncated periodic structure, surface solitons may form around one of the edge waveguides. At low amplitudes, they are broad and expand into the lattice region, but with the increase of the norm the light localizes in the surface channel. Nevertheless, in contrast to the medium with the uniform nonlinearity, the further growth of the amplitude results in a faster increase of the nonlinear contribution to the refractive index in the space between maxima of the linear lattice. Since nonlinear effects dominate when the soliton's amplitude is high, the large-amplitude soliton shifts into the region between the first and second waveguide rows. This effect induced solely by the surface stems from the modulation of the nonlinearity only in the half-space, giving rise to a preferable direction of the soliton's shift. The competition between the linear refraction and self-action results in a nontrivial $N(\mu)$ dependence, that predicts the existence of minimal, $N_{\min}$, and maximal, $N_{\max}$, soliton norms [Fig. 31(a)]. The corresponding Hamiltonian-versus-norm dependence exhibits two cuspidal points, which is typical for solitons in 2D lattices with the nonlinearity modulation [Fig. 31(b)]. Stability domains of the surface solitons are shown in the planes of $(\sigma, N)$ and $(p, N)$, in Figs. 31(c) and 31(d), respectively. One can see that the stability domain may completely vanish when the depth of the nonlinearity modulation, $\sigma$, exceeds a certain critical value, or when the depth of the linear lattice, $p$, becomes too small. Interesting dynamics may be observed when the surface soliton is excited by a Gaussian beam. If the input power is too low, the beam diffracts and almost all light diffracts into the bulk of the lattice. At intermediate values of $N$, one achieves an effective excitation of a surface soliton, when almost all the input power remains trapped in a vicinity of the launch channel. If the power is too high, the input beam drifts into the space between the first and second lattice rows, where it collapses. This suggests a possibility of engineering an all-optical limiter incorporating the interface of the lattice with the spatially modulated nonlinearity.

Notice that very recently a rigorous proof of existence of fundamental surface gap solitons at $n$-dimensional nonlinear interfaces with periodic variations of nonlinearity and refractive index at both sides of the interface was given by Dohnal, Plum, and Reichel (2010) using variational methods.

## VII. Solitons in lattices with quadratic nonlinearities

The theoretical results outlined in the above sections were obtained in NL models with the cubic or saturable nonlinearity. Theoretical predictions were also made for the existence and stability of optical spatial solitons in models of 1D photonic crystals with the quadratic, alias $\chi^{(2)}$ (i.e., second-harmonic-generating), nonlinearity [reviews of solitons in uniform $\chi^{(2)}$ media were presented by Etrich *et al.* (2000) and Buryak, Di Trapani, Skryabin, and Trillo (2002)].

### A. Discrete models

The simplest model of the 1D photonic crystal with the $\chi^{(2)}$ nonlinearity was introduced by Sukhorukov, Kivshar, Bang, and Soukoulis (2000), in the form of an array of infinitely narrow quadratically-nonlinear stripes embedded into a host linear medium where the light propagation is described by the following coupled-mode equations:

$$
\begin{aligned}
&i\frac{\partial u}{\partial z} + \frac{\partial^2 u}{\partial x^2} + V_1(x)u + \sum_n \delta(x - hn)(\beta_1 u + u^* v) = 0, \\
&i\frac{\partial v}{\partial z} + \frac{1}{2}\frac{\partial^2 v}{\partial x^2} + V_2(x)v + \sum_n \delta(x - hn)(\beta_2 v + u^2) = 0.
\end{aligned}
\tag{55}
$$

Here $u(x,z)$ and $v(x,z)$ are amplitudes of the fundamental-frequency (FF) wave and its second harmonic (SH), $V_{1,2}(x)$ account for the linear lattice potential that may exist in the medium, $\delta(x)$ is the delta-function, $h$ is the spacing of the grating formed by the narrow stripes of the $\chi^{(2)}$ nonlinearity, while the respective $\chi^{(2)}$ coefficient is scaled to be 1, and coefficients $\beta_1, \beta_2$ account for the possibility that the comb of the delta-functions generates an additional linear potential. Similar to the model with the array of infinitely narrow Kerr-nonlinear stripes embedded into the linear host medium [Sukhorukov and Kivshar (2002a and 2002b)], Eqs. (55) can be explicitly integrated in the linear segments, which makes the remaining equations for stationary modes equivalent to those for a *discrete* $\chi^{(2)}$ system. After rescaling, the latter equations take the following form:

$$\begin{aligned} \eta_1 U_n + U_{n+1} + U_{n-1} + \chi_1 U_n^* V_n &= 0, \\ \eta_2 V_n + V_{n+1} + V_{n-1} + \chi_2 U_n^2 &= 0, \end{aligned} \tag{56}$$

with the propagation constants for the FH and SH, $k$ and $2k$, hidden in coefficients $\eta_{1,2}$. It is relevant to mention that the full (nonstationary) model of the discrete $\chi^{(2)}$ lattice was introduced by Peschel, Peschel, and Lederer (1998), and by Darmanyan, Kobyakov, and Lederer (1998), who had also found basic types of discrete solitons in that model. In particular, the discrete $\chi^{(2)}$ solitons, as well as their counterparts in the DNLSE with the cubic nonlinearity, may be classified into *staggered* and *unstaggered* ones, in the cases when, respectively, the discrete field features alternating signs at adjacent sites of the lattice, or keeps the same sign. For the two-component discrete model based on Eqs. (56), a situation is also possible when the FH discrete field is staggered, while its SH counterpart is not. Figure 32 displays generic examples of basic types of fundamental on-site discrete solitons generated by Eqs. (56). This figure also includes predictions for the shapes of the solitons produced by the VA based on the simplest ansatz relevant to the description of discrete solitons (Malomed and Weinstein, 1996), viz., $\{U_n, V_n\} = \{A a^{-|n|}, B b^{-|n|}\}$, where $A, B$ and $a, b$ are real constants (coefficients $a$ and/or $b$ corresponding to the staggered components are negative). In the model assuming a single infinitely narrow $\chi^{(2)}$ layer embedded into the linear medium, soliton solutions can be found in an exact form, as shown by Sukhorukov, Kivshar, and Bang (1999).

Another form of a (semi-)discrete system with the $\chi^{(2)}$ nonlinearity was proposed by Panoiu, Osgood, and Malomed (2006). They considered a complex waveguide built in the form of a slab substrate, with an array of guiding ribs mounted on top of it, both parts being made of a $\chi^{(2)}$ material [cf. a similar system with the Kerr nonlinearity described by Eqs. (44)]. Selecting the guiding characteristics of this setting, one may design the system in which the slab and array support, respectively, the transmission of the SH and FF waves only (*system A*), or vice versa (*system B*). The corresponding models are based on the following systems of coupled-mode equations:

$$i\frac{d\phi_n}{dz} + \rho(\phi_{n-1} + \phi_{n+1}) + \phi_n^*\Psi(\eta = n) = 0,$$
$$i\frac{\partial \Psi}{\partial z} + \frac{1}{2}\frac{\partial^2 \Psi}{\partial \eta^2} + \beta\Psi + \frac{1}{2}\sum_n \phi_n^2 \delta(\eta - n) = 0,$$
(57)

for *system A*, and

$$i\frac{\partial \Phi}{\partial z} + \frac{1}{2}\frac{\partial^2 \Phi}{\partial \eta^2} + \Phi^* \sum_n \psi_n \delta(\eta - n) = 0,$$
$$i\frac{d\psi_n}{dz} + \rho(\psi_{n-1} + \psi_{n+1}) + \beta\psi_n + \frac{1}{2}\Phi^2(\eta = n) = 0,$$
(58)

for *system B*, where the small and capital letters, $\phi_n / \Phi(\eta)$ and $\psi_n / \Psi(\eta)$, stand, respectively, for the discrete and continuous amplitudes of the FF and SH components, $\eta$ is the transverse coordinate, $\rho$ is the effective coefficient of the coupling between discrete waveguides in the array via evanescent fields, and $\beta$ determines the phase mismatch between the FF and SH waves. Systems (57) and (58) conserve the respective norms, $P_A = \sum_n |\phi_n|^2 + \int_{-\infty}^{\infty} |\Psi(\eta)|^2 d\eta$ and $P_B = \int_{-\infty}^{\infty} |\Phi(\eta)|^2 d\eta + \sum_n |\psi_n|^2$.

Stationary *semi-discrete* solutions to Eqs. (57) and (58) were found in the form of $\{\phi_n, \psi_n, \Phi, \Psi\} = \{u_n, v_n, U(\eta), V(\eta)\}\exp(ikz)$. Typical examples of the on-site- and inter-site-centered (alias odd and even) solitons found in systems of both A and B types are displayed in Fig. 33. Numerical results demonstrate that, in the A-type system the families of both odd and even semi-discrete solitons have no existence threshold in terms of the total power, and both families are *completely stable*, as predicted by the VK criterion and verified in direct simulations. On the other hand, in the B-type system there are power thresholds necessary for the existence of solitons of both types, while branches of the odd and even solitons are stable at values of $k$ exceeding those corresponding to the threshold points (also in full agreement with the prediction of the VK criterion). The stability of the inter-site-centered semi-discrete solitons in these two (semi-)discrete systems is a significant finding, as their counterparts in the fully discrete $\chi^{(2)}$ system are unstable.

## B. The continuous model

A model of a $\chi^{(2)}$ photonic crystal in a fully continuous form was recently elaborated by Pasquazi and Assanto (2009). The propagation of light in such a crystal is described by the following system of coupled-mode equations for the FF and SH components:

$$i\frac{\partial u}{\partial z} = \frac{1}{2}\frac{\partial^2 u}{\partial x^2} - u + 2\cos(\gamma x)u^*v,$$
$$i\frac{\partial v}{\partial z} = \frac{1}{4}\frac{\partial^2 v}{\partial x^2} - \alpha v + 2\cos(\gamma x)u^2, \qquad (59)$$

where $\alpha$ determines the FF-SH phase mismatch, and $2\pi/\gamma$ is the period of the spatially periodic modulation of the nonlinearity coefficient [$\chi^{(2)}$ *grating*], with the zero mean value. In stationary soliton solutions generated by this model, the SH component is more sensitive to the presence of the nonlinear grating than its FF counterpart, see an example in Fig. 34. The stability of the full family of the soliton solutions was investigated via the computations of the corresponding eigenvalues for small perturbations. The result was that the solitons tend to become stable with the increase of $\alpha$ and decrease of $\gamma$ in Eqs. (59).

Notice that spatial solitons emerging due to twin-beam second-harmonic generation in hexagonal lattices of purely nonlinear origin (i.e. those created by modulation of only $\chi^{(2)}$ susceptibility) created by poling lithium niobate planar waveguides were recently observed experimentally by Gallo et al (2008). It was demonstrated that such solitons can be steered by acting on power, direction, and wavelength of the fundamental frequency input.

## VIII. Three-dimensional solitons

In this section we address 3D optical solitons (alias *light bullets*, LBs), in settings with two spatial and one temporal dimensions in purely nonlinear or mixed linear-nonlinear lattices. LBs are spatiotemporal solitons that form when a suitable nonlinearity may be in balance with both spatial diffraction and temporal GVD (see the seminal paper by Silberberg, 1990, and the review by Malomed, Mihalache, Wise, and Torner, 2005). Stable 3D matter-wave solitons were also predicted in BEC in the presence of attractive and repulsive interactions in suitable linear trapping potentials (Baizakov, Malomed, and Salerno, 2003; Yang and Musslimani, 2003; Mihalache *et al.*, 2005).

In principle, LBs may be supported by a variety of nonlinear mechanisms, but their experimental realization usually faces two cardinal challenges, namely, the identification of a type of the nonlinearity which is capable to support *stable* LBs, and realization of a physical setting where the appropriate nonlinearity, diffraction, and dispersion are all present with suitable strengths, without producing conspicuous losses. Different approaches were suggested to resolve these problems. Below we briefly discuss theoretical predictions for the formation of stable LBs in systems with inhomogeneous nonlinearities. In particular, we consider the "bullets" predicted to exist in radial tandem structures consisting of alternating rings made of highly dispersive but weakly nonlinear, and strongly nonlinear but weakly dispersive materials, and "bullets" forming in mixed Bessel OLs with out-of-phase modulations of the refractive index and nonlinearity.

The formation of LBs in radial tandem structures was studied by Torner and Kartashov (2009). Such radial tandems represent engineered structures composed of different materials featuring either strong saturable nonlinearity or strong GVD, but not necessarily both present at a given wavelength. Thus, each material is used at its best to obtain high average values of the dispersion and nonlinearity in this composite structure, which is designed to guide the transmission of relatively broad modes covering several rings of the tandem. The evolution of wavepackets in the structure obeys the NLSE:

$$i\frac{\partial q}{\partial z} = -\frac{1}{2}\left(\frac{\partial^2 q}{\partial x^2} + \frac{\partial^2 q}{\partial y^2}\right) + \frac{\beta(x,y)}{2}\frac{\partial^2 q}{\partial t^2} + \sigma(x,y)\frac{q|q|^2}{1+S|q|^2}, \quad (60)$$

where it is assumed that a radially-symmetric structure is composed of periodically alternating rings of width $d$ exhibiting anomalous dispersion and zero nonlinearity, with $\beta = -2$, $\sigma = 0$, and weakly dispersive, but highly nonlinear domains, with $\beta = -0.1$, $\sigma = -1$. Two types of geometries were considered – those with the central domain exhibiting the nonlinearity, or vice versa. Linear propagation patterns in such structures indicate that, for large domain widths $d$, the local diffraction resembles that in uniform media (depending on whether the central domain is strongly or weakly dispersive, the wavepacket expands more in time or in space). However, when the domain's width is sufficiently small, the beam experiences the action of the average dispersion of the structure, and the expansion becomes comparable in space and time. The addition of the nonlinearity, which may compensate the diffraction and GVD, results in the formation of LBs. Since solutions approach those in uniform media with the average dispersion and nonlinearity in the limit of $d \to 0$, the saturation of the nonlinearity is necessary to avoid the collapse that occurs in the Kerr media. For suitable parameters, "bullets" may cover several rings, featuring pronounced shape modulations. They expand substantially at low and high amplitudes, the latter being a consequence of the nonlinearity saturation. The total energy of the LB is a non-monotonous function of propagation constant $\mu$, see Fig. 35(a), so that in the NL defined by the radial tandems LBs always exist above a threshold value of the energy, that diminishes with the increase of the domain's width, $d$ [see Fig. 35(b)]. One can see from this plot how the difference between the corresponding $N(\mu)$ curves diminishes as the domain's width, $d$, becomes smaller, which confirms the expectation that the transmission of light in the structure with sufficiently small domains mimics the transmission in uniform media with the respective average parameters. The stability analysis shows that LBs in the tandems with the linear central core have much wider stability domains than in the tandems with the nonlinear core, cf. Figs. 35(c) and 35(d). In the latter case, the bullets may be prone to the azimuthal modulational instability since they develop ring-like spatial intensity distributions, although this instability may be suppressed by the further decrease of the radial width of the domain.

Light bullets in Bessel OLs with an out-of-phase modulations of the linear refractive index and nonlinearity were studied by Ye *et al.* (2009). The evolution of such states is described by the NLSE in the form of $iq_z = -(1/2)(q_{xx} + q_{yy} - \beta q_{tt}) - (1 - \sigma R)|q|^2 q - pRq$, which represents an extension of model (14) to the 3D setting with an axially symmetric mixed lattice, where $R(r) = J_0[(2b_{\text{lin}})^{1/2} r]$ and $r \equiv (x^2 + y^2)^{1/2}$. Low-amplitude solitons in such a lattice behave similarly to their counterparts in the linear lattice, i.e., they strongly expand in both space and time. The increase of the soliton's amplitude leads to the concentration of light near the central guiding core. In this regime, effects of the linear and nonlinear lattices are comparable, hence the soliton's maximum is located at $r = 0$. Further growth of the soliton's amplitude results in a transformation of the spatial shape of the LBs due to the inhomogeneous nonlinearity landscape: The solitons in this regime develop ring-like spatial profiles, an effect which is most pronounced around the peak of the pulse, while the temporal distribution does not change qualitatively, remaining bell-shaped. This shape transformation is the cause of the non-monotonous dependencies featured by the respective $N(\mu)$ curves ($N$ first grows with $\mu$, but then diminishes at $\mu \to \infty$), and leads to the loss of the stability at high values of the amplitude and for large nonlinearity-modulation depths, when azimuthal perturbations become even more destructive than radial ones, leading to the destabilization of solitons on the branch with $dN/d\mu < 0$. It was found that the width of the stability domain for the LBs, in terms of $\mu$, first expands with the increase of the nonlinearity modulation depth, $\sigma$, but then starts decreasing. In contrast, the largest possible energy of stable bullets is a monotonously increasing function of $\sigma$. Increasing the depth of the linear lattice typically causes an expansion of the stability domain in terms of $\mu$, and a reduction of the largest possible energy of stable LBs.

## IX. Concluding remarks

The first aim of this review is to provide a coherent survey of the remarkable progress that has been made in theoretical studies of solitons and other nonlinear-wave patterns supported by effective periodic potentials induced by NLs (nonlinear lattices), as well as by combinations of linear and nonlinear lattices. Most of these results have been reported for one-dimensional geometries, but a considerable number of results are available in two-dimensional settings too, and some – even in three dimensions. The analysis of the accumulated results makes it possible to draw general conclusions concerning the core properties of solitons in these systems.

In the 1D case, a property which makes solitons drastically different from their well-studied counterparts in uniform media, and in media equipped with linear-lattice potentials, is the existence of the finite *threshold value* of the soliton norm (total power), below which the solitons do not exist. In 2D, a challenging issue is the identification of stability conditions for the solitons in purely nonlinear lattices. It has been found that a crucial condition affecting the stability of 2D solitons is the *sharpness* of the nonlinearity-modulation functions supporting the solitons. Another generic property of the solitons in the settings that involve competing linear and nonlinear lattices is their enhanced *mobility* and *power-dependent location* of soliton peaks, as well as the strong dependence of the intrinsic structure of the solitons on the peak power. Generic scenarios of the creation of such solitons are completely different too from what was known for their counterparts created in linear lattices. In particular, the solitons supported by the periodically modulated nonlinearity *do not* emerge by bifurcating from Bloch bands.

In spite of the great progress made in the theoretical studies there remain many problems awaiting further development and analysis. Many of such problems are suggested by the possibilities to extend results that have been established in 1D settings into 2D geometries. These include, in particular, the spontaneous symmetry breaking in symmetric double- or four-well nonlinear potentials in 2D, search for stable soliton complexes and vortices in the 2D nonlinear lattices, the study of the soliton mobility and collisions in such media, effects of the commensurability and incommensurability in mixed linear-nonlinear lattices (also in 2D), the soliton formation in random and quasi-periodic nonlinear landscapes, stabilization of 2D multicomponent (vectorial) soliton states, etc. Some relevant problems are awaiting the analysis in the 1D setting too – for instance, the spontaneous symmetry breaking of a two-component mixture in the nonlinear double-well potential.

Although many suggestions about potential physical realizations of the theoretical findings for the solitons in nonlinear lattices have been put forward in optics, nanophotonics, and matter waves in BEC, most predictions are still awaiting experimental implementation. Thus far, experimental observations that are relevant to nonlinear lattices have been reported only in photorefractive crystals with photoinduced lattices, and in photonic-crystal fibers filled with an index-matching liquid. No specific experimental studies of nonlinear lattices have been reported, as yet, in the realm of BEC, as well as in nanophotonics systems, such as nanowire arrays. It is expected that the theoretical predictions that may be most plausible for the experimental realization are those involving 1D settings. These include the creation of solitons and their bound states above the predicted existence threshold, the demonstration of their mobility and collisions, the realization of the predicted spontaneous symmetry breaking in nonlinear double-well potentials, etc. An essentially more challenging problem for the experimental implementation is the creation of 2D solitons that may be supported by nonlinear lattices. All in all, one concludes that a whole field is awaiting experimental exploration.

## X. Acknowledgements

We thank all our colleagues and collaborators for fruitful interactions over the past few years, in conducting various research projects whose results are reported, *inter alia*, here, and for their help in completing this review. We are especially indebted to members of our groups, co-authors of joint papers which were used in this review, and colleagues who helped us with valuable advices: F. Abdullaev, G. Assanto, B. Baizakov, S. Carrasco, R. Carretero-González, L.-C. Crasovan, K. Chow, D. Christodoulides, J. Cuevas, A. Desyatnikov, R. Driben, N. Efremidis, A. Ferrando, G. Fibich, D. Frantzeskakis, A. Gubeskys, L. Hadzievski, K. Hizanidis, P. Kevrekidis, D. Kip, Y. Kivshar, Y. Kominis, V. Konotop, W. Krolikowski, F. Lederer, A. Maluckov, T. Mayteevarunyoo, H. Michinel, D. Mihalache, D. Mazilu, D. Neshev, V. Perez-Garcia, N. Panoiu, T. Pertsch, H. Sakaguchi, M. Salerno, M. Segev, D. Skryabin, G. I. Stegeman, A. Sukhorukov, A. Szameit, J. P. Torres, M. Trippenbach, V. Vysloukh, M. I. Weinstein, Z. Xu, and F. Ye.

This work was supported, in a part, by the Generalitat de Catalunya and by the Government of Spain through the Ramon-y-Cajal program, and also benefited from the generous support provided by the Fundació Cellex Barcelona and the Fundacio Mir-Puig. B.A.M. appreciates hospitality of ICFO during the preparation of this review.

**Figure captions**

Figure 1. (a) The norm of the stationary soliton solution of Eq. (17) versus its amplitude at $r_0 = 0$, in the fundamental model of the 1D nonlinear lattice. Rhombuses connected by the continuous line are values found from the direct numerical solution, while the dashed curve shows the prediction of the variational approximation. (b) The chemical potential versus the number of atoms for numerically found solitons [from Sakaguchi and Malomed (2005a)].

Figure 2. (Color online) Spatially localized solitons of the NLSE with inhomogeneous nonlinearity coefficient $R(x) = r_0 / \cosh^3(x)$, obtained by mapping of periodic solutions of Eq. (19), $W(X) = C \operatorname{sn}(\mu X, k) / \operatorname{dn}(\mu X, k)$, with specifically selected values of modulus $k$, in the model devised to produce exact soliton solutions [from Belmonte-Beitia, Perez-Garcia, and Vekslerchik, 2007].

Figure 3. (Color online) Profiles of (a) fundamental, (b) even, (c) dipole-mode, and (d) triple-mode solitons residing at the center of layered thermal sample and described by Eq. (23). (e) Profiles of fundamental solitons shifted from the center of the sample. (f) The distribution of the thermally-induced perturbation of the refractive index, $\sigma T$, for a fundamental soliton with $\mu = 0.9$ residing at the center of the sample (red curve), and for a shifted fundamental soliton with $\mu = 6.5$ (black curve). In shaded regions $\sigma > 0$, while in white regions $\sigma < 0$ [from Kartashov, Vysloukh, and Torner (2008a)].

Figure 4. (Color online) The number of bosons (norm) versus energy (chemical potential) $\mathcal{E}$ in the first gap $\mathcal{E}_1^{(+)} < \mathcal{E} < \mathcal{E}_2^{(-)}$ in the dynamical model of the $^{87}\text{Rb} - ^{40}\text{K}$ (boson-fermion) mixture. Panels (a) and (b) correspond to different signs of average nonlinearity coefficient $\chi$ near the $\mathcal{E}_1^{(+)}$ edge of the gap. Panels (c), (d), and (e) show explicit shapes of the modes corresponding to different points on the solution branches, while panel (f) shows the dynamics of mode G (in the insets the initial and final shapes are shown by solid lines) [from Bludov and Konotop, 2006].

Figure 5. (Color online) The phase-space construction of asymptotic (solitary) solutions for odd $n$ (a) and even $n$ (b), in the model with the nonlinear lattice of the Kronig-Penney type. Black dots depict the transition at the boundary be-

| | |
|---|---|
| | tween linear and a nonlinear layers. Examples of solitons corresponding to $n=2$ (c) and 3 (d). Shaded areas indicate nonlinear layers [from Kominis, 2006]. |
| Figure 6. | (Color online) The critical angle versus propagation constant $\mu$ for (a) odd and (b) even solitons in lattices with out-of-phase modulation of linear refractive index and nonlinearity. (c) The propagation dynamics of odd solitons with $\mu=9.8$, launched into the lattice at two different angles. Distributions of the absolute value of the field corresponding to different input angles are superimposed. In all cases, $\sigma=0.4$ [from Kartashov, Vysloukh and Torner (2008b)]. |
| Figure 7. | (Color online) (a) A stable three-peak soliton in the second bandgap at $\mu=0.4$, in the model of the 1D photonic-crystal waveguide of the Kronig-Penney type. (b) A weakly unstable nearly flat-top soliton in the second bandgap corresponding to $\mu=0$. (c) An unstable counterpart of the three-peak soliton with the inverted shape, corresponding to $\mu=-0.22$. In all the cases, the nonlinearity is defocusing and $D/L=0.5$. Left panels show the soliton shape, while right panels show the spectral plane of stability eigenvalues for the respective solitons, $\lambda=\lambda_r+i\lambda_i$ (the soliton is stable if $\lambda_i=0$) [from Mayteevarunyoo and Malomed (2008)]. |
| Figure 8. | Top panels: effective potential $V(x_0)$ as a function of the soliton's central coordinate, $x_0$, with the trap's strength $\Omega=0.05$, for a fundamental bright soliton of the unit amplitude, initially placed at the trap's center $[x_0(0)=0]$, in the model of the collisionally inhomogeneous BEC. Different values of the gradient modify the character of the potential: in (a) it is purely attractive, while in (b) it is either purely gravitational $\left(\delta=\sqrt{3}\Omega\right)$ or expulsive $(\delta=2\Omega)$. Bottom panels: the evolution of the center of the bright soliton for (c) an attractive effective potential, and (d) for gravitational or expulsive potentials [from Theocharis *et al.*, (2005)]. |
| Figure 9. | (a) Three lowest branches of soliton solutions with $R_{\pm}=\pm 1$, in the model of the nonlinearity management for BEC. Shown is the number of particles $N$ versus chemical potential $\mu$. The bolder regions of curves $\Gamma^n$ correspond to stable solutions, while the lighter ones correspond to unstable ones. Shaded |

are regions of the multistability. The explicit shapes of stable modes close to the linear limit are shown in the insets. (b) The same as in (a) but for $R_+ = 5$, $R_- = -1$ [from Zezyulin et al., (2007)].

Figure 10. (a) and (c): Chemical potential $\mu$ versus norm $N$ at different values of the commensurability factor, $q$, for ordinary and gap solitons, respectively, in the BEC model including mutually commensurate or incommensurate linear and nonlinear lattices. (b) and (d): The stability boundary for ordinary and gap solitons, defined, as per the VK criterion, by condition $d\mu/dN = 0$. Note that the threshold value of the norm necessary for the existence of the solitons, of both the ordinary and gap types, vanishes at two points, $q = 0$ and $q = 2$, which correspond, respectively, to the models with the constant nonlinearity coefficient, and with the direct commensurability between the linear and nonlinear lattices [from Sakaguchi and Malomed (2010)].

Figure 11. (a) An example of the evolution of a perturbed cn-type wave near the edge of its stability area at $g_0 = -1$, $b = -0.7$, $k = 0.95$, in the model of the nonlinear lattice admitting exact periodic solutions in terms of the elliptic functions. (b) An example of stable evolution of a perturbed dn-type wave with $g_0 = +1$, $g_1 = -2$, $b = r = 1$, and $k = 0.9$ [from Tsang, Malomed and Chow (2009)].

Figure 12. (Color online) The profile of an even-dipole vector soliton at $\mu_1 = 2.67$, $\mu_2 = 3$, $C = 1$, in the model of the two-component system with the nonlinear lattice. (b) Energy sharing between components of the even-dipole vector solitons versus $\mu_1$ at $\mu_2 = 3$, $C = 1$. Domains of stability (white) and instability (shaded) in the $(\mu_1, \mu_2)$ plane at $C = 1$ (c), and in the $(C, \mu_1)$ plane at $\mu_2 = 3$ (d) [from Kartashov, Malomed, Vysloukh, and Torner, (2009b)].

Figure 13. A set of bifurcation diagrams describing the symmetry breaking of pinned modes in the one-dimensional double-well pseudopotential at different values of the wells' width $a$ ($a = 0$ corresponds to the limit form of the model with the delta-functions) [from Mayteevarunyoo, Malomed and Dong (2008)].

Figure 14. (Color online) (a),(b): Intensity profiles for typical examples of spatial sub-wavelength solitons of the TE and TM types (black and red curves, respec-

tively), in the model of the array of nanowires. Shaded and white areas correspond, severally, to the strips made of silicon and silica. Panels (a) and (b) display solitons of the on-site and off-site types. (c),(d): The total power versus the nonlinear shift of the propagation constant of the subwavelength solitons, whose examples, corresponding to points A and B, are shown in panels (a),(b). Panels (c) and (d) display families of the TE and TM solitons, respectively. Black (red) curves correspond to the on-site (off-site) solitons, while solid (dashed) curves designate stable (unstable) soliton families. [from Gorbach and Skryabin (2009)].

Figure 15. The dashed-dotted curves show examples of stable exact solutions obtained for trapped solitons within the framework of the model of the nonlinear defect, based on one-dimensional equation (39) with $f(z) = \delta(z)$. The solid curves show counterparts of these states, generated by the numerical solution of the underlying 3D Gross-Pitaevskii equation, for the same values of parameters, $\varepsilon$ and $n_0$ [from Abdullaev, Gammal, and Tomio (2004)].

Figure 16. Typical examples of stationary discrete solutions corresponding to $W_2 = 0$, in the model of the discrete linear-nonlinear lattice based on Eq. (41): (a) bright solitons; (b) kinks (dark solitons); (c) states sitting on top of a finite background ("anti-dark solitons"). Modes (a) and (b) may be stable, while those of type (c) are always unstable [from Abdullaev *et al.*, (2008)].

Figure 17. Stationary fundamental (a),(b) and dipole (c),(d) modes supported by the interface between discrete lattices with the attractive and repulsive on-site cubic nonlinearity. Modes shown in (a),(b) correspond to $C = 0.1$, while modes shown in (c),(d) correspond to $C = 0.135$ [from Machacek *et al.*, (2006)].

Figure 18. (Color online) Examples of stable staggered discrete solitons in the lattice with interlaced focusing nonlinear and linear sites. These solitons were found in the finite bandgap. (a),(b) The evolution of stable soliton and antisymmetric bound state of two solitons, with random initial perturbations added to them. (c),(d) Profiles of the respective stationary solutions [from Hizanidis, Kominis, and Efremidis (2008)].

Figure 19. (Color online) Shapes of typical odd, even, and twisted two-component solitons in the semi-discrete model based on Eqs. (44). The values of the XPM coupling constant are indicated in panels (a), (b), and (c). Equal propagation constants are taken here for the discrete and continuous components, $\lambda_1 = \lambda_2 \equiv \lambda$ [from Panoiu, Malomed, and Osgood (2008)].

Figure 20. (Color online) The release of soliton trains in the model of the matter-wave laser. Panels (a)-(f) pertain to the following values of the negative scattering length in Eq. (45): $R_0 = 0.9 R_{\mathrm{cr}}$, $R_0 = 2.0 R_{\mathrm{cr}}$, $R_0 = 3.3 R_{\mathrm{cr}}$, $R_0 = 4.8 R_{\mathrm{cr}}$, $R_0 = 6.5 R_{\mathrm{cr}}$, and $R_0 = 8.7 R_{\mathrm{cr}}$. Recall that the equilibrium position for the soliton exists at $|R_0| < |R_{\mathrm{cr}}|$ [from Rodas-Verde, Michinel, and Pérez-García (2005)].

Figure 21. (Color online) (a) The domain (dashed area) in the plane of the strength of the linear lattice, $V$, and amplitude of the nonlinearity modulation, $G$, where the condition for the stability of the *Bloch oscillations* of the gap soliton in the framework of Eq. (47) is satisfied. In this case, the constant part of the nonlinearity coefficient is $g = -0.777$. Also shown are examples of stable (c) and unstable (d) Bloch oscillations of the gap solitons under the action of the constant driving force. Parameters in panels (b) and (c) correspond, respectively, to points B and A in (a) [from Salerno, Konotop, and Bludov (2008)].

Figure 22. (a) Examples of stable 2D soliton solutions supported by the circular modulation of the local nonlinearity, with $\rho_0 = 0$, $\mu = -0.0399$ (solid line) and $\rho_0 = 0.5$, $\mu = -0.0648$ (dashed line). (b) The chemical potential versus the norm for soliton families found at $\rho_0 = 0$, 0.2, and 0.5. (c) The stability diagram for the soliton solutions. In the regions between the two borders, the solitons are stable simultaneously according to the VK criterion (i.e., against radial perturbations) and against azimuthal modulations [from Sakaguchi and Malomed (2006a)].

Figure 23. (a) The norm (total power) of 2D solitons versus propagation constant $\mu$ in the model of the 2D nonlinear lattice, built as an array of self-focusing circles, for several values of $w_{\mathrm{s}}$ in the medium with the cubic nonlinearity. (b) The minimum norm versus the lattice spacing, $w_{\mathrm{s}}$. The horizontal dashed lines in (a) and (b) correspond to the critical norm, $N_T = 5.85$. (c) The real part of

Figure 24. (Color online) (a) The geometry of the liquid-infiltrated PCF. (b) The intensity distribution in a numerically calculated gap soliton for power $P = 4 \times 10^{-5}$ and $s = 10$. (c),(d): Experimentally observed output diffraction pattern and soliton localization in the PCF at low (3 mW) and high (100 mW) input powers, respectively [from Rasmussen *et al.*, (2009)].

Figure 25. (Color online) (a) Intensity distributions for fundamental solitons created in the liquid-infiltrated PCF. The panels correspond to (from left to right) $\gamma \to 0$, $\gamma = 0.0010$, $\gamma = 0.0015$, and $\gamma = 0.0017$, in the PCF with *pitch* (the spacing between parallel voids) $\Lambda = 23 \, \mu\text{m}$, radius $a = 4 \, \mu\text{m}$, and $\lambda = 1.55 \, \mu\text{m}$. (b) Nodal solitons with different orientation of nodal lines in the PCF with $\Lambda = 23 \, \mu\text{m}$, $a = 8 \, \mu\text{m}$, and $\lambda = 1.06 \, \mu\text{m}$ obtained for $\gamma = 0.006$. The first two panels show distributions of the absolute value of the field, while the last two panels show the corresponding phase patterns [from Ferrando et al., (2003) and Ferrando et al., (2005a)].

Figure 26. The lowest-order eigenfunctions of nonlinear operator $\mathcal{L}$ generated by the soliton solution in the fundamental representation of $\mathcal{C}_{6v}$, in the model of the PCF with the respective symmetry of the intrinsic structure. The symmetry of the full operator is $\mathcal{C}_{6v}$, i.e., $[\mathcal{L}, \mathcal{C}_{6v}] = 0$. Modes in the two middle rows correspond to vortices with charges $\pm 1$ and $\pm 2$ [from Ferrando, Zacarés, and Garcia-March, (2005)].

Figure 27. (Color online) Examples of different types of vortex-like solitons in the dual-core PCF: (a) a double-vortex state; (b) a combined state of vortex and fundamental solitons; (c) a double-triple vortex state; (d) a combined state of a triple vortex and fundamental soliton [from Salgueiro and Kivshar, (2009)].

Figure 28. (Color online) The existence domain for the vector solitons in the PCF model is shown in the plane of $(\mu_1, \mu_2)$, for two values of coupling constant $C$ [from Salgueiro et al., (2005)].

Figure 29.  (Color online) (a)-(c) The phase space for each part of the structure (in the one-dimensional model based on the Kronig-Penney lattice with the intrinsic surface), as per Kominis, Papadopoulos and Hizanidis, (2007): (a) the nonlinear part for $\mu > \varepsilon_1$; (b) the linear part for $\mu < \varepsilon_i$ ($i = 2$ or 3); (c) the linear part for $\mu > \varepsilon_i$ ($i = 2$ or 3). (d)-(f) The phase-space representation of the soliton solutions for $n$ even. (d) The nonlinear homogeneous part; (e) the linear homogeneous part at $\mu < \varepsilon_3$; (f) the linear homogeneous part at $\mu > \varepsilon_3$. The dotted line denotes the solution in the lattice part, and the solid line denotes the solution in the uniform part.

Figure 30.  (Color online) Profiles of (a) fundamental, (b) dipole-mode, and (c) tri-pole solitons with different values of $\mu$, in the model of the thermal layered medium with a surface. (d) Distributions of the refractive index for fundamental solitons with $\mu = 5$ (black curve) and $\mu = 1$ (red curve). In all the cases, the sign of the nonlinearity alternates between different layers of the thermal medium. In gray regions, $\sigma > 0$ holds [from Kartashov, Vysloukh, and Torner, (2009b)].

Figure 31.  (Color online) (a) The norm of the surface soliton residing at the interface of lattice with out-of-phase modulation of refractive index and nonlinearity versus the propagation constant; (b) the Hamiltonian versus the norm at $p = 3$, $\sigma = 0.6$. Black curves show stable soliton branches, while red curves correspond to unstable ones. Stability domains for surface solitons: (c) in the plane of $(\sigma, N)$ at $p = 3$; (d) in the plane of $(p, N)$ at $\sigma = 0.7$ [from Kartashov *et al.*, (2008)].

Figure 32.  Typical examples of stable on-site-centered (alias odd) discrete solitons generated by Eqs. (56), in the model of the lattice with the $\chi^{(2)}$ nonlinearity. Triangles and squares show numerically found profiles of the fundamental and second harmonics, respectively. The dashed and continuous curves represent the respective profiles as predicted by the variational approximation. (a) Solitons with staggered fundamental and unstaggered second-harmonic components; (b) both components staggered; (c) both unstaggered [from Sukhorukov *et al.*, (2000)].

Figure 33.  The top and bottom panels display typical examples of stable *semi-discrete* $\chi^{(2)}$ solitons generated by Eqs. (57) and (58), respectively, whereas the left and right panels show the odd and even species of the solitons. Vertical lines designate the location of the discrete waveguides in the system. These examples pertain to zero mismatch, [from Panoiu, Osgood and Malomed (2006)].

Figure 34.  A soliton generated by Eqs. (59) in the case of the large wavenumber of the nonlinear grating, $\gamma = 20$, and large mismatch, $\alpha = 10$, in the model of the photonic crystal with the $\chi^{(2)}$ nonlinearity. The grating is represented by the gray pattern in the background. The profile of the SH component follows the form of the nonlinear grating, while the FF component features a sech-type profile, with only small distortions induced by the grating, as additionally shown in the inset, where the grating is indicated by the dashed sinusoid [from Pasquazi and Assanto (2009)].

Figure 35.  The norm versus the propagation constant for (a) different values of $S$ at $d = 0.4$ and (b) different values of $d$ at $S = 0.5$, in the model of the radial tandem structure supporting "light bullets". In panel (b), values of the domain's width are $d = 0.8$, 0.6, 0.4, 0.2, and 0.1, from the lower to upper curve. (c) The perturbation growth rate versus $\mu$, for the azimuthal perturbation index $k = 0$ and $S = 0.5$. In panels (a)-(c), the central domain is linear. (d) The growth rate versus $\mu$ at $k = 1$, $d = 1.2$, $S = 0.5$ in the structure with linear (1) and nonlinear (2) central domains [from Torner and Kartashov (2009)].

Cuevas, J., B. A. Malomed, P. G. Kevrekidis, and D. J. Frantzeskakis, 2009, "Solitons in quasi-one-dimensional Bose-Einstein condensates with competing dipolar and local interactions", Phys. Rev. A **79**, 053608.

Darmanyan, S., A. Kobyakov, and F. Lederer, 1998, "Strongly localized modes in discrete systems with quadratic nonlinearity", Phys. Rev. E **57**, 2344.

Davies, E. B., 1979, "Symmetry breaking in a non-linear Schrödinger equation", Comm. Math. Phys. **64**, 191.

Davis, K. B., M.-O. Mewes, M. R. Andrews, N. J. van Druten, D. S. Durfee, D. M. Kurn, and W. Ketterle, 1995, "Bose-Einstein condensation in a gas of sodium atoms", Phys. Rev. Lett. **75**, 3969.

Davis, K., K. Miura, N. Sugimoto, and K. Hirao, 1996, "Writing waveguides in glass with a fs-laser," Opt. Lett. **21**, 1729.

De Nicola, S., B. A. Malomed, and R. Fedele, 2006, "An effective potential for one-dimensional matter-wave solitons in an axially inhomogeneous trap", Phys. Lett. A **360**, 164.

Desaix, M., D. Anderson, and M. Lisak, 1991, "Variational approach to collapse of optical pulses", J. Opt. Soc. Am. B **8**, 2082.

Dohnal, T. and D. Pelinovsky, 2008, "Surface gap solitons at a nonlinearity interface," SIAM J. Appl. Dyn. Syst. **7**, 249.

Dohnal, T., M. Plum, and W. Reichel, 2010, "Surface gap soliton ground states for the nonlinear Schrödinger equation," arXiv:1011.2886v1.

Dong, G., B. Hu, and W. Lu, 2006, "Ground-state properties of a Bose-Einstein condensate tuned by a far-off-resonant optical field," Phys. Rev. A **74**, 063601.

Dong, L., and H. Li, 2010, "Surface solitons in nonlinear lattices", J. Opt. Soc. Am. B **27**, 1179.

Driben, R., and B. A. Malomed, 2008, "Stabilization of two-dimensional solitons and vortices against supercritical collapse by lattice potentials", Eur. Phys. J. D **50**, 317.

Driben, R., B. A. Malomed, A. Gubeskys, and J. Zyss, 2007, "Cubic-quintic solitons in the checkerboard potential", Phys. Rev. E **76**, 066604.

Dror, N., and B. A. Malomed, 2010, "Stable solitons pinned to a point-like nonlinearity in the presence of periodic potentials", to be published.

Dudley, J. M., G. Genty, and S. Coen, 2006, "Supercontinuum generation in photonic crystal fiber," Rev. Mod. Phys. **78**, 1135.
117

Guerin, W., J.-F. Riou, J. P. Gaebler, V. Josse, P. Bouyer, and A. Aspect, 2006, "Guided quasicontinuous atom laser", Phys. Rev. Lett. **97**, 200402.

Hang, C., and V. V. Konotop, 2010, "All-optical steering of light via spatial Bloch oscillations in a gas of three-level atoms", Phys. Rev. A **81**, 053849.

Hang, C., V. V. Konotop, and G. Huang, 2009, "Spatial solitons and instabilities of light beams in a three-level atomic medium with a standing-wave control field," Phys. Rev. A **79**, 033826.

Hao, R., R. Yang, L. Li, and G. Zhou, 2008, "Solutions for the propagation of light in nonlinear optical media with spatially inhomogeneous nonlinearities," Opt. Commun. **281**, 1256.

Hasegawa, A., and F. Tappert, 1973, "Transmission of stationary nonlinear optical pulses in dispersive dielectric fibers. I. Anomalous dispersion," Appl. Phys. Lett. **23**, 142.

Harrison, W. A., 1966, *Pseudopotentials in the Theory of Metals* (Benjamin, New York).

He, Y., D. Mihalache, and B. Hu, 2010, "Soliton drift, rebound, penetration, and trapping at the interface between media with uniform and spatially modulated nonlinearities," Opt. Lett. **35**, 1716.

Henderson, K., C. Ryu, C. MacCormick, and M. G. I Boshier, 2009, "Experimental demonstration of painting arbitrary and dynamic potentials for Bose–Einstein condensates", New J. Phys. **11**, 043030.

Herrmann, J., U. Griebner, N. Zhavoronkov, A. Husakou, D. Nickel, J. C. Knight, W. J. Wadsworth, P. S. J. Russell, and G. Korn, 2002, "Experimental evidence for supercontinuum generation by fission of higher-order solitons in photonic fibers", Phys. Rev. Lett. **88**, 173901.

Hizanidis, K., Y. Kominis, and N. K. Efremidis, 2008, "Interlaced linear-nonlinear optical waveguide arrays", Opt. Exp. **16**, 18296.

Hukriede, J., D. Runde, and D. Kip, 2003, "Fabrication and application of holographic Bragg gratings in lithium niobate channel waveguides," J. Phys. D: Appl. Phys. **36**, R1.

Hung, N. V., P. Zin, M. Trippenbach, and B. A. Malomed, 2010, "Two-dimensional solitons in media with the stripe-shaped nonlinearity modulation," Phys. Rev. E **82**, 046602.

Ilan, B., and M. I. Weinstein, 2010, "Band-edge solitons, nonlinear Schrödinger/Gross-Pitaevskii equations and effective media", arXiv:1002.1986v2 (to be published in SIAM Journal of Multiscale Modeling and Simulation).
121

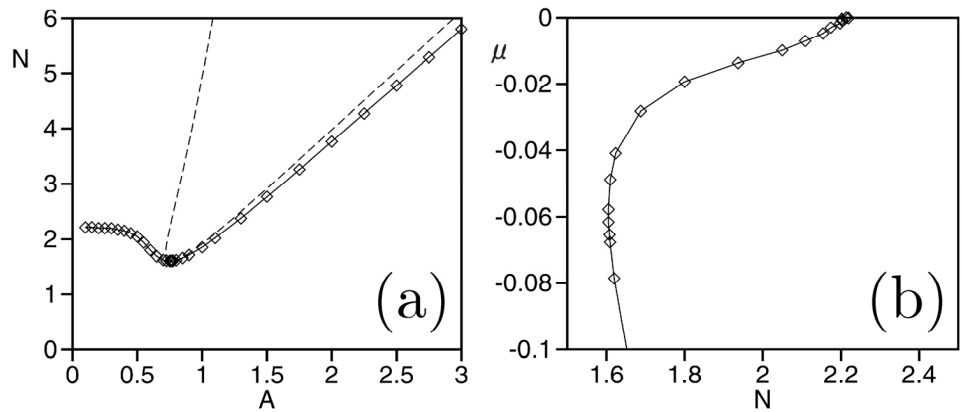

Figure 1.  (a) The norm of the stationary soliton solution of Eq. (17) versus its amplitude at $r_0 = 0$, in the fundamental model of the 1D nonlinear lattice. Rhombuses connected by the continuous line are values found from the direct numerical solution, while the dashed curve shows the prediction of the variational approximation. (b) The chemical potential versus the number of atoms for numerically found solitons [from Sakaguchi and Malomed (2005a)].

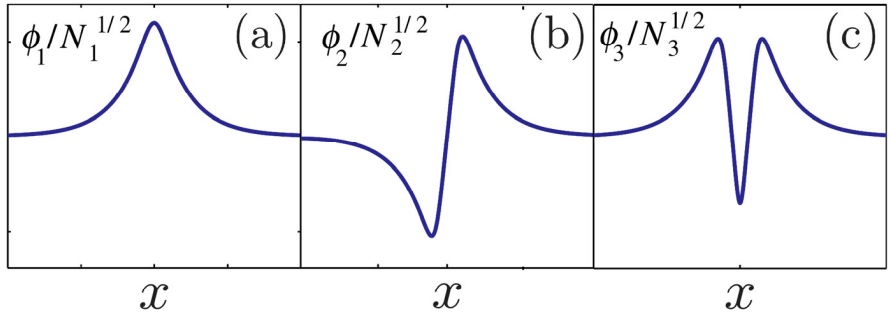

Figure 2. (Color online) Spatially localized solitons of the NLSE with inhomogeneous nonlinearity coefficient $R(x) = r_0 / \cosh^3(x)$, obtained by mapping of periodic solutions of Eq. (19), $W(X) = C \operatorname{sn}(\mu X, k) / \operatorname{dn}(\mu X, k)$, with specifically selected values of modulus $k$, in the model devised to produce exact soliton solutions [from Belmonte-Beitia, Perez-Garcia, and Vekslerchik, 2007].

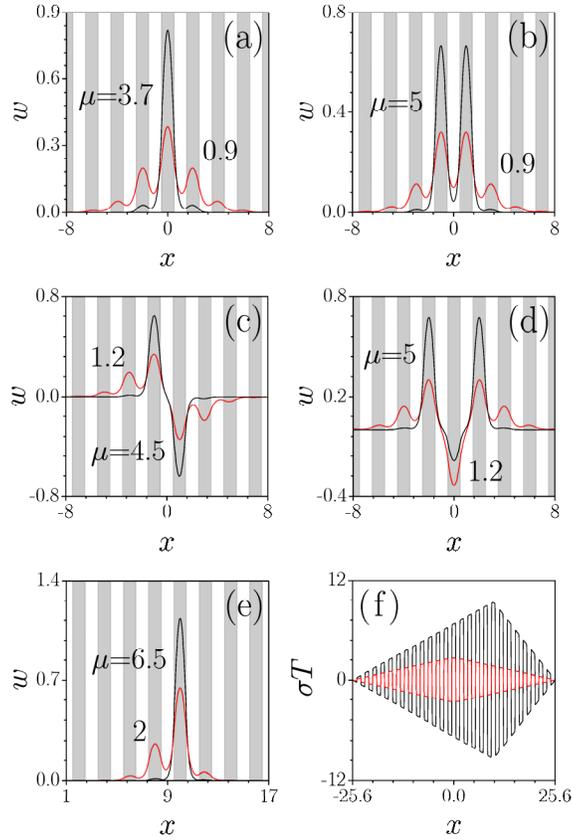

Figure 3. (Color online) Profiles of (a) fundamental, (b) even, (c) dipole-mode, and (d) triple-mode solitons residing at the center of layered thermal sample and described by Eq. (23). (e) Profiles of fundamental solitons shifted from the center of the sample. (f) The distribution of the thermally-induced perturbation of the refractive index, $\sigma T$, for a fundamental soliton with $\mu = 0.9$ residing at the center of the sample (red curve), and for a shifted fundamental soliton with $\mu = 6.5$ (black curve). In shaded regions $\sigma > 0$, while in white regions $\sigma < 0$ [from Kartashov, Vysloukh, and Torner (2008a)].

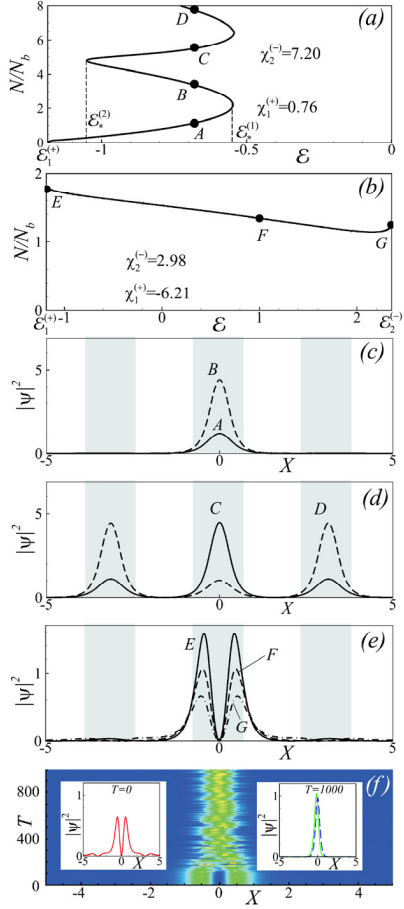

Figure 4. (Color online) The number of bosons (norm) versus energy (chemical potential) $\mathcal{E}$ in the first gap $\mathcal{E}_1^{(+)} < \mathcal{E} < \mathcal{E}_2^{(-)}$ in the dynamical model of the $^{87}\text{Rb}-^{40}\text{K}$ (boson-fermion) mixture. Panels (a) and (b) correspond to different signs of average nonlinearity coefficient $\chi$ near the $\mathcal{E}_1^{(+)}$ edge of the gap. Panels (c), (d), and (e) show explicit shapes of the modes corresponding to different points on the solution branches, while panel (f) shows the dynamics of mode G (in the insets the initial and final shapes are shown by solid lines) [from Bludov and Konotop, 2006].

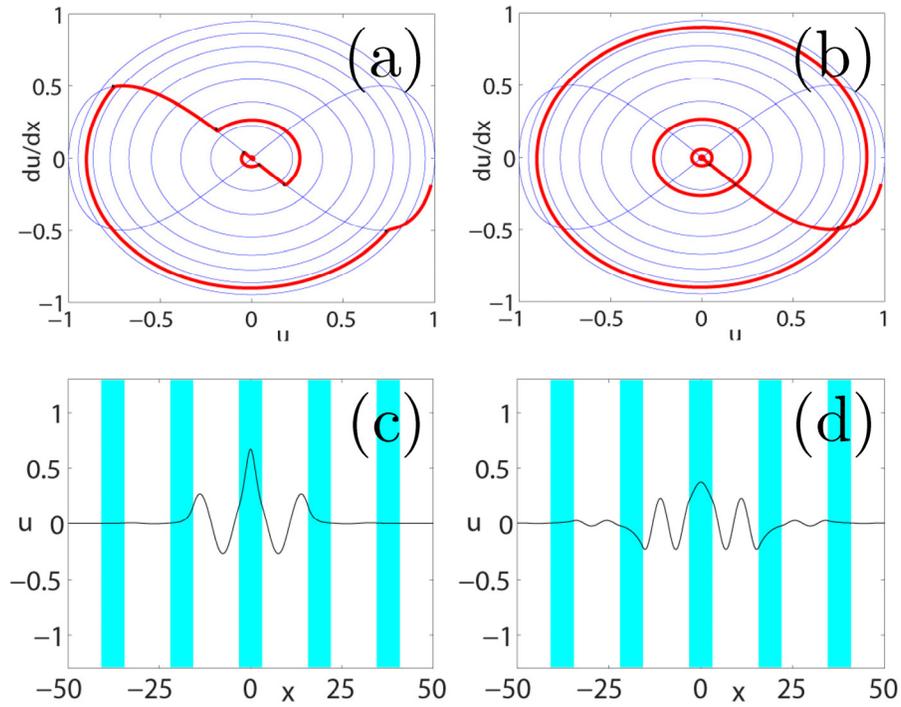

Figure 5. (Color online) The phase-space construction of asymptotic (solitary) solutions for odd $n$ (a) and even $n$ (b), in the model with the nonlinear lattice of the Kronig-Penney type. Black dots depict the transition at the boundary between linear and a nonlinear layers. Examples of solitons corresponding to $n = 2$ (c) and 3 (d). Shaded areas indicate nonlinear layers [from Kominis, 2006].

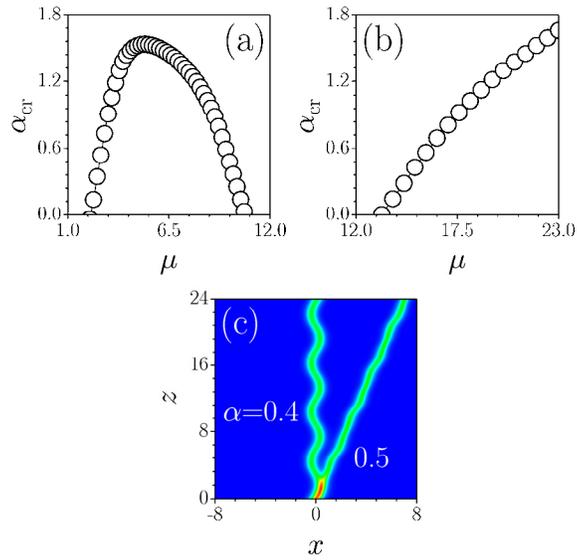

Figure 6. (Color online) The critical angle versus propagation constant $\mu$ for (a) odd and (b) even solitons in lattices with out-of-phase modulation of linear refractive index and nonlinearity. (c) The propagation dynamics of odd solitons with $\mu = 9.8$, launched into the lattice at two different angles. Distributions of the absolute value of the field corresponding to different input angles are superimposed. In all cases, $\sigma = 0.4$ [from Kartashov, Vysloukh and Torner (2008b)].

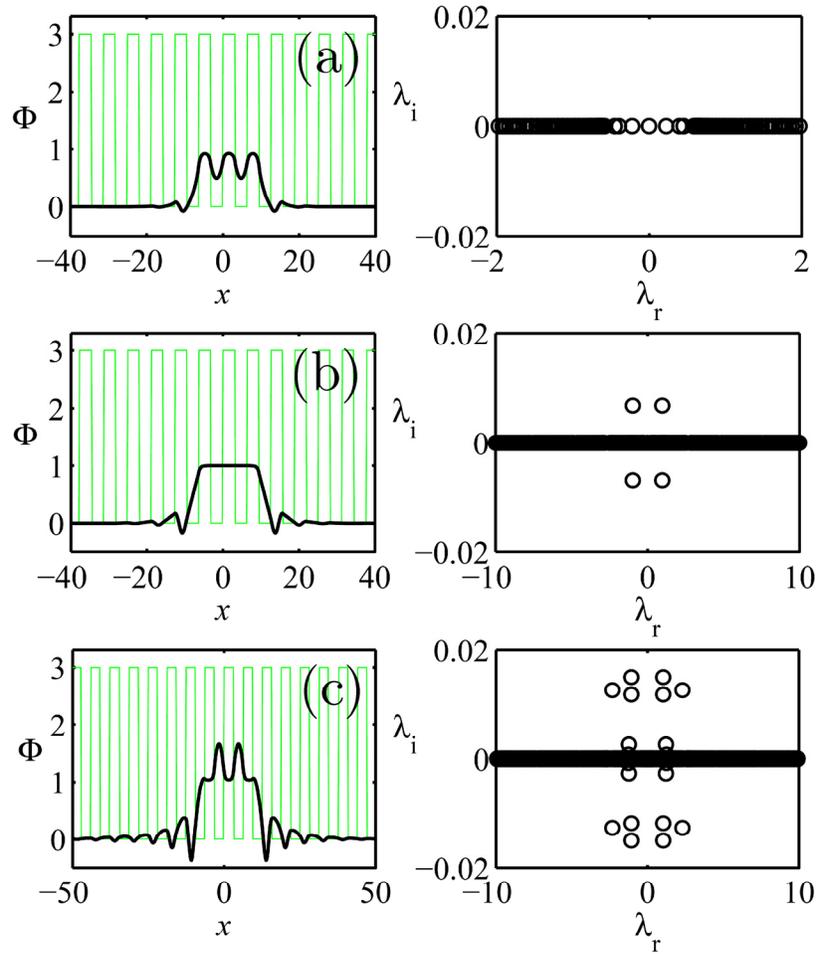

Figure 7. (Color online) (a) A stable three-peak soliton in the second bandgap at $\mu = 0.4$, in the model of the 1D photonic-crystal waveguide of the Kronig-Penney type. (b) A weakly unstable nearly flat-top soliton in the second bandgap corresponding to $\mu = 0$. (c) An unstable counterpart of the three-peak soliton with the inverted shape, corresponding to $\mu = -0.22$. In all the cases, the nonlinearity is defocusing and $D/L = 0.5$. Left panels show the soliton shape, while right panels show the spectral plane of stability eigenvalues for the respective solitons, $\lambda = \lambda_r + i\lambda_i$ (the soliton is stable if $\lambda_i = 0$) [from Mayteevarunyoo and Malomed (2008)].

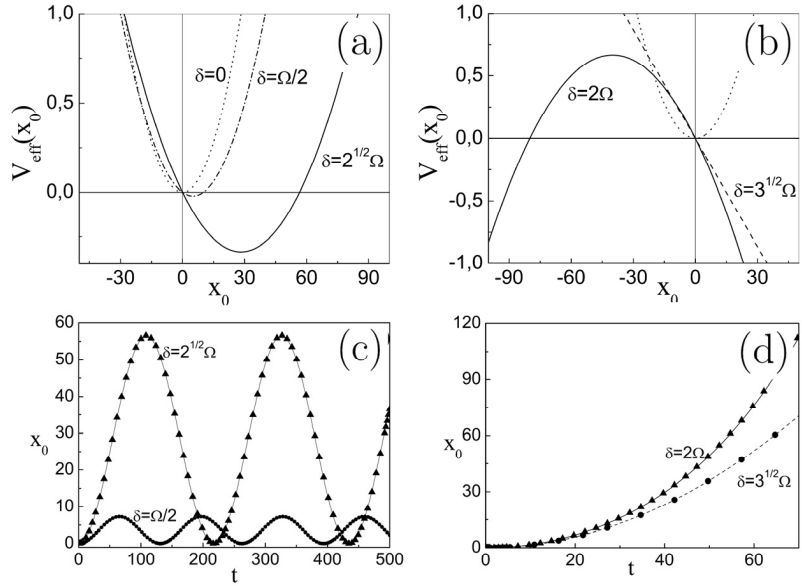

Figure 8. Top panels: effective potential $V(x_0)$ as a function of the soliton's central coordinate, $x_0$, with the trap's strength $\Omega = 0.05$, for a fundamental bright soliton of the unit amplitude, initially placed at the trap's center $[x_0(0) = 0]$, in the model of the collisionally inhomogeneous BEC. Different values of the gradient modify the character of the potential: in (a) it is purely attractive, while in (b) it is either purely gravitational ($\delta = 3^{1/2}\Omega$) or expulsive ($\delta = 2\Omega$). Bottom panels: the evolution of the center of the bright soliton for (c) an attractive effective potential, and (d) for gravitational or expulsive potentials [from Theocharis et al., (2005)].

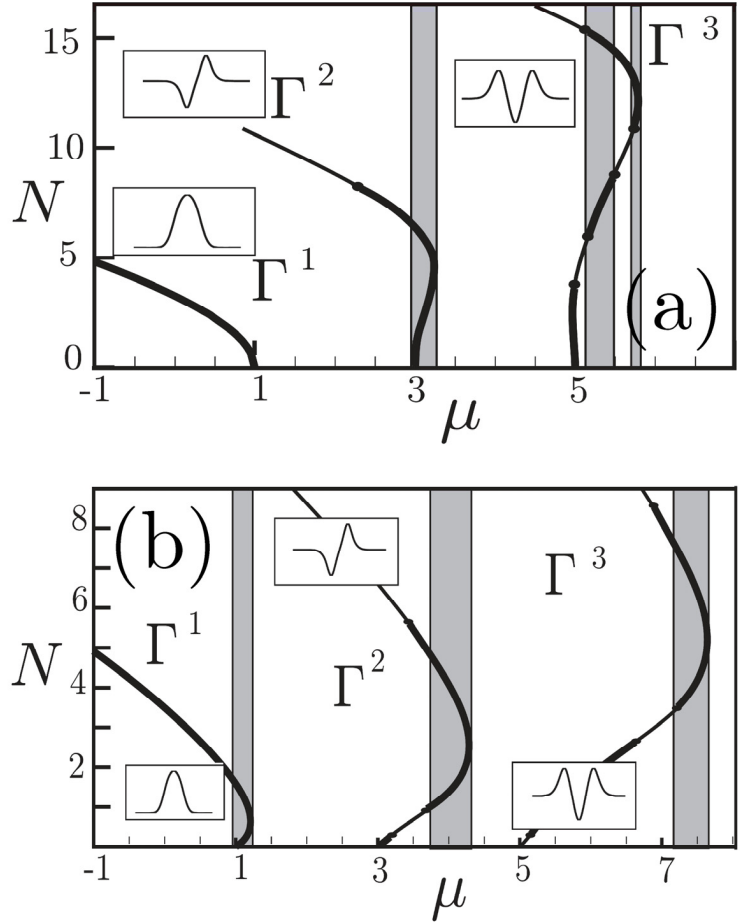

Figure 9. (a) Three lowest branches of soliton solutions with $R_{\pm} = \pm 1$, in the model of the nonlinearity management for BEC. Shown is the number of particles $N$ versus chemical potential $\mu$. The bolder regions of curves $\Gamma^n$ correspond to stable solutions, while the lighter ones correspond to unstable ones. Shaded are regions of the multistability. The explicit shapes of stable modes close to the linear limit are shown in the insets. (b) The same as in (a) but for $R_+ = 5$, $R_- = -1$ [from Zezyulin et al., (2007)].

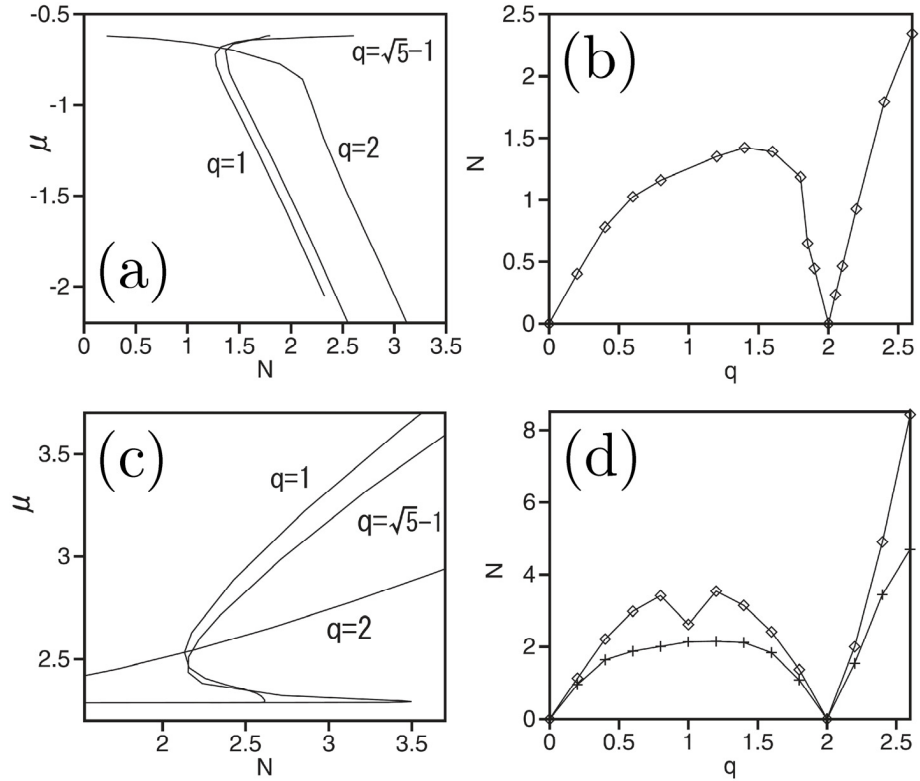

Figure 10. (a) and (c): Chemical potential $\mu$ versus norm $N$ at different values of the commensurability factor, $q$, for ordinary and gap solitons, respectively, in the BEC model including mutually commensurate or incommensurate linear and nonlinear lattices. (b) and (d): The stability boundary for ordinary and gap solitons, defined, as per the VK criterion, by condition $d\mu/dN = 0$. Note that the threshold value of the norm necessary for the existence of the solitons, of both the ordinary and gap types, vanishes at two points, $q=0$ and $q=2$, which correspond, respectively, to the models with the constant nonlinearity coefficient, and with the direct commensurability between the linear and nonlinear lattices [from Sakaguchi and Malomed (2010)].

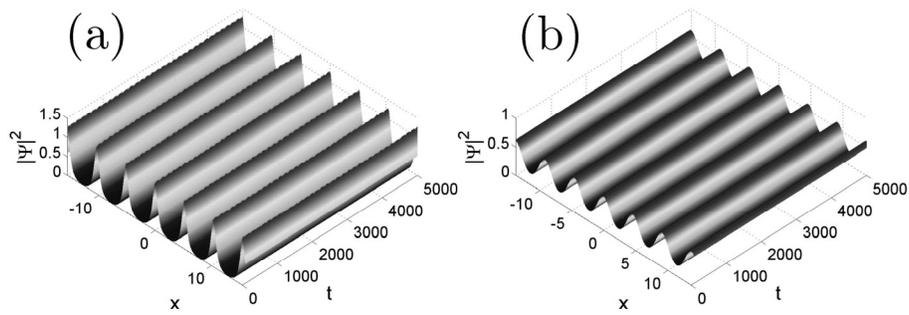

Figure 11.  (a) An example of the evolution of a perturbed cn-type wave near the edge of its stability area at $g_0 = -1$, $b = -0.7$, $k = 0.95$, in the model of the nonlinear lattice admitting exact periodic solutions in terms of the elliptic functions. (b) An example of stable evolution of a perturbed dn-type wave with $g_0 = +1$, $g_1 = -2$, $b = r = 1$, and $k = 0.9$ [from Tsang, Malomed and Chow (2009)].

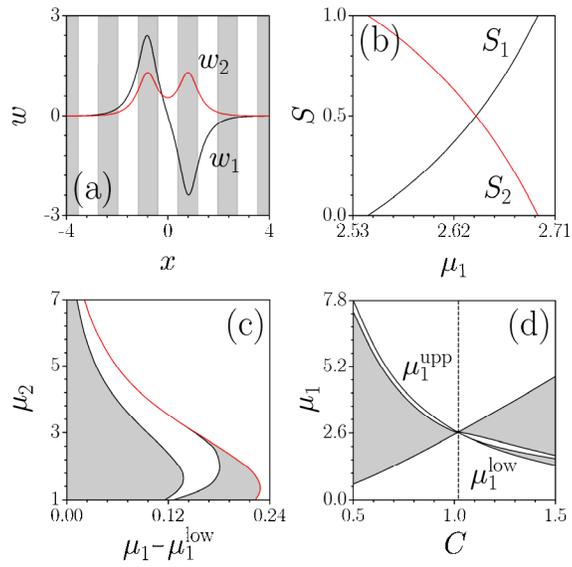

Figure 12. (Color online) The profile of an even-dipole vector soliton at $\mu_1 = 2.67$, $\mu_2 = 3$, $C = 1$, in the model of the two-component system with the nonlinear lattice. (b) Energy sharing between components of the even-dipole vector solitons versus $\mu_1$ at $\mu_2 = 3$, $C = 1$. Domains of stability (white) and instability (shaded) in the $(\mu_1, \mu_2)$ plane at $C = 1$ (c), and in the $(C, \mu_1)$ plane at $\mu_2 = 3$ (d) [from Kartashov, Malomed, Vysloukh, and Torner, (2009b)].

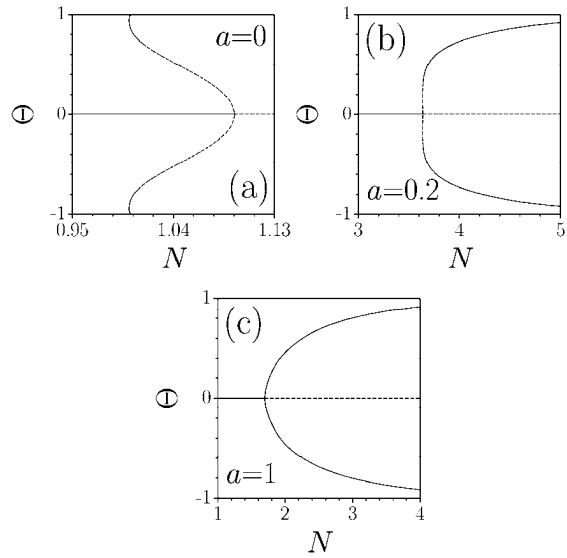

Figure 13.   A set of bifurcation diagrams describing the symmetry breaking of pinned modes in the one-dimensional double-well pseudopotential at different values of the wells' width $a$ ($a = 0$ corresponds to the limit form of the model with the delta-functions) [from Mayteevarunyoo, Malomed and Dong (2008)].

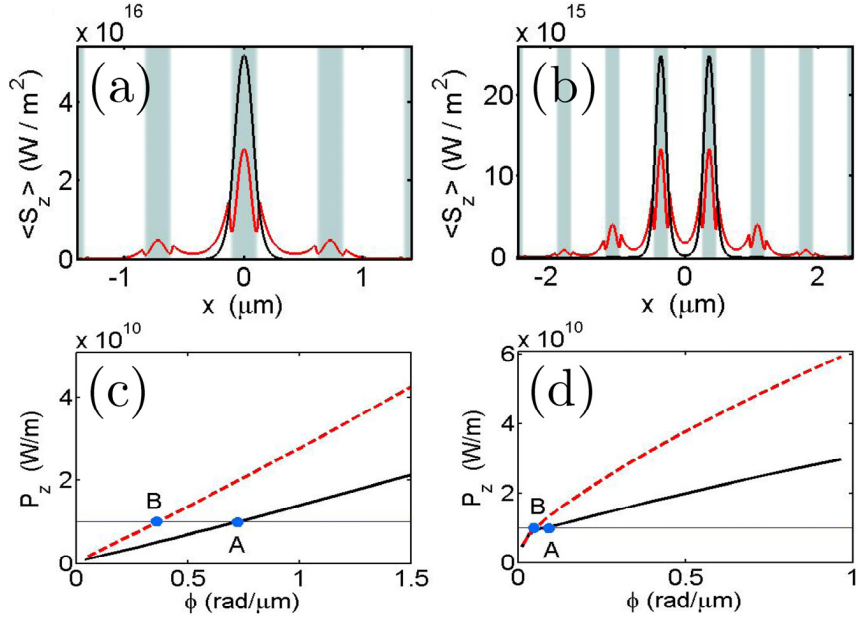

Figure 14. (Color online) (a),(b): Intensity profiles for typical examples of spatial subwavelength solitons of the TE and TM types (black and red curves, respectively), in the model of the array of nanowires. Shaded and white areas correspond, severally, to the strips made of silicon and silica. Panels (a) and (b) display solitons of the on-site and off-site types. (c),(d): The total power versus the nonlinear shift of the propagation constant of the subwavelength solitons, whose examples, corresponding to points A and B, are shown in panels (a),(b). Panels (c) and (d) display families of the TE and TM solitons, respectively. Black (red) curves correspond to the on-site (off-site) solitons, while solid (dashed) curves designate stable (unstable) soliton families. [from Gorbach and Skryabin (2009)].

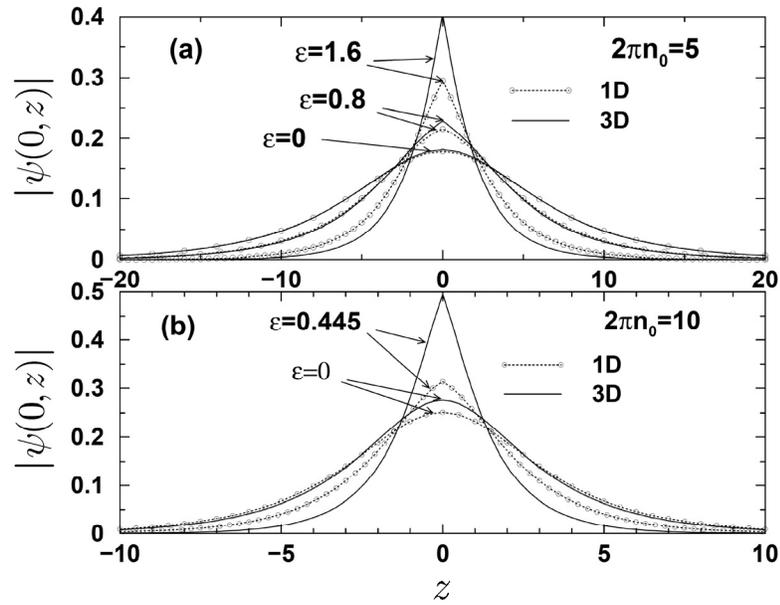

Figure 15. The dashed-dotted curves show examples of stable exact solutions obtained for trapped solitons within the framework of the model of the nonlinear defect, based on one-dimensional equation (39) with $f(z) = \delta(z)$. The solid curves show counterparts of these states, generated by the numerical solution of the underlying 3D Gross-Pitaevskii equation, for the same values of parameters, $\varepsilon$ and $n_0$ [from Abdullaev, Gammal, and Tomio (2004)].

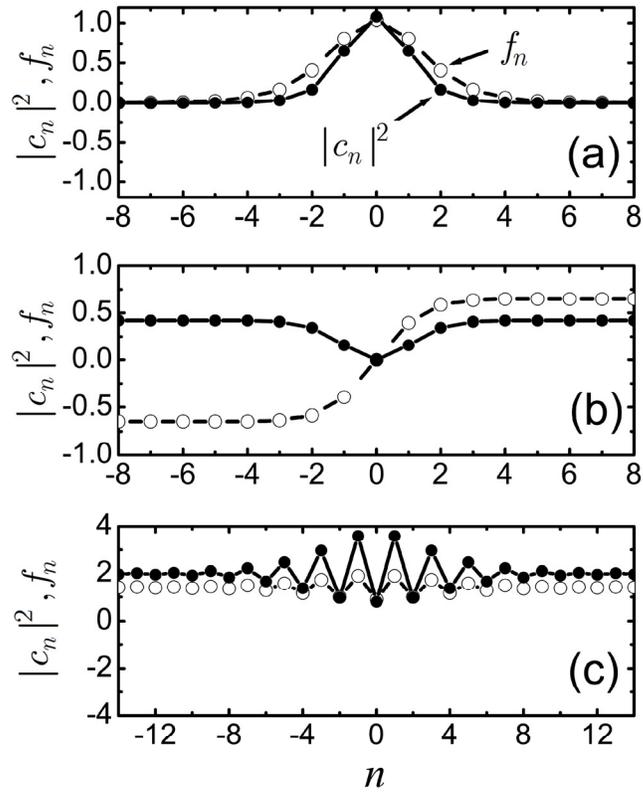

Figure 16. Typical examples of stationary discrete solutions corresponding to $W_2 = 0$, in the model of the discrete linear-nonlinear lattice based on Eq. (41): (a) bright solitons; (b) kinks (dark solitons); (c) states sitting on top of a finite background ("anti-dark solitons"). Modes (a) and (b) may be stable, while those of type (c) are always unstable [from Abdullaev et al., (2008)].

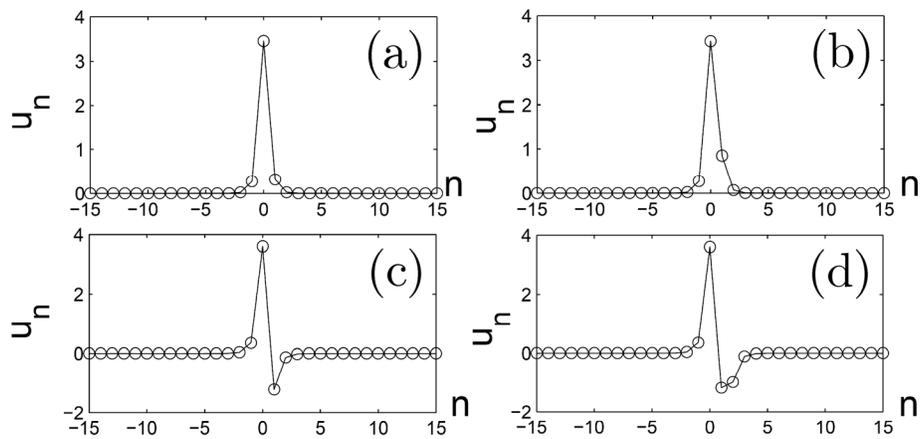

Figure 17.  Stationary fundamental (a),(b) and dipole (c),(d) modes supported by the interface between discrete lattices with the attractive and repulsive on-site cubic nonlinearity. Modes shown in (a),(b) correspond to $C = 0.1$, while modes shown in (c),(d) correspond to $C = 0.135$ [from Machacek *et al.*, (2006)].

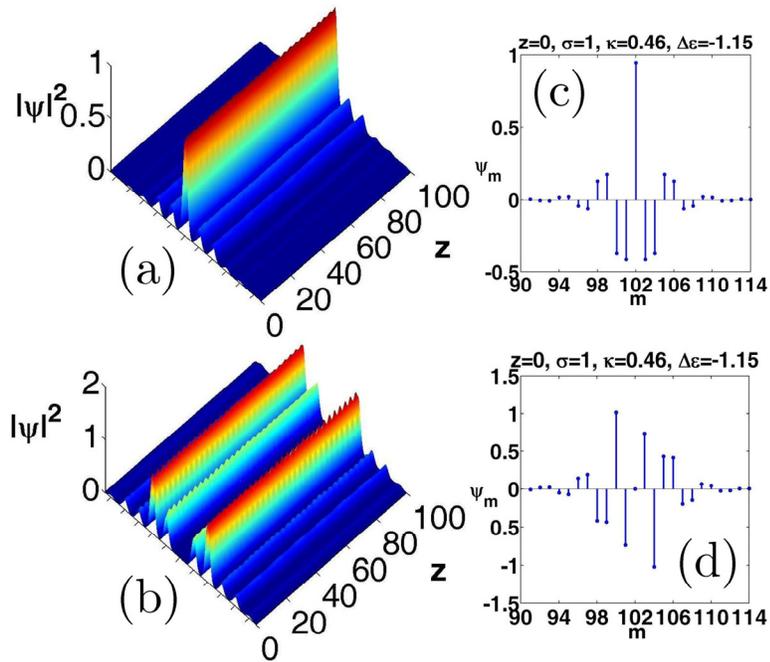

Figure 18. (Color online) Examples of stable staggered discrete solitons in the lattice with interlaced focusing nonlinear and linear sites. These solitons were found in the finite bandgap. (a),(b) The evolution of stable soliton and antisymmetric bound state of two solitons, with random initial perturbations added to them. (c),(d) Profiles of the respective stationary solutions [from Hizanidis, Kominis, and Efremidis (2008)].

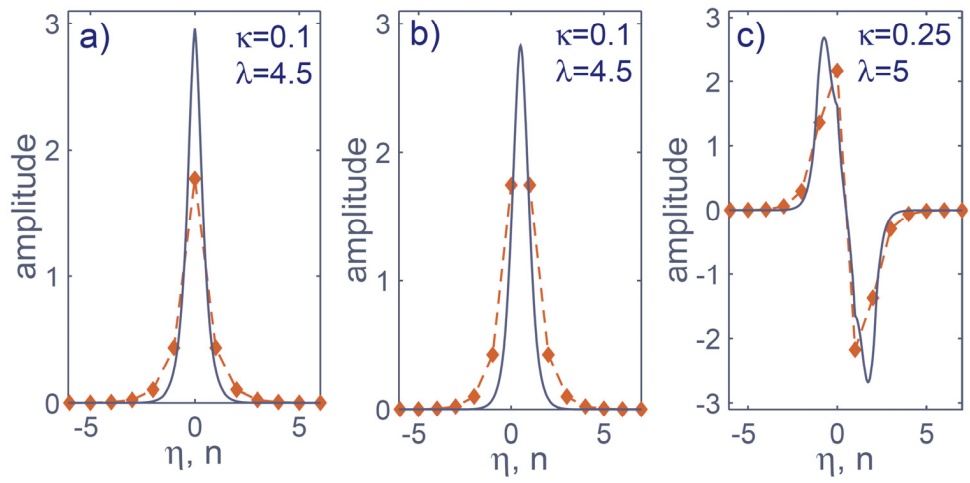

Figure 19. (Color online) Shapes of typical odd, even, and twisted two-component solitons in the semi-discrete model based on Eqs. (44). The values of the XPM coupling constant are indicated in panels (a), (b), and (c). Equal propagation constants are taken here for the discrete and continuous components, $\lambda_1 = \lambda_2 \equiv \lambda$ [from Panoiu, Malomed, and Osgood (2008)].

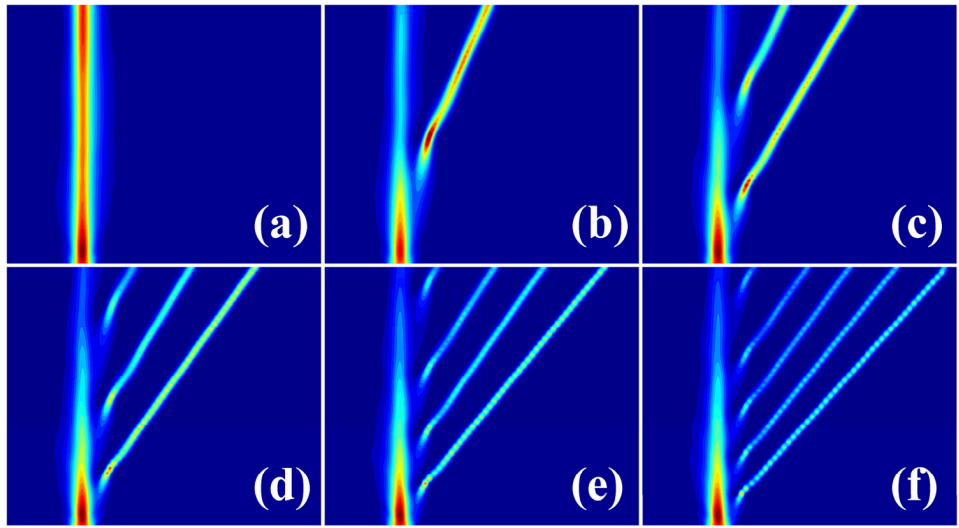

Figure 20. (Color online) The release of soliton trains in the model of the matter-wave laser. Panels (a)-(f) pertain to the following values of the negative scattering length in Eq. (45): $R_0 = 0.9R_{\mathrm{cr}}$, $R_0 = 2.0R_{\mathrm{cr}}$, $R_0 = 3.3R_{\mathrm{cr}}$, $R_0 = 4.8R_{\mathrm{cr}}$, $R_0 = 6.5R_{\mathrm{cr}}$, and $R_0 = 8.7R_{\mathrm{cr}}$. Recall that the equilibrium position for the soliton exists at $|R_0| < |R_{\mathrm{cr}}|$ [from Rodas-Verde, Michinel, and Pérez-García (2005)].

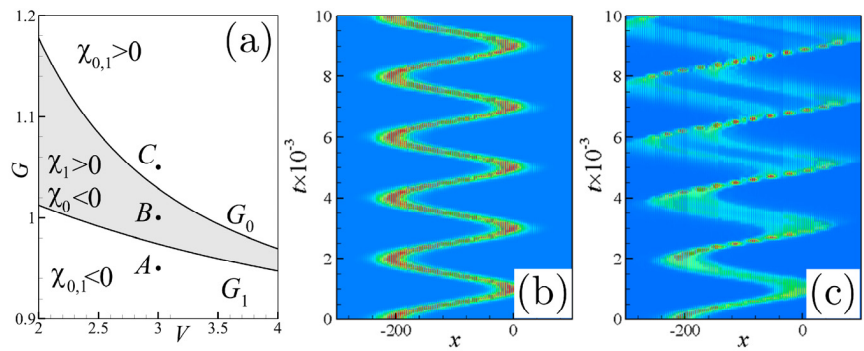

Figure 21. (Color online) (a) The domain (dashed area) in the plane of the strength of the linear lattice, $V$, and amplitude of the nonlinearity modulation, $G$, where the condition for the stability of the *Bloch oscillations* of the gap soliton in the framework of Eq. (47) is satisfied. In this case, the constant part of the nonlinearity coefficient is $g = -0.777$. Also shown are examples of stable (c) and unstable (d) Bloch oscillations of the gap solitons under the action of the constant driving force. Parameters in panels (b) and (c) correspond, respectively, to points B and A in (a) [from Salerno, Konotop, and Bludov (2008)].

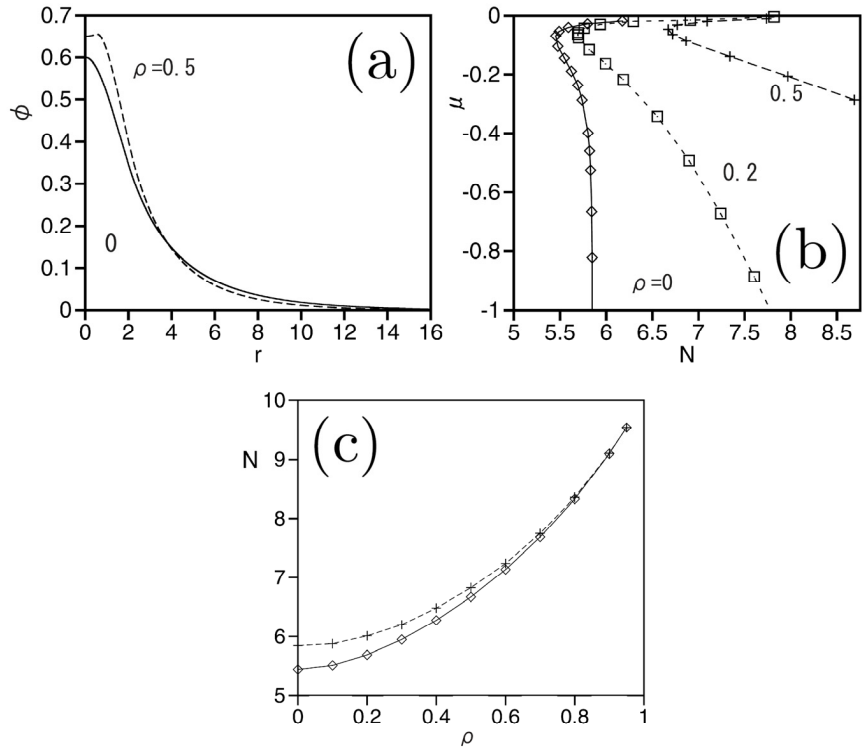

Figure 22. (a) Examples of stable 2D soliton solutions supported by the circular modulation of the local nonlinearity, with $\rho_0 = 0$, $\mu = -0.0399$ (solid line) and $\rho_0 = 0.5$, $\mu = -0.0648$ (dashed line). (b) The chemical potential versus the norm for soliton families found at $\rho_0 = 0$, $0.2$, and $0.5$. (c) The stability diagram for the soliton solutions. In the regions between the two borders, the solitons are stable simultaneously according to the VK criterion (i.e., against radial perturbations) and against azimuthal modulations [from Sakaguchi and Malomed (2006a)].

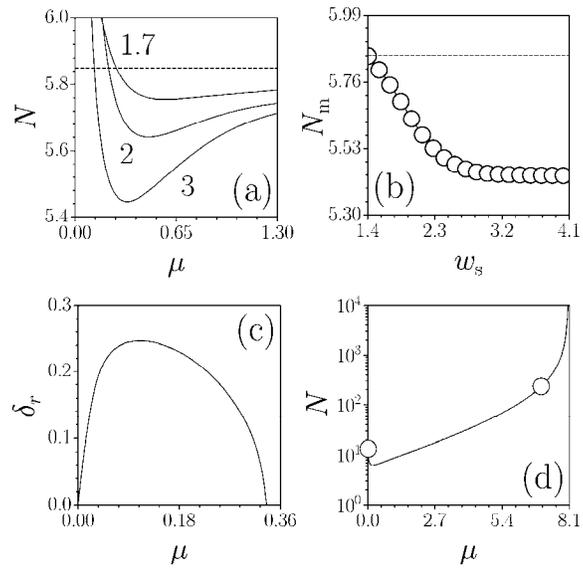

Figure 23. (a) The norm (total power) of 2D solitons versus propagation constant $\mu$ in the model of the 2D nonlinear lattice, built as an array of self-focusing circles, for several values of $w_s$ in the medium with the cubic nonlinearity. (b) The minimum norm versus the lattice spacing, $w_s$. The horizontal dashed lines in (a) and (b) correspond to the critical norm, $N_T = 5.85$. (c) The real part of the perturbation growth rate versus $\mu$ at $w_s = 3$. (d) The norm versus $\mu$ in the *saturable* medium [from Kartashov et al., (2009a)].

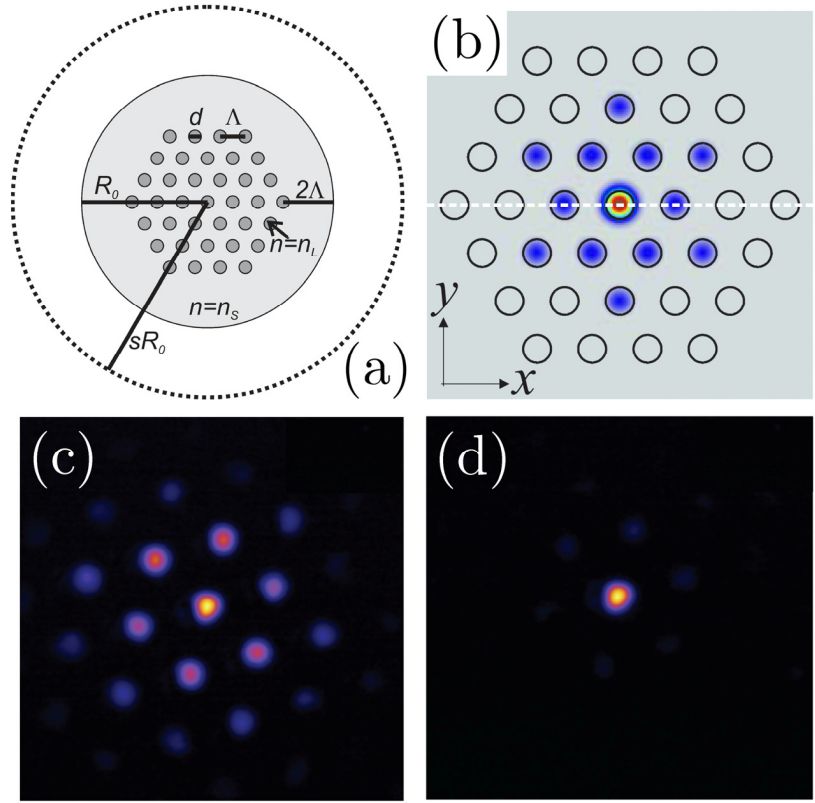

Figure 24. (Color online) (a) The geometry of the liquid-infiltrated PCF. (b) The intensity distribution in a numerically calculated gap soliton for power $P = 4\times10^{-5}$ and $s=10$. (c),(d): Experimentally observed output diffraction pattern and soliton localization in the PCF at low (3 mW) and high (100 mW) input powers, respectively [from Rasmussen *et al.*, (2009)].

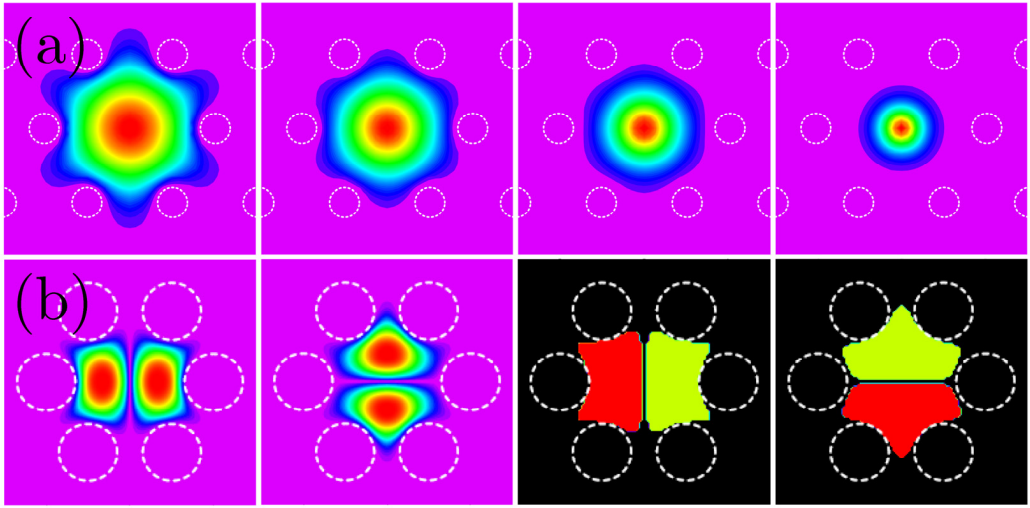

Figure 25. (Color online) (a) Intensity distributions for fundamental solitons created in the liquid-infiltrated PCF. The panels correspond to (from left to right) $\gamma \to 0$, $\gamma = 0.0010$, $\gamma = 0.0015$, and $\gamma = 0.0017$, in the PCF with *pitch* (the spacing between parallel voids) $\Lambda = 23\ \mu\mathrm{m}$, radius $a = 4\ \mu\mathrm{m}$, and $\lambda = 1.55\ \mu\mathrm{m}$. (b) Nodal solitons with different orientation of nodal lines in the PCF with $\Lambda = 23\ \mu\mathrm{m}$, $a = 8\ \mu\mathrm{m}$, and $\lambda = 1.06\ \mu\mathrm{m}$ obtained for $\gamma = 0.006$. The first two panels show distributions of the absolute value of the field, while the last two panels show the corresponding phase patterns [from Ferrando et al., (2003) and Ferrando et al., (2005a)].

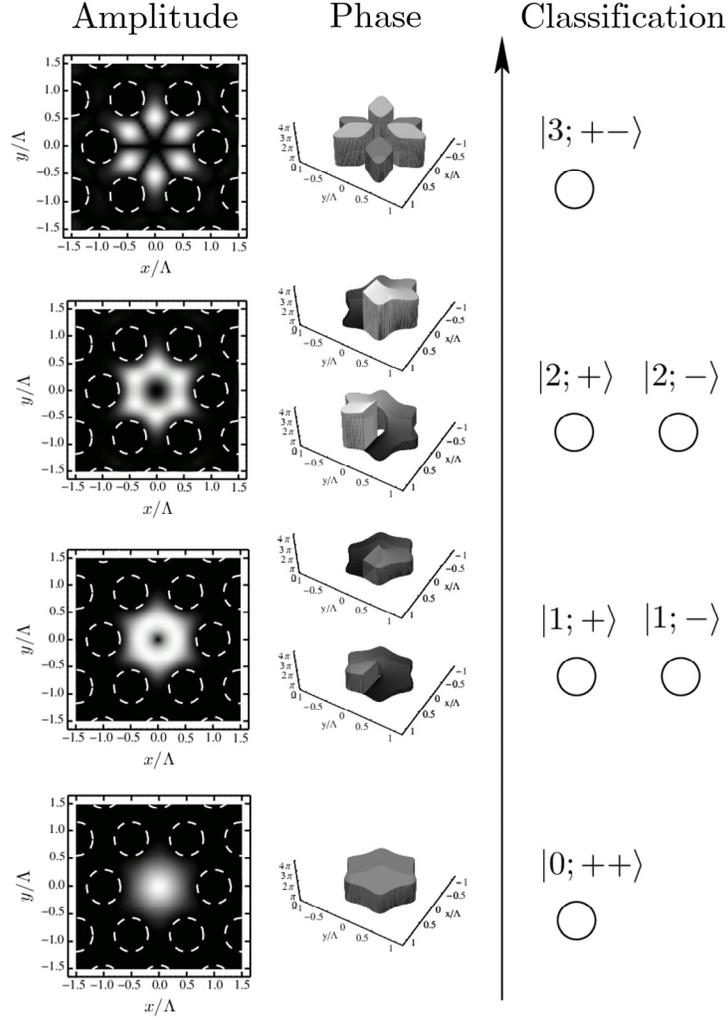

Figure 26. The lowest-order eigenfunctions of nonlinear operator $\mathcal{L}$ generated by the soliton solution in the fundamental representation of $\mathcal{C}_{6v}$, in the model of the PCF with the respective symmetry of the intrinsic structure. The symmetry of the full operator is $\mathcal{C}_{6v}$, i.e., $[\mathcal{L}, \mathcal{C}_{6v}] = 0$. Modes in the two middle rows correspond to vortices with charges $\pm 1$ and $\pm 2$ [from Ferrando, Zacarés, and Garcia-March, (2005)].

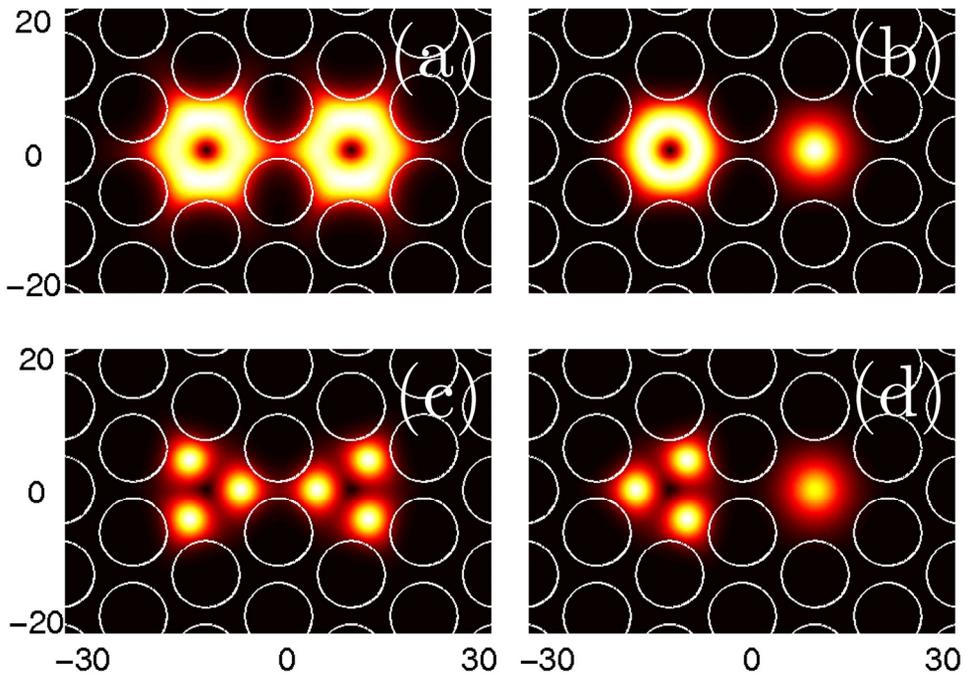

Figure 27. (Color online) Examples of different types of vortex-like solitons in the dual-core PCF: (a) a double-vortex state; (b) a combined state of vortex and fundamental solitons; (c) a double-triple vortex state; (d) a combined state of a triple vortex and fundamental soliton [from Salgueiro and Kivshar, (2009)].

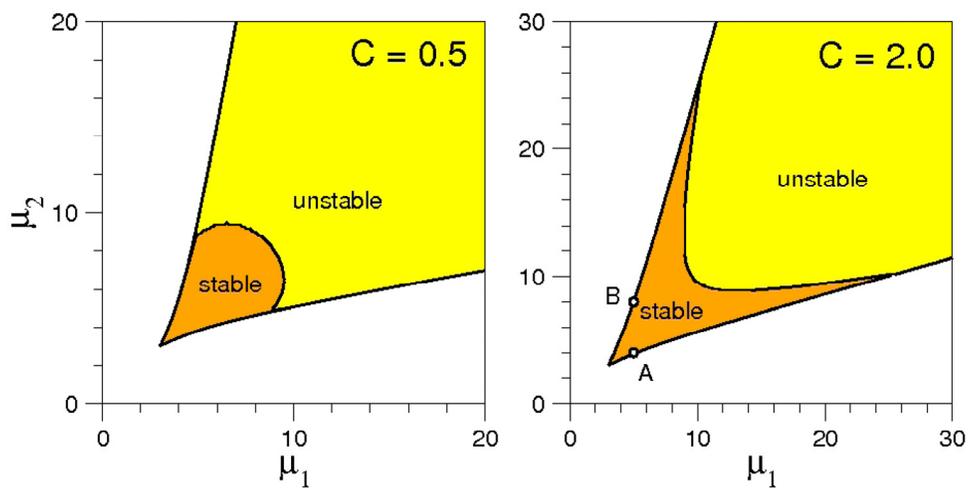

Figure 28. (Color online) The existence domain for the vector solitons in the PCF model is shown in the plane of $(\mu_1, \mu_2)$, for two values of coupling constant $C$ [from Salgueiro et al., (2005)].

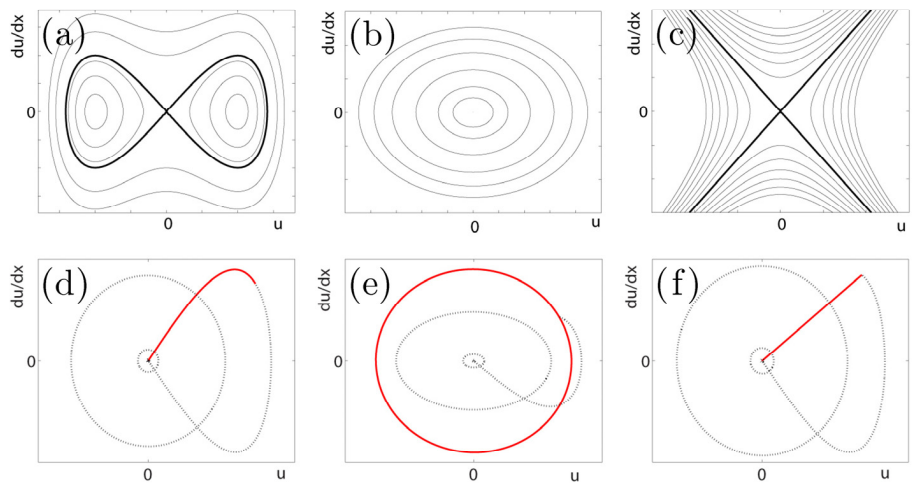

Figure 29. (Color online) (a)-(c) The phase space for each part of the structure (in the one-dimensional model based on the Kronig-Penney lattice with the intrinsic surface), as per Kominis, Papadopoulos and Hizanidis, (2007): (a) the nonlinear part for $\mu > \varepsilon_1$; (b) the linear part for $\mu < \varepsilon_i$ ($i=2$ or 3); (c) the linear part for $\mu > \varepsilon_i$ ($i=2$ or 3). (d)-(f) The phase-space representation of the soliton solutions for $n$ even. (d) The nonlinear homogeneous part; (e) the linear homogeneous part at $\mu < \varepsilon_3$; (f) the linear homogeneous part at $\mu > \varepsilon_3$. The dotted line denotes the solution in the lattice part, and the solid line denotes the solution in the uniform part.

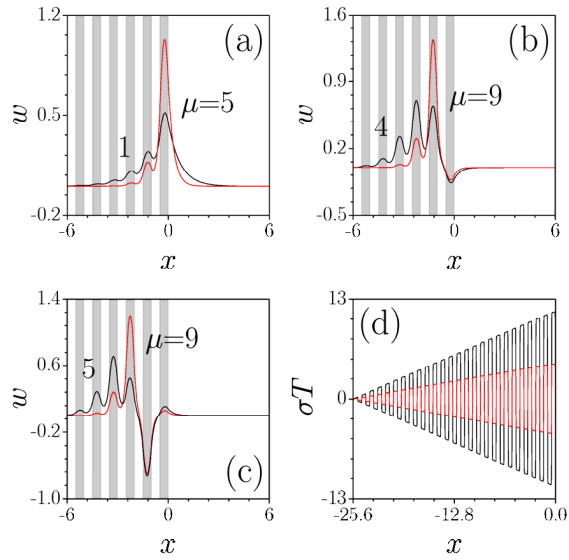

Figure 30.  (Color online) Profiles of (a) fundamental, (b) dipole-mode, and (c) tri-pole solitons with different values of $\mu$, in the model of the thermal layered medium with a surface. (d) Distributions of the refractive index for fundamental solitons with $\mu=5$ (black curve) and $\mu=1$ (red curve). In all the cases, the sign of the nonlinearity alternates between different layers of the thermal medium. In gray regions, $\sigma>0$ holds [from Kartashov, Vysloukh, and Torner, (2009b)].

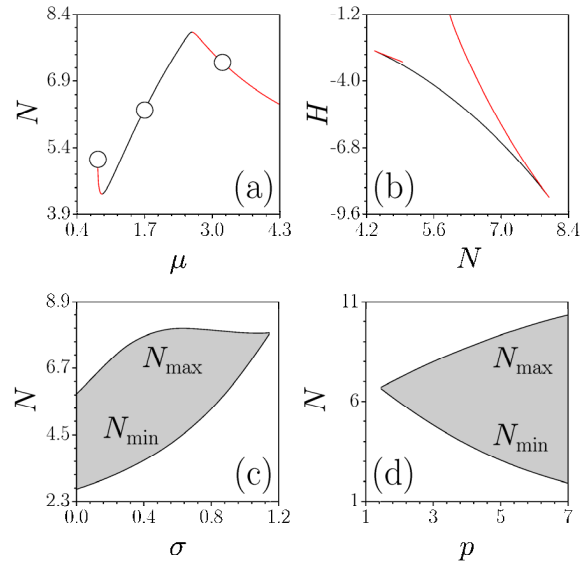

Figure 31. (Color online) (a) The norm of the surface soliton residing at the interface of lattice with out-of-phase modulation of refractive index and nonlinearity versus the propagation constant; (b) the Hamiltonian versus the norm at $p=3$, $\sigma=0.6$. Black curves show stable soliton branches, while red curves correspond to unstable ones. Stability domains for surface solitons: (c) in the plane of $(\sigma, N)$ at $p=3$; (d) in the plane of $(p, N)$ at $\sigma=0.7$ [from Kartashov *et al.*, (2008)].

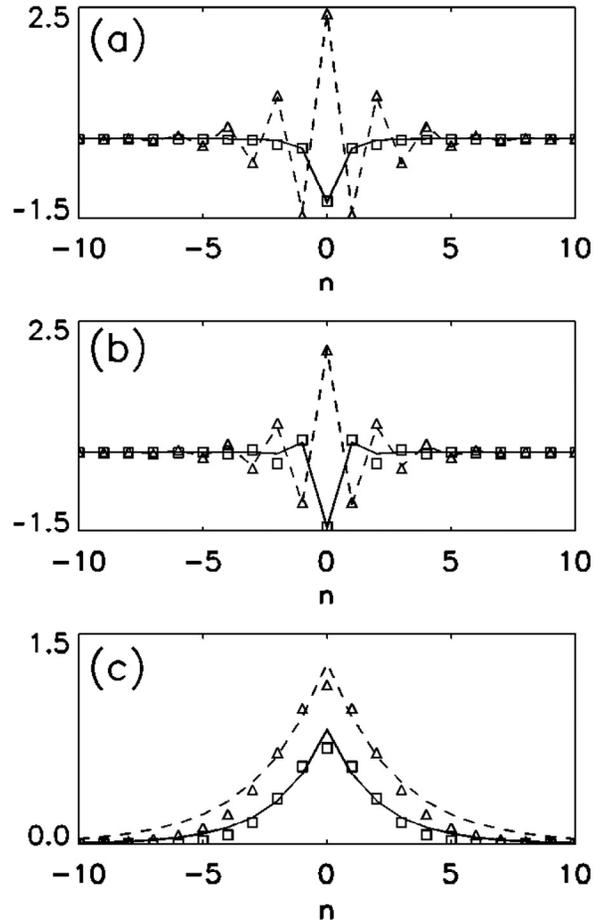

Figure 32. Typical examples of stable on-site-centered (alias odd) discrete solitons generated by Eqs. (56), in the model of the lattice with the $\chi^{(2)}$ nonlinearity. Triangles and squares show numerically found profiles of the fundamental and second harmonics, respectively. The dashed and continuous curves represent the respective profiles as predicted by the variational approximation. (a) Solitons with staggered fundamental and unstaggered second-harmonic components; (b) both components staggered; (c) both unstaggered [from Sukhorukov et al., (2000)].

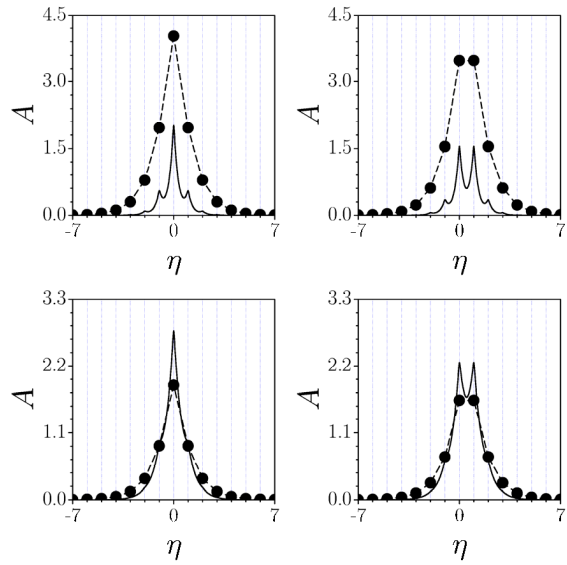

Figure 33. The top and bottom panels display typical examples of stable *semi-discrete* $\chi^{(2)}$ solitons generated by Eqs. (57) and (58), respectively, whereas the left and right panels show the odd and even species of the solitons. Vertical lines designate the location of the discrete waveguides in the system. These examples pertain to zero mismatch, [from Panoiu, Osgood and Malomed (2006)].

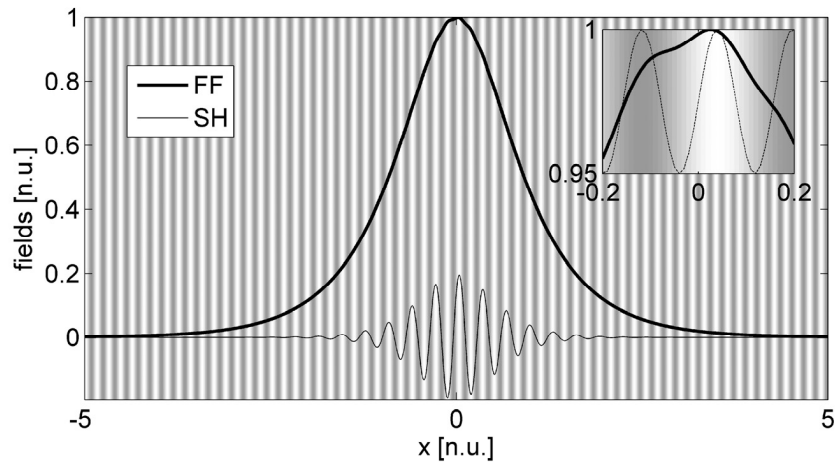

Figure 34. A soliton generated by Eqs. (59) in the case of the large wavenumber of the nonlinear grating, $\gamma = 20$, and large mismatch, $\alpha = 10$, in the model of the photonic crystal with the $\chi^{(2)}$ nonlinearity. The grating is represented by the gray pattern in the background. The profile of the SH component follows the form of the nonlinear grating, while the FF component features a sech-type profile, with only small distortions induced by the grating, as additionally shown in the inset, where the grating is indicated by the dashed sinusoid [from Pasquazi and Assanto (2009)].

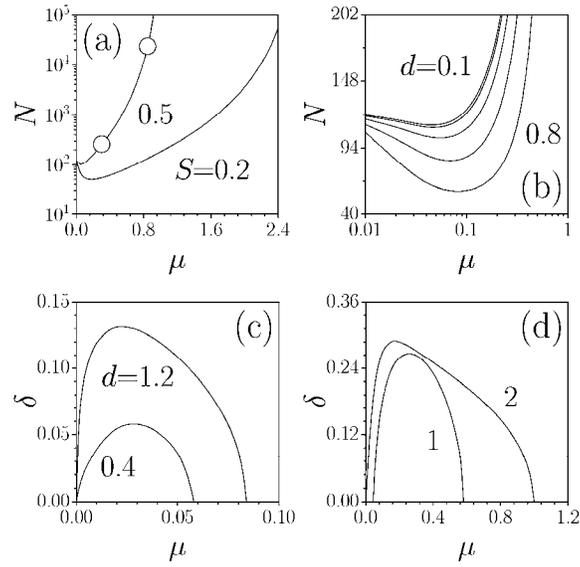

Figure 35. The norm versus the propagation constant for (a) different values of $S$ at $d = 0.4$ and (b) different values of $d$ at $S = 0.5$, in the model of the radial tandem structure supporting "light bullets". In panel (b), values of the domain's width are $d = 0.8$, $0.6$, $0.4$, $0.2$, and $0.1$, from the lower to upper curve. (c) The perturbation growth rate versus $\mu$, for the azimuthal perturbation index $k = 0$ and $S = 0.5$. In panels (a)-(c), the central domain is linear. (d) The growth rate versus $\mu$ at $k = 1$, $d = 1.2$, $S = 0.5$ in the structure with linear (1) and nonlinear (2) central domains [from Torner and Kartashov (2009)].